\pdfoutput=1
\documentclass[prd,onecolumn,notitlepage,eqsecnum,nofootinbib]{revtex4-1}
\usepackage{amssymb}
\usepackage{amsmath}
\usepackage{amsfonts} 
\usepackage{bm}
\usepackage{graphicx}
\usepackage{comment} 
\newcommand{\A}{{\cal A}}

\newcommand{\E}{{\cal E}} 
\newcommand{\B}{{\cal B}} 
\newcommand{\stf}[1]{{\langle {#1} \rangle}}
\newcommand{\Aa}[1]{{\cal A}^{\scriptscriptstyle \sf #1}} 
\newcommand{\EE}[1]{{\cal E}^{\scriptscriptstyle \sf #1}} 
\newcommand{\EEbar}[1]{\bar{\cal E}^{\scriptscriptstyle \sf #1}} 
  
\newcommand{\EEd}[1]{\dot{\cal E}^{\scriptscriptstyle \sf #1}}  
  
\newcommand{\EEdd}[1]{\ddot{\cal E}^{\scriptscriptstyle \sf #1}} 
\newcommand{\BB}[1]{{\cal B}^{\scriptscriptstyle \sf #1}} 
\newcommand{\BBbar}[1]{\bar{\cal B}^{\scriptscriptstyle \sf #1}} 
  
\newcommand{\BBd}[1]{\dot{\cal B}^{\scriptscriptstyle \sf #1}}  
  
\newcommand{\BBdd}[1]{\ddot{\cal B}^{\scriptscriptstyle \sf #1}} 
\newcommand{\PP}[1]{{\cal P}^{\scriptscriptstyle \sf #1}}

\newcommand{\QQ}[1]{{\cal Q}^{\scriptscriptstyle \sf #1}}

\newcommand{\GG}[1]{{\cal G}^{\scriptscriptstyle \sf #1}}

\newcommand{\HH}[1]{{\cal H}^{\scriptscriptstyle \sf #1}}

\newcommand{\ee}[2]{e^{\scriptscriptstyle \sf #1}_{#2}} 
\newcommand{\eedot}[2]{\dot{e}^{\scriptscriptstyle \sf #1}_{#2}} 
\newcommand{\eeddot}[2]{\ddot{e}^{\scriptscriptstyle \sf #1}_{#2}} 
\newcommand{\bb}[2]{b^{\scriptscriptstyle \sf #1}_{#2}} 
\newcommand{\bbdot}[2]{\dot{b}^{\scriptscriptstyle \sf #1}_{#2}} 
\newcommand{\bbddot}[2]{\ddot{b}^{\scriptscriptstyle \sf #1}_{#2}} 
\newcommand{\pp}[2]{p^{\scriptscriptstyle \sf #1}_{#2}} 
\newcommand{\qq}[2]{q^{\scriptscriptstyle \sf #1}_{#2}} 
\newcommand{\g}[2]{g^{\scriptscriptstyle \sf #1}_{#2}} 
\newcommand{\hh}[2]{h^{\scriptscriptstyle \sf #1}_{#2}} 
\newcommand{\dilog}{\,\mbox{dilog}\,}
\newcommand{\ce}[2]{{\sf e}^{\scriptscriptstyle \sf #1}_{#2}} 
\newcommand{\cb}[2]{{\sf b}^{\scriptscriptstyle \sf #1}_{#2}} 
\newcommand{\cp}[1]{{\sf p}^{\scriptscriptstyle \sf #1}} 
\newcommand{\cq}[1]{{\sf q}^{\scriptscriptstyle \sf #1}} 
\newcommand{\cg}[1]{{\sf g}^{\scriptscriptstyle \sf #1}} 
\newcommand{\ch}[1]{{\sf h}^{\scriptscriptstyle \sf #1}} 
\newcommand{\pn}{\mbox{\sc pn}}  
\allowdisplaybreaks
\begin{document}
\title{Nonrotating black hole in a post-Newtonian tidal environment II} 
\author{Eric Poisson and Eamonn Corrigan}  
\affiliation{Department of Physics, University of Guelph, Guelph,
  Ontario, N1G 2W1, Canada} 
\date{May 30, 2018} 
\begin{abstract} 
In the first part of the paper we construct the metric of a tidally
deformed, nonrotating black hole. The metric is presented as an
expansion in powers of $r/b \ll 1$, in which $r$ is the distance to
the black hole and $b$ the characteristic length scale of the tidal
field --- the typical distance to the remote bodies responsible for
the tidal environment. The metric is expanded through order $(r/b)^4$
and written in terms of a number of tidal multipole moments, the
gravitoelectric moments $\E_{ab}$, $\E_{abc}$, $\E_{abcd}$ and the
gravitomagnetic moments $\B_{ab}$, $\B_{abc}$, $\B_{abcd}$. It differs
from the similar construction of Poisson and Vlasov in that the tidal
perturbation is presented in Regge-Wheeler gauge instead of the
light-cone gauge employed previously. In the second part of the 
paper we determine the tidal moments by matching the black-hole metric
to a post-Newtonian metric that describes a system of bodies with weak
mutual gravity. This extends the previous work of Taylor and Poisson
(paper I in this sequence), which computed only the leading-order
tidal moments, $\E_{ab}$ and $\B_{ab}$. The matching is greatly
facilitated by the Regge-Wheeler form of the black-hole metric, and
this motivates the work carried out in the first part of the
paper. The tidal moments are calculated accurately through the first
post-Newtonian approximation, and at this order they are independent
of the precise nature of the compact body. The moments therefore apply
equally well to a rotating black hole, or to a (rotating or
nonrotating) neutron star. As an application of this formalism, we
examine the intrinsic geometry of a tidally deformed event horizon,
and describe it in terms of a deformation function that represents a
quadrupolar and octupolar tidal bulge.    
\end{abstract} 
\pacs{  }
\maketitle

\section{Introduction and summary} 
\label{sec:intro} 

The tidal interaction between neutron stars in a close binary system
has recently been the subject of intense investigation, following the
observation \cite{flanagan-hinderer:08} that the tidal deformation of
each body could have a measurable impact on the emitted gravitational
waves. The effect depends on the tidal deformability of each neutron
star, and a large effort has been deployed to the computation of this
quantity for realistic models of neutron stars, and to ascertain the
importance of the tidal deformation on the gravitational-wave signal
\cite{hinderer-etal:10, baiotti-etal:11, vines-flanagan-hinderer:11,
  pannarale-etal:11, lackey-etal:12, damour-nagar-villain:12,
  read-etal:13, vines-flanagan:13, maselli-gualtieri-ferrari:13,
  lackey-etal:14, favata:14, yagi-yunes:14, wade-etal:14,
  bernuzzi-etal:15, lackey-etal:17, ferrari-gualtieri-maselli:12,
  maselli-etal:12, maselli-etal:13, hinderer-etal:16,
  steinhoff-etal:16, xu-lai:17, cullen-etal:17,
  harry-hinderer:18}. The recent observation of GW170817 by 
the LIGO and Virgo instruments \cite{GW170817:17}, with its first
attempt to measure the tidal deformability of neutron stars,
inaugurated a new era of gravitational-wave astronomy that is likely,
in the fullness of time, to reveal some aspects of nuclear matter
equation of state and neutron-star internal structure.  

Detailed modeling of the tidal dynamics of compact objects in general
relativity requires the precise specification of the tidal environment
in which the compact object resides. In a context in which the tidal
field is weak and varies slowly compared with the dynamical timescale
of the compact body, the tidal environment can be described in terms
of a number of tidal multipole moments. These come in two guises. The
gravitoelectric moments $\E_{ab}$, $\E_{abc}$, $\E_{abcd}$, and so on,
are produced by mass densities external to the compact body, and have
direct analogues in Newtonian gravity. The gravitomagnetic moments    
$\B_{ab}$, $\B_{abc}$, $\B_{abcd}$, and so on, are produced by external
mass currents, and have no analogues in Newtonian gravity. A
specification of the tidal environment amounts to a determination of
the tidal moments in terms of the state of motion of the two-body
system. This task is the central concern of this paper.  

\begin{figure} 
\includegraphics[width=0.6\linewidth]{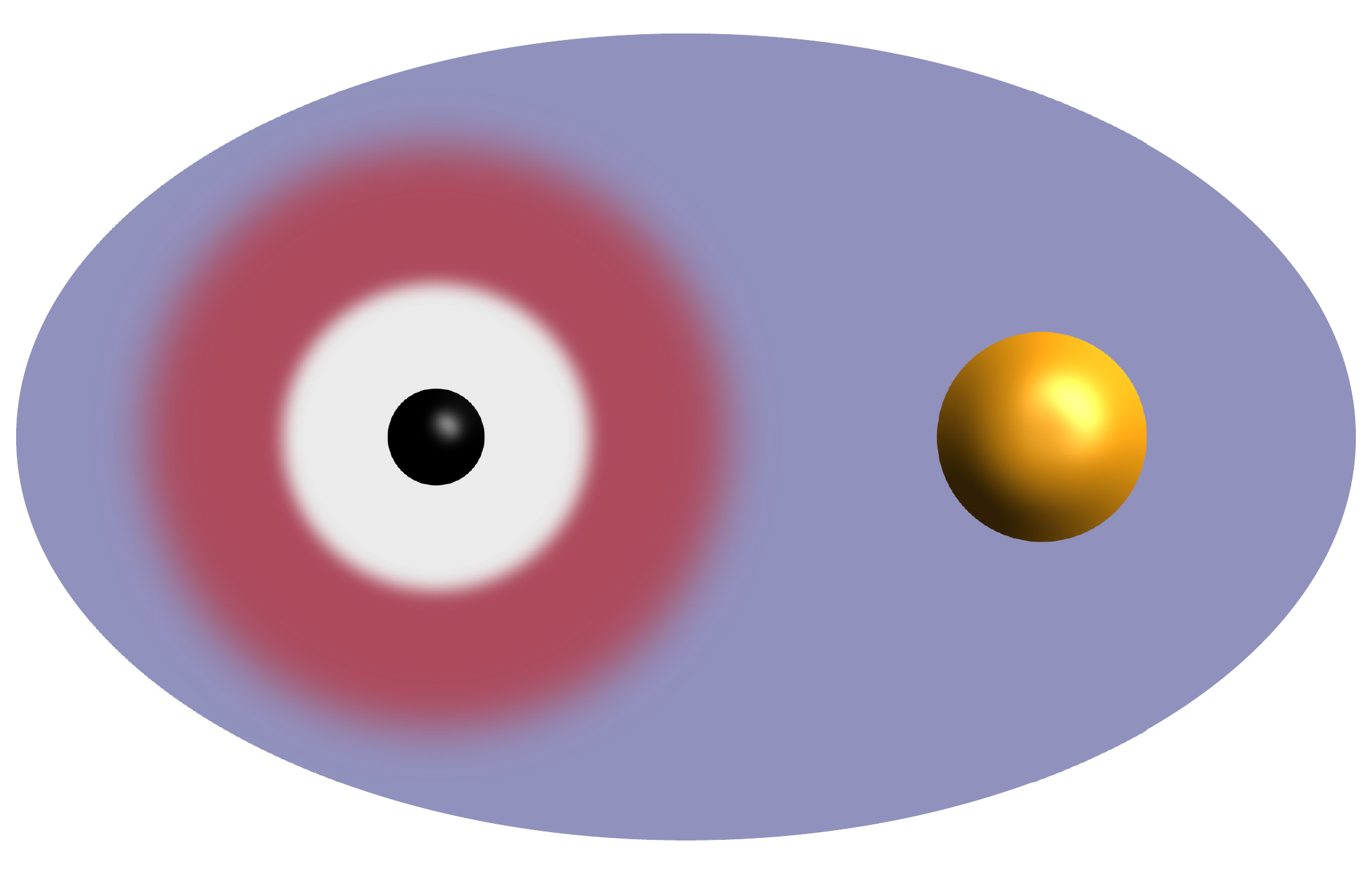}
\caption{A post-Newtonian system consisting of a black hole (left,
  black) and a normal star (right, yellow online). The post-Newtonian
  domain is pictured as an ellipse (blue online), and it excludes the
  fuzzy white region surrounding the black hole. The black-hole domain
  is pictured as the dark fuzzy region (red online), which extends all
  the way down to the black hole. The matching of the black-hole and
  post-Newtonian metrics is carried out in the overlap between the
  black-hole and post-Newtonian domains.}  
\label{fig:domains} 
\end{figure}

A method to determine tidal moments, in a context in which the compact
object is a member of a post-Newtonian binary system, was developed by
Taylor and Poisson in Ref.~\cite{taylor-poisson:08}, the first paper
in this sequence --- hereafter referred to as Paper I. The method, a
generalization to compact objects of a previous implementation limited
to weakly self-gravitating bodies \cite{damour-soffel-xu:91,
  damour-soffel-xu:92, damour-soffel-xu:93}, relies on the matching of
two distinct metrics in an overlapping domain of validity. The
situation is illustrated in Fig.~\ref{fig:domains}.  

The first metric is that of a tidally deformed compact object,
presented as a perturbed version of the Schwarzschild metric, and
expressed in terms of the tidal multipole moments. This metric is
valid in a small domain that surrounds the compact body, and the tidal 
moments appear in it as freely specifiable functions of time. These
cannot be determined by integrating the Einstein field equations in
the small domain, because the companion body is excluded and the
integration cannot incorporate its precise influence on the
gravitational field of the compact object.    

The second metric is a post-Newtonian metric given in a large
domain that includes both bodies, but leaves out a small region around
the compact object, in which gravity is too strong to be adequately
captured by a post-Newtonian approximation. In this way, the internal
gravity of the compact object is allowed to be strong, while the
mutual gravity between bodies is assumed to be weak. Because the
domain excludes the compact body, the post-Newtonian metric also
contains freely specifiable functions of time.  

There exists an overlap between the small domain of the first metric
and the large domain of the second metric. Matching the metrics in
this overlap determines the tidal moments and the missing details of
the post-Newtonian metric. In this manner, the tidal environment of
the compact body is determined, up to a desired number of moments, and
up to a desired post-Newtonian order. In Paper I
\cite{taylor-poisson:08} the method was exploited to calculate the leading
order tidal moments, $\E_{ab}$ and $\B_{ab}$, through the first
post-Newtonian ($1\pn$) approximation. In this paper we extend this
calculation to the next two tidal moments, $\E_{abc}$ and $\E_{abcd}$,
as well as $\B_{abc}$ and $\B_{abcd}$; this calculation also is
carried out to $1\pn$ order. 

The multipole moments are defined precisely in Sec.~\ref{sec:moments},
and their scaling with the mass $M_2$ of the companion body, the
interbody distance $b$, and the orbital velocity $V$ is also described
in this section. For concreteness we choose the compact
object to be a nonrotating black hole of mass $M_1$. The precise
nature of the body, however, is of no consequence in the determination
of the tidal moments at $1\pn$ order: the effects of spin enter at
$1.5\pn$ order, and finite-size effects enter at $5\pn$ order. Our
moments, therefore, determine the tidal environment of a neutron star
just as well as that of a black hole, and these objects are allowed to
rotate.  

The construction of the metric of a tidally deformed black hole begins
in Sec.~\ref{sec:potentials} with the introduction of tidal potentials
constructed from the tidal moments. The metric is next obtained in
$(v,r,\theta,\phi)$ coordinates in Sec.~\ref{sec:metric1}, with $v$
denoting advanced time, and transformed to $(t,r,\theta,\phi)$
coordinates in Sec.~\ref{sec:metric2}. The goal is to obtain a metric
accurate through fourth order in an expansion in powers of $r/b$,
where $r$ is the distance to the black hole and $b$ is the interbody
distance. This metric incorporates terms that involve 
the tidal quadrupole moments $\E_{ab}$ and $\B_{ab}$ and their time 
derivatives, the octupole moments $\E_{abc}$ and $\B_{abc}$ and their
time derivatives, and the hexadecapole moments $\E_{abcd}$ and
$\B_{abcd}$. The nonlinearity of the field equation implies that
terms at order $(r/b)^4$ also include bilinear combinations 
of $\E_{ab}$ and $\B_{ab}$. To achieve all this we rely heavily on the
formalism of Poisson and Vlasov \cite{poisson-vlasov:10}, which
provides an essential foundation for this work. 

In fact, the metric of a tidally deformed black hole, accurate through
order $(r/b)^4$, was already constructed by Poisson and Vlasov, and in
principle, this metric could have been imported directly without
having to perform the additional work described in
Secs.~\ref{sec:metric1} and \ref{sec:metric2}. The need for this 
work comes from the fact that the Poisson-Vlasov metric is given in a
form that does not facilitate a matching with the post-Newtonian
metric. First, the metric is expressed in light-cone coordinates, and
its post-Newtonian expansion does not reduce to the standard
post-Newtonian form that is required for matching. Second, the
Poisson-Vlasov metric is expressed in terms of tidal moments that were
specifically defined to simplify the description of the deformed event
horizon; these definitions make the post-Newtonian expression of the
metric more complicated than it has to be.  

In our developments in Secs.~\ref{sec:metric1} and \ref{sec:metric2}, 
we endeavor to arrive at a form for the black-hole metric that will
simplify, to the fullest extent possible, the task of matching it to a
post-Newtonian metric. And we aim to achieve this not just at the
$1\pn$ order of the calculations carried out in this paper, but also
at higher post-Newtonian orders, in preparation for future work. In
this regard, the most important property of the black-hole metric is
that it becomes compatible with the standard form of the
post-Newtonian metric after expansion through $1\pn$ order. This is
achieved with two technical devices. First, the metric of the
unperturbed black hole, given by the Schwarzschild solution, is
presented in the harmonic radial coordinate $\bar{r}$, related to the
usual areal radius $r$ by $\bar{r} = r - M$. This ensures that the
unperturbed metric reduces to the standard post-Newtonian metric after
expansion. Second, the tidal perturbation is presented in
Regge-Wheeler gauge, which happily produces a perturbed metric that
continues to respect the standard post-Newtonian form. Such simplicity
could not be achieved with the Poisson-Vlasov metric. It also was not
achieved in Paper I \cite{taylor-poisson:08}, with a perturbed metric
presented initially in the light-cone gauge. Matching with the
post-Newtonian metric required a transformation of the black-hole
metric to harmonic coordinates, a technically demanding step that was
(in retrospect) unnecessary. Our new construction allows us to avoid
this step altogether, and provides a solid infrastructure that will
facilitate future extensions of this work. 

As stated previously, the metric of a tidally deformed black hole can
be expressed in terms of tidal multipole moments that appear as freely
specifiable functions of time. The moments, however, are not uniquely
defined, and they admit redefinitions that leave the form
of the metric unchanged up to integration constants. It was this
freedom that was exploited by Poisson and Vlasov
\cite{poisson-vlasov:10} to simplify the description of the perturbed
horizon. In this work we exploit it differently, to simplify the
post-Newtonian expansion of the metric. The redefinition of tidal
moments is explored in Sec.~\ref{sec:calibrations}, and sets of tidal
moments that belong to different ``calibrations'' are related to one
another.  

The task of matching the black-hole and post-Newtonian metrics begins
in Sec.~\ref{sec:PN} with the post-Newtonian expansion of the
black-hole metric and the extraction of the gravitational potentials
(a Newtonian potential $U$, a vector potential $U_a$, and a
post-Newtonian potential $\Psi$). The matching itself is carried out
in Sec.~\ref{sec:matching}. The final products are the tidal moments
expressed in terms of external potentials that represent the
gravitational field of the companion body. These are evaluated
concretely in Sec.~\ref{sec:2body}, and the tidal moments are finally
obtained in terms of the orbital degrees of freedom of the two-body
system. At this stage our task is complete: the quadrupole, octupole,
and hexadecapole tidal moments are all determined through the first
post-Newtonian order. We note that the octupole tidal moments were
previously given in Ref.~\cite{johnsonmcdaniel-etal:09} for the
specific case of circular orbits; our expressions agree with
theirs. The hexadecapole moments are presented here for the first
time.   

As an application of our results, in Sec.~\ref{sec:horizon} we examine
the intrinsic geometry of the tidally deformed event horizon. To
describe the deformation it is useful to introduce a fictitious
two-dimensional surface embedded in a three-dimensional flat
space, to describe this surface by the equation $r = 2M_1(1 
+ \varepsilon)$, and to choose the displacement function $\varepsilon$
in such a way that the embedded surface possesses the same intrinsic 
geometry as a two-dimensional cross-section of the event horizon. In
relativistic units in which $G = c =1$, used throughout the paper, we
find that $\varepsilon$ is given by 
\begin{align} 
\varepsilon &= \frac{1}{2} q_1^2 q_2 \biggl(\frac{M}{b}\biggr)^3 
\Biggl\{ -\biggl[ 1 + \frac{1}{2} q_1 V^2 \biggr] (3\cos^2\theta-1) 
+ 3 \biggl[ 1 + \frac{1}{2} (q_1 - 4)V^2 \biggr] \sin^2\theta \cos(2\psi) 
\nonumber \\ & \quad \mbox{} 
- \frac{3}{5} q_1 V^2 \sin\theta(5\cos^2\theta-1) \cos(\psi) 
+ q_1 V^2 \sin^3\theta\cos(3\psi) + 1.5\pn \Biggr\}, 
\end{align} 
where $M := M_1 + M_2$, $q_1 := M_1/M$, $q_2 := M_2/M$, 
$V = (M/b)^{1/2}$ is the orbital velocity, and $(\theta,\phi)$ are
polar angles that specify a position on the embedded surface, with the
polar axis taken to be normal to the orbital plane. The phase function
is defined by $\psi := \phi - \bar{\omega} v$, with $v$ representing
advanced-time on the horizon, and 
\begin{equation}  
\bar{\omega} = \sqrt{\frac{M}{b^3}} \biggl[ 1 
- \frac{1}{2}(3+q_1 q_2) V^2 + 2\pn \biggr]. 
\end{equation} 
is the angular frequency of the tidal field. 

The displacement function can be decomposed into a quadrupole term
that is independent of the phase $\psi$, and another quadrupole term
that oscillates at twice the frequency $\bar{\omega}$; both terms
originate from $\E_{ab}$. The remaining terms at $1\pn$ order consist
of octupole deformations that oscillate at once and three times the tidal
frequency; both terms originate from $\E_{abc}$, and they were
omitted in the earlier statement of this result given in Paper I. The
hexadecapole tidal moment $\E_{abcd}$ makes no appearance in
$\varepsilon$ at $1\pn$ order, and the gravitomagnetic moments are
excluded on the grounds that they have the wrong parity to contribute
to a scalar quantity such as $\varepsilon$. The tidal displacement
represents a bulge aligned with $\phi = \bar{\omega} v$. 

Some of the calculations presented in the main text rely on tables and 
technical developments relegated to three appendices. In
Appendix~\ref{sec:irreducible} we provide decompositions of Cartesian  
tensors into irreducible pieces, and in Appendix~\ref{sec:distance} we
record various derivatives of a distance function that appears in
gravitational potentials. In Appendix~\ref{sec:tables} we display a
large number of tables that contain definitions of various quantities
used in our calculations.

\section{Tidal moments and scales} 
\label{sec:moments} 

We consider a nonrotating black hole of mass $M_1$ immersed in a tidal
environment described by the gravitoelectric tidal moments $\E_{ab}$,
$\E_{abc}$, $\E_{abcd}$, and the gravitomagnetic tidal moments
$\B_{ab}$, $\B_{abc}$, $\B_{abcd}$; all tidal moments are defined as 
three-dimensional Cartesian tensors, and are functions of time
only. The gravitational field of the tidally deformed black hole is 
described by a metric $g_{\alpha\beta}$ presented as an expansion in
powers of $r/b \ll 1$, where $r$ is the distance to the black hole and
$b$ is the characteristic length scale of the tidal field --- the
typical distance to the remote bodies responsible for the tidal
environment. The tidal moments determine the behavior of the metric
when $r \gg M_1$, and their time dependence is arbitrary; it cannot 
be determined by integrating the field equations locally, in a
neighborhood of the black hole limited to $r \ll b$. To determine the
tidal moments, the local metric $g_{\alpha\beta}$ must be matched to
a global metric that extends beyond $r = b$ and includes the remote
bodies.  

The tidal moments can be loosely interpreted as describing the Weyl
curvature of the tidally deformed black hole in a distance interval
$M_1 \ll r \ll b$, in which the tidal field dominates over the hole's
own gravity. Formally, however, the tidal moments are defined in terms
of the Weyl tensor of a different, but related, spacetime. The metric
of this spacetime is the limit of $g_{\alpha\beta}$ when $M_1$ is taken
to zero while keeping the tidal moments fixed. The Weyl tensor and its
covariant derivatives are then computed for this metric, and their
components in local Lorentzian coordinates $(x^0, x^a)$ are evaluated
at $r = 0$, now describing a regular world line of the spacetime. 
Additional details regarding the definition of the tidal
moments are provided in Sec.~II of Ref.~\cite{poisson-vlasov:10}.            

The tidal quadrupole moments are defined by 
\begin{subequations}
\label{tidalmoment_2} 
\begin{align}  
\E_{ab} &:= \bigl( C_{0a0b} \bigr)^{\rm STF}, \\ 
\B_{ab} &:= \frac{1}{2} \bigl( \epsilon_{apq} C^{pq}_{\ \ b0}
\bigr)^{\rm STF}
\end{align} 
\end{subequations} 
in terms of the Weyl tensor $C_{\alpha\beta\gamma\delta}$ of the
related spacetime. Here $\epsilon_{abc}$ is the permutation symbol,
and the STF sign instructs us to symmetrize all free indices and
remove all traces. This operation is also indicated in an
angular-bracket notation, such as  
$A_\stf{abcd} := (A_{abcd})^{\rm STF}$. The tidal moments are
functions of coordinate time $x^0$ (either $t$ or advanced-time $v$),
and we use overdots to indicate differentiation with respect to this
coordinate. 

The tidal octupole moments are defined by 
\begin{subequations}
\label{tidalmoment_3} 
\begin{align}  
\E_{abc} &:= \bigl( C_{0a0b|c} \bigr)^{\rm STF}, \\ 
\B_{abc} &:= \frac{3}{8} \bigl( \epsilon_{apq} C^{pq}_{\ \ b0|c} 
\bigr)^{\rm STF}, 
\end{align} 
\end{subequations} 
in which a vertical stroke indicates covariant differentiation. These
also are functions of coordinate time, and time derivatives are again
indicated with overdots.   

The tidal hexadecapole moments are defined by 
\begin{subequations} 
\label{tidalmoment_4} 
\begin{align} 
\E_{abcd} &:= \frac{1}{2} \bigl( C_{0a0b|cd} \bigr)^{\rm STF}, \\
\B_{abcd} &:= \frac{3}{20} \bigl( \epsilon_{apq} C^{pq}_{\ \ b0|cd}  
\bigr)^{\rm STF}.
\end{align} 
\end{subequations} 
The numerical factors inserted in these equations are inherited from
Zhang's choice of normalization for the tidal moments \cite{zhang:86}. 

The tidal environment is characterized by an external mass scale $M_2$, a
distance scale $b$, a velocity scale $V \sim \sqrt{M/b}$, where $M :=
M_1 + M_2$, and an angular-velocity scale $\omega \sim V/b$. The
gravitoelectric moments scale as 
\begin{equation} 
\E_{ab} \sim \frac{M_2}{b^3}, \qquad 
\E_{abc} \sim \frac{M_2}{b^4}, \qquad 
\E_{abcd} \sim \frac{M_2}{b^5},
\end{equation} 
and the gravitomagnetic moments scale as 
\begin{equation} 
\B_{ab} \sim \frac{M_2 V}{b^3}, \qquad 
\B_{abc} \sim \frac{M_2 V}{b^4}, \qquad 
\B_{abcd} \sim \frac{M_2 V}{b^5}. 
\end{equation} 
Time derivatives of the gravitoelectric tidal moments scale as 
\begin{equation} 
\dot{\E}_{ab} \sim \frac{M_2 V}{b^4}, \qquad 
\ddot{\E}_{ab} \sim \frac{M_2 V^2}{b^5}, \qquad 
\dot{\E}_{abc} \sim \frac{M_2 V}{b^5},
\end{equation} 
while the derivatives of the gravitomagnetic moments scale as  
\begin{equation} 
\dot{\B}_{ab} \sim \frac{M_2 V^2}{b^4}, \qquad 
\ddot{\B}_{ab} \sim \frac{M_2 V^3}{b^5}, \qquad 
\dot{\B}_{abc} \sim \frac{M_2 V^2}{b^5}. 
\end{equation} 
We assume that $b \gg M_1$, so that the source of the tidal field 
is situated far from the black hole. This implies that 
$V^2 \ll (1+M_2/M_1)$, so that the orbital velocity is much smaller
than the speed of light. It also follows that 
\begin{equation} 
M_1 \omega \sim (1 + M_2/M_1)^{1/2} (M_1/b)^{3/2} \ll 1, 
\end{equation} 
so that the time scale $\omega^{-1}$ associated with variations of the
tidal field is much longer than $M_1$, the time scale associated
with the black hole. The tidal field is thus assumed to be weak and to
vary slowly.  

The metric of a tidally deformed black hole shall be presented as an 
expansion in powers of $r/b$, assuming that $r$, the distance to the
black hole, is much smaller than $b$, the distance scale of the tidal
field. The metric shall be expanded through order $(r/b)^4$, and
expressed in terms of tidal potentials constructed from the tidal
moments.   

\section{Tidal potentials} 
\label{sec:potentials} 

The metric of a tidally deformed black hole can be expressed in terms
of a number of tidal potentials that are constructed from the tidal
moments. Details are provided in Sec.~II of
Ref.~\cite{poisson-vlasov:10}, and we summarize the main points here.  

The tidal moments are combined with $\Omega^a := x^a/r$, with $r$
denoting the usual Euclidean distance, so as to form scalar and vector
potentials that form an irreducible representation of the rotation
group labeled by multipole order $\ell$. (Tensor potentials are not
required, by virtue of the gauge choice adopted in 
Secs.~\ref{sec:metric1} and \ref{sec:metric2}.) Each vector potential
is required to be purely transverse, in the sense of being orthogonal
to $\Omega^a$. The required potentials are displayed in
Table~\ref{tab:potentials}.   

A transformation from Cartesian coordinates $x^a$ to spherical
coordinates $(r,\theta,\phi)$ is effected by 
$x^a = r \Omega^a(\theta^A)$, in which $\Omega^a$ is now parameterized
by two polar angles $\theta^A = (\theta,\phi)$.  Explicitly, we have
that $\Omega^a 
= [\sin\theta\cos\phi,\sin\theta\sin\phi,\cos\theta]$. The Jacobian
matrix is given by 
\begin{equation} 
\frac{\partial x^a}{\partial r} = \Omega^a, \qquad 
\frac{\partial x^a}{\partial \theta^A} = r \Omega^a_A, 
\label{Jacob} 
\end{equation} 
with $\Omega^a_A := \partial \Omega^a/\partial \theta^A$. We have that 
$\Omega_{AB} := \delta_{ab} \Omega^a_A \Omega^b_B 
= \mbox{diag}[1,\sin^2\theta]$ is the metric on the
unit two-sphere, and $\Omega^{AB}$ is its inverse. The inverse of the
Jacobian matrix is 
\begin{equation} 
\frac{\partial r}{\partial x^a} = \Omega_a, \qquad 
\frac{\partial \theta^A}{\partial x^a} = \frac{1}{r} \Omega^A_a, 
\label{Jacob_inverse} 
\end{equation} 
where $\Omega^A_a := \Omega^{AB} \delta_{ab} \Omega^b_B$. 

We introduce $D_A$ as the covariant-derivative operator compatible
with $\Omega_{AB}$, and $\epsilon_{AB}$ as the Levi-Civita tensor on
the unit two-sphere (with nonvanishing components
$\epsilon_{\theta\phi} = -\epsilon_{\phi\theta} = \sin\theta$). We
adopt the convention that uppercase Latin indices are raised and
lowered with $\Omega^{AB}$ and $\Omega_{AB}$, respectively. Finally,
we note that $D_C \Omega_{AB} = D_C \epsilon_{AB} = 0$.  

We convert the vector potentials from their initial Cartesian forms to
angular-coordinate versions by making use of the matrix
$\Omega^a_A$. We thus define  
\begin{equation} 
\BB{q}_A := \BB{q}_a \Omega^a_A,
\label{conversion} 
\end{equation} 
and apply the same rule to all other vector potentials. After this 
conversion the tidal potentials become scalar and vector fields on the
unit two-sphere, and they become independent of $r$. It is easy to
show that the conversion is reversed with 
$\BB{q}_a = \BB{q}_A \Omega^A_a$. 

The tidal potentials can all be expressed in terms of spherical
harmonics. Let $Y^{\ell m}$ be real-valued spherical-harmonic
functions (as defined in Table~\ref{tab:Ylm}). The relevant vectorial
and tensorial harmonics of even parity are 
\begin{subequations} 
\label{Ylm_even} 
\begin{align} 
& Y^{\ell m}_A := D_A Y^{\ell m}, \\
& Y^{\ell m}_{AB} := \Bigl[ D_A D_B 
+ \frac{1}{2} \ell(\ell+1) \Omega_{AB} \Bigr] Y^{\ell m};   
\end{align} 
\end{subequations}
notice that $Y^{\ell m}_{AB}$ is tracefree, 
$\Omega^{AB} Y^{\ell m}_{AB}= 0$, by virtue of the
eigenvalue  equation satisfied by the spherical harmonics. The
vectorial and tensorial harmonics of odd parity are 
\begin{subequations} 
\label{Ylm_odd} 
\begin{align} 
& X^{\ell m}_A := -\epsilon_A^{\ B} D_B Y^{\ell m}, \\
& X^{\ell m}_{AB} := -\frac{1}{2} \Bigl( \epsilon_A^{\ C} D_B 
+ \epsilon_B^{\ C} D_A \Bigr) D_C Y^{\ell m} = 0; 
\end{align} 
\end{subequations} 
the tensorial harmonics $X^{\ell m}_{AB}$ also are tracefree:
$\Omega^{AB} X^{\ell m}_{AB} = 0$. The decomposition of the tidal
potentials in spherical harmonics is presented in Tables~\ref{tab:E_ang},
\ref{tab:B_ang},  \ref{tab:EE_ang}, \ref{tab:BB_ang},
\ref{tab:EBeven_ang}, and \ref{tab:EBodd_ang}. 

\section{Metric in $(v,r,\theta,\phi)$ coordinates} 
\label{sec:metric1} 

In this section and the next we construct the metric of a tidally
deformed black hole, expanded through order $(r/b)^4$. In these
sections and the ones that follow, we denote the mass of the black hole
by $M$ instead of $M_1$. The original practice will be resumed in
Sec.~\ref{sec:2body}. 

In the standard $(v,r,\theta^A)$ coordinates, the metric of an
unperturbed black hole is given by the Schwarzschild solution, which
has the nonvanishing components  
\begin{equation} 
g_{vv} = -f, \qquad 
g_{vr} = 1, \qquad  
g_{AB} = r^2 \Omega_{AB}, 
\label{Schw_vr} 
\end{equation} 
where $f := 1-2M/r$. The tidal perturbation introduces additional
terms in the metric. These are constructed by incorporating a tidal
potential such as $\EE{q}(v,\theta^A)$, multiplied by a radial
function such as $\ee{q}{vv}(r)$, in the metric. All potentials
listed in Table~\ref{tab:potentials} participate, along with their
derivatives with respect to $v$. Including all such combinations
produces a complete metric ansatz that can then be inserted within the
Einstein field equations to determine the radial functions. 

The perturbation is presented in the Regge-Wheeler
gauge (see, for example, Ref.~\cite{martel-poisson:05} for definitions
and a review of black-hole perturbation theory), which requires
even-parity terms to be confined to the $vv$, $vr$, $rr$, and $AB$
components of the metric, and odd-parity terms to be confined to the
$vA$ and $rA$ components of the metric. In addition, the Regge-Wheeler
gauge requires $g_{AB}$ to be proportional to $\Omega_{AB}$. The gauge
is uniquely defined for $\ell \geq 2$, but it is not defined when
$\ell = 0$ and $\ell = 1$. The choices made in the monopole and dipole
cases will be described below.

The metric of a tidally deformed, nonrotating black hole takes the
form of
\begin{subequations} 
\label{blackhole_metric_vr} 
\begin{align} 
g_{vv} &= -f + r^2 \ee{q}{vv}\, \EE{q} + r^3 \eedot{q}{vv}\, \EEd{q} 
+ r^4 \eeddot{q}{vv}\, \EEdd{q} 
+ r^3 \ee{o}{vv}\, \EE{o} + r^4 \eedot{o}{vv}\, \EEd{o} 
+ r^4 \ee{h}{vv}\, \EE{h} 
+ r^4 \bigl( \pp{m}{vv}\, \PP{m} + \qq{m}{vv}\, \QQ{m} \bigr) 
\nonumber \\ & \quad \mbox{} 
+ r^4 \bigl( \pp{q}{vv}\, \PP{q} + \qq{q}{vv}\, \QQ{q} \bigr) 
+ r^4 \bigl( \pp{h}{vv}\, \PP{h} + \qq{h}{vv}\, \QQ{h} \bigr) 
+ r^4 \bigl( \g{d}{vv}\, \GG{d} + \g{o}{vv}\, \GG{o} \bigr) + O(r^5), \\ 
g_{vr} &= 1 + r^2 \ee{q}{vr}\, \EE{q} + r^3 \eedot{q}{vr}\, \EEd{q} 
+ r^4 \eeddot{q}{vr}\, \EEdd{q} 
+ r^3 \ee{o}{vr}\, \EE{o} + r^4 \eedot{o}{vr}\, \EEd{o} 
+ r^4 \ee{h}{vr}\, \EE{h} 
+ r^4 \bigl( \pp{m}{vr}\, \PP{m} + \qq{m}{vr}\, \QQ{m} \bigr) 
\nonumber \\ & \quad \mbox{} 
+ r^4 \bigl( \pp{q}{vr}\, \PP{q} + \qq{q}{vr}\, \QQ{q} \bigr) 
+ r^4 \bigl( \pp{h}{vr}\, \PP{h} + \qq{h}{vr}\, \QQ{h} \bigr) 
+ r^4 \bigl( \g{d}{vr}\, \GG{d} + \g{o}{vr}\, \GG{o} \bigr) + O(r^5), \\ 
g_{rr} &= r^2 \ee{q}{rr}\, \EE{q} + r^3 \eedot{q}{rr}\, \EEd{q} 
+ r^4 \eeddot{q}{rr}\, \EEdd{q} 
+ r^3 \ee{o}{rr}\, \EE{o} + r^4 \eedot{o}{rr}\, \EEd{o} 
+ r^4 \ee{h}{rr}\, \EE{h} 
+ r^4 \bigl( \pp{m}{rr}\, \PP{m} + \qq{m}{rr}\, \QQ{m} \bigr) 
\nonumber \\ & \quad \mbox{} 
+ r^4 \bigl( \pp{q}{rr}\, \PP{q} + \qq{q}{rr}\, \QQ{q} \bigr) 
+ r^4 \bigl( \pp{h}{rr}\, \PP{h} + \qq{h}{rr}\, \QQ{h} \bigr) 
+ r^4 \bigl( \g{d}{rr}\, \GG{d} + \g{o}{rr}\, \GG{o} \bigr) + O(r^5), \\ 
g_{vA} &= r^3 \bb{q}{v}\, \BB{q}_A + r^4 \bbdot{q}{v}\, \BBd{q}_A 
+ r^5 \bbddot{q}{v}\, \BBdd{q}_A 
+ r^4 \bb{o}{v}\, \BB{o}_A + r^5 \bbdot{o}{v}\, \BBd{o}_A 
+ r^5 \bb{h}{v}\, \BB{h}_A 
+ r^5 \bigl( \hh{q}{v}\, \HH{q}_A + \hh{h}{v}\, \HH{h}_A \bigr) 
+ O(r^6), \\ 
g_{rA} &= r^3 \bb{q}{r}\, \BB{q}_A + r^4 \bbdot{q}{r}\, \BBd{q}_A 
+ r^5 \bbddot{q}{r}\, \BBdd{q}_A 
+ r^4 \bb{o}{r}\, \BB{o}_A + r^5 \bbdot{o}{r}\, \BBd{o}_A 
+ r^5 \bb{h}{r}\, \BB{h}_A 
+ r^5 \bigl( \hh{q}{r}\, \HH{q}_A + \hh{h}{r}\, \HH{h}_A \bigr) 
+ O(r^6), \\ 
g_{AB} &= r^2 \Omega_{AB} \Bigl[ 1 + r^2 \ee{q}{}\, \EE{q} 
+ r^3 \eedot{q}{}\, \EEd{q} + r^4 \eeddot{q}{}\, \EEdd{q} 
+ r^3 \ee{o}{}\, \EE{o} + r^4 \eedot{o}{}\, \EEd{o} 
+ r^4 \ee{h}{}\, \EE{h} 
+ r^4 \bigl( \pp{m}{}\, \PP{m} + \qq{m}{}\, \QQ{m} \bigr) 
\nonumber \\ & \quad \mbox{} 
+ r^4 \bigl( \pp{q}{}\, \PP{q} + \qq{q}{}\, \QQ{q} \bigr) 
+ r^4 \bigl( \pp{h}{}\, \PP{h} + \qq{h}{}\, \QQ{h} \bigr) 
+ r^4 \bigl( \g{d}{}\, \GG{d} + \g{o}{}\, \GG{o} \bigr) + O(r^5) \Bigr],  
\end{align}
\end{subequations} 
where, for example, $\EE{q}$ is the tidal potential introduced in
Table~\ref{tab:E_ang}, and $\ee{q}{vv}$, $\ee{q}{vr}$, $\ee{q}{rr}$,
and $\ee{q}{}$ are the radial functions that come with it. An overdot
on a tidal potential indicates differentiation with respect to $v$ ---
the time dependence is contained in the tidal moments --- and the
radial functions attached to time-differentiated potentials are also
adorned with overdots (though these functions are not
differentiated with respect to $v$).   

The radial functions are determined by integrating the Einstein field
equations in vacuum. The solutions are presented in Tables
\ref{tab:e_functions_vr}, \ref{tab:b_functions_vr},
\ref{tab:pq_functions_vr}, and \ref{tab:gh_functions_vr}. Integration
for $\ell = 0$ and $\ell = 1$ requires some gauge
choices. For $\ell = 0$ we let $\pp{m}{vv}$ and $\pp{m}{}$ be the two
independent functions, and we set $\pp{m}{vr} = -\pp{m}{vv}/f$ and
$\pp{m}{rr} = 2\pp{m}{vv}/f^2$, a choice inspired from the structure
of the solutions for $\ell \geq 2$. For $\ell = 1$ we set
$\g{d}{}=0$, and find that $\g{d}{vv}=0$ as a consequence of the
field equations.  

The radial functions depend on a number of integration constants,
which are denoted with a sans-serif symbol such as $\ce{q}{1}$. (Other
constants of integration are fixed by demanding that the metric be
well behaved at $r=2M$. An exception to this rule concerns the terms
proportional to $\cp{m}$ and $\cq{m}$; this point is discussed below.)
These constants correspond to the freedom to redefine the tidal
moments according to  
\begin{subequations} 
\label{moment-transf1} 
\begin{align}  
\E_{ab} &\to \E_{ab} + \ce{q}{1}\, M \dot{\E}_{ab} 
+ \ce{q}{2}\, M^2 \ddot{\E}_{ab} 
+ \cp{q}\, M^2 \E_{p\langle a} \E^p_{\ b\rangle}  
+ \cq{p}\, M^2 \B_{p\langle a} \B^p_{\ b\rangle}, \\ 
\B_{ab} &\to \B_{ab} + \cb{q}{1}\, M \dot{\B}_{ab} 
+ \cb{q}{2}\, M^2 \ddot{\B}_{ab} 
+ \ch{q}\, M^2 \E_{p\langle a} \B^p_{\ b\rangle}, \\ 
\E_{abc} &\to \E_{abc} + \ce{o}{1}\, M \dot{\E}_{abc} 
+ \cg{o}\, M \epsilon_{pq\langle a} \E^p_{\ b} \B^q_{\ c\rangle}, \\ 
\B_{abc} &\to \B_{abc} + \cb{o}{1}\, M \dot{\B}_{abc}
\end{align} 
\end{subequations} 
and 
\begin{subequations} 
\label{moment-transf2} 
\begin{align} 
2 \E_{abcd} &\to 2 \E_{abcd} 
+ \cp{h}\, \E_{\langle ab} \E_{cd \rangle} 
+ \cq{h}\, \B_{\langle ab} \B_{cd \rangle}, \\
\frac{10}{3} \B_{abcd} &\to \frac{10}{3} \B_{abcd} 
+ \ch{h}\, \E_{\langle ab} \B_{cd \rangle},  
\end{align} 
\end{subequations} 
as well as the freedom to redefine the mass parameter according to  
\begin{equation} 
M \to M + \cp{m} M^5 \E_{pq} \E^{pq} 
+ \cq{m} M^5 \B_{pq} \B^{pq}. 
\label{mass_transf} 
\end{equation} 
More precisely stated, the radial functions presented in the Tables
can be obtained from bare radial functions --- the same functions
with all integration constants set to zero --- by applying the
redefinitions of Eqs.~(\ref{moment-transf1}), (\ref{moment-transf2}),
and (\ref{mass_transf}).   

The freedom described by Eqs.~(\ref{moment-transf1}) can be used to
calibrate the tidal moments in a number of ways. For example, in 
Ref.~\cite{poisson-vlasov:10} the freedom to redefine the tidal moments
was exploited to ensure that in the light-cone gauge employed there, the
event horizon of the deformed black hole continues to be situated at
$r=2M$. It was also exploited to ensure that the intrinsic geometry of
the deformed horizon is independent of all $v$-derivatives of the
tidal moments. In this ``horizon calibration'', the integration
constants are fixed to  
\begin{subequations} 
\label{Hcal1} 
\begin{align} 
& \ce{q}{1} = -\frac{92}{15}, \qquad 
\ce{q}{2} = \frac{5569}{225}, \qquad 
\ce{o}{1} = -\frac{188}{21}, \qquad 
\cb{q}{1} = -\frac{76}{15}, \qquad 
\cb{q}{2} = \frac{18553}{1050}, \qquad  
\cb{o}{1} = -\frac{919}{105}, \\ 
& 
\cp{q} = -\frac{2}{7}, \qquad 
\cq{q} = \frac{18}{7}, \qquad  
\cg{o} = -10, \qquad 
\ch{q} = -\frac{44}{7}.  
\end{align} 
\end{subequations} 
The redefinitions of Eqs.~(\ref{moment-transf2}) do not involve the
black-hole mass $M$, and the freedom contained in these equations must  
be exploited to ensure that the tidal moments $\E_{abcd}$ and
$\B_{abcd}$ that appear in the metric are those that are actually
defined by Eqs.~(\ref{tidalmoment_4}). This is achieved with  
\begin{equation} 
\cp{h} = \frac{355}{3}, \qquad   
\cq{h} = -\frac{5}{3}, \qquad  
\ch{h} = -10. 
\label{Hcal2} 
\end{equation} 
Finally, the freedom described by Eq.~(\ref{mass_transf}) was also used
by Poisson and Vlasov \cite{poisson-vlasov:10} to simplify the
description of the event horizon. In this ``horizon calibration'' of
the mass parameter,     
\begin{equation} 
\cp{m} = 0, \qquad 
\cq{m} = 0. 
\label{Hcal3} 
\end{equation} 
The results of Eqs.~(\ref{Hcal1}), (\ref{Hcal2}), and (\ref{Hcal3})
can all be established by showing that the perturbation presented here
in Regge-Wheeler gauge can be obtained from the light-cone gauge of
Poisson and Vlasov by a gauge transformation; this requires these
specific values for the integration constants.    

The ``horizon calibration'' is convenient for the purposes of
examining the intrinsic geometry of the deformed horizon, but it may
not be convenient for other purposes. An alternative calibration,
designed to simplify the form of the radial functions, would be
obtained by setting all integration constants (except for 
$\cp{h}$, $\cq{h}$, and $\ch{h}$) to zero. Yet another choice is the
``post-Newtonian calibration''  to be introduced in
Sec.~\ref{sec:metric2}.    

Some of the radial functions listed in the Tables have factors of 
$f := 1-2M/r$ appearing in denominators, and this might suggest that
these functions are not regular at $r=2M$. Closer scrutiny, however,
reveals that the functions are in fact regular. It can indeed be
shown that $\eeddot{q}{vr} = O(f)$, $\eeddot{q}{rr} = O(1)$, and 
$\bbddot{q}{r} = O(1)$ when $r \to 2M$. The functions $\pp{m}{vr}$,
$\pp{m}{rr}$, $\qq{m}{vr}$, and $\qq{m}{rr}$, however, are genuinely
singular at $r=2M$. The singular terms in the metric come with the
integration constants $\cp{m}$ and $\cq{m}$, which represent the shift
in mass parameter described by Eq.~(\ref{mass_transf}). These
contributions to the monopole perturbation can easily be recast in a
nonsingular form with a gauge transformation. But the (singular) gauge
adopted here emerges as a natural choice when the metric is expressed
in the $(t,r,\theta^A)$ coordinates of Sec.~\ref{sec:metric2}. In any
event, the choices of Eq.~(\ref{Hcal3}) ensure that the singular terms
are eliminated from the metric.  

\section{Metric in $(t,r,\theta,\phi)$ coordinates} 
\label{sec:metric2} 

The metric in $(t,r,\theta^A)$ coordinates can be obtained from the
metric of Eq.~(\ref{blackhole_metric_vr}) by performing the coordinate
transformation  
\begin{equation} 
v = t + r \Delta, \qquad 
\Delta := 1 + \frac{2M}{r} \ln\biggl( \frac{r}{2M} - 1 \biggr), 
\end{equation} 
which implies that $dv = dt + f^{-1}\, dr$. In Sec.~\ref{sec:metric1},
a tidal potential such as $\EE{q}$ was considered to be a function of 
$v$, and it must now be re-expressed in terms of $t$. Given
our assumption that the tidal moments vary slowly, this can
be done with the help of a Taylor expansion, 
\begin{equation} 
\EE{q}(v) = \EE{q}(t) + r \Delta \EEd{q}(t) 
+ \frac{1}{2}  r^2 \Delta^2 \EEdd{q}(t) + O(r^3). 
\end{equation} 
After making such substitutions and performing the coordinate
transformation, we find that the metric of a tidally deformed,
nonrotating black hole can also be expressed as 
\begin{subequations} 
\label{blackhole_metric_tr} 
\begin{align} 
g_{tt} &= -f + r^2 \ee{q}{tt}\, \EE{q} + r^3 \eedot{q}{tt}\, \EEd{q} 
+ r^4 \eeddot{q}{tt}\, \EEdd{q} 
+ r^3 \ee{o}{tt}\, \EE{o} + r^4 \eedot{o}{tt}\, \EEd{o} 
+ r^4 \ee{h}{tt}\, \EE{h} 
+ r^4 \bigl( \pp{m}{tt}\, \PP{m} + \qq{m}{tt}\, \QQ{m} \bigr) 
\nonumber \\ & \quad \mbox{} 
+ r^4 \bigl( \pp{q}{tt}\, \PP{q} + \qq{q}{tt}\, \QQ{q} \bigr) 
+ r^4 \bigl( \pp{h}{tt}\, \PP{h} + \qq{h}{tt}\, \QQ{h} \bigr) 
+ r^4 \bigl( \g{d}{tt}\, \GG{d} + \g{o}{tt}\, \GG{o} \bigr) + O(r^5), \\ 
g_{tr} &= r^3 \eedot{q}{tr}\, \EEd{q} + r^4 \eeddot{q}{tr}\, \EEdd{q} 
+ r^4 \eedot{o}{tr}\, \EEd{o}  
+ r^4 \bigl( \g{d}{tr}\, \GG{d} + \g{o}{tr}\, \GG{o} \bigr) + O(r^5), \\ 
g_{rr} &= f^{-1} + r^2 \ee{q}{rr}\, \EE{q} + r^3 \eedot{q}{rr}\, \EEd{q} 
+ r^4 \eeddot{q}{rr}\, \EEdd{q} 
+ r^3 \ee{o}{rr}\, \EE{o} + r^4 \eedot{o}{rr}\, \EEd{o} 
+ r^4 \ee{h}{rr}\, \EE{h} 
+ r^4 \bigl( \pp{m}{rr}\, \PP{m} + \qq{m}{rr}\, \QQ{m} \bigr) 
\nonumber \\ & \quad \mbox{} 
+ r^4 \bigl( \pp{q}{rr}\, \PP{q} + \qq{q}{rr}\, \QQ{q} \bigr) 
+ r^4 \bigl( \pp{h}{rr}\, \PP{h} + \qq{h}{rr}\, \QQ{h} \bigr) 
+ r^4 \bigl( \g{d}{rr}\, \GG{d} + \g{o}{rr}\, \GG{o} \bigr) + O(r^5), \\ 
g_{tA} &= r^3 \bb{q}{t}\, \BB{q}_A + r^4 \bbdot{q}{t}\, \BBd{q}_A 
+ r^5 \bbddot{q}{t}\, \BBdd{q}_A 
+ r^4 \bb{o}{t}\, \BB{o}_A + r^5 \bbdot{o}{t}\, \BBd{o}_A 
+ r^5 \bb{h}{t}\, \BB{h}_A 
+ r^5 \bigl( \hh{q}{t}\, \HH{q}_A + \hh{h}{t}\, \HH{h}_A \bigr) 
+ O(r^6), \\ 
g_{rA} &= r^4 \bbdot{q}{r}\, \BBd{q}_A 
+ r^5 \bbddot{q}{r}\, \BBdd{q}_A + r^5 \bbdot{o}{r}\, \BBd{o}_A  
+ r^5 \bigl( \hh{q}{r}\, \HH{q}_A + \hh{h}{r}\, \HH{h}_A \bigr) 
+ O(r^6), \\ 
g_{AB} &= r^2 \Omega_{AB} \Bigl[ 1 + r^2 \ee{q}{}\, \EE{q} 
+ r^3 \eedot{q}{}\, \EEd{q} + r^4 \eeddot{q}{}\, \EEdd{q} 
+ r^3 \ee{o}{}\, \EE{o} + r^4 \eedot{o}{}\, \EEd{o} 
+ r^4 \ee{h}{}\, \EE{h} 
+ r^4 \bigl( \pp{m}{}\, \PP{m} + \qq{m}{}\, \QQ{m} \bigr) 
\nonumber \\ & \quad \mbox{} 
+ r^4 \bigl( \pp{q}{}\, \PP{q} + \qq{q}{}\, \QQ{q} \bigr) 
+ r^4 \bigl( \pp{h}{}\, \PP{h} + \qq{h}{}\, \QQ{h} \bigr) 
+ r^4 \bigl( \g{d}{}\, \GG{d} + \g{o}{}\, \GG{o} \bigr) + O(r^5) \Bigr],  
\end{align}
\end{subequations} 
where all tidal potentials are now given as functions of $t$. The
new radial functions are listed in Tables
\ref{tab:e_functions_tr}, \ref{tab:b_functions_tr}, 
\ref{tab:pq_functions_tr}, and \ref{tab:gh_functions_tr}.

The freedom to redefine the tidal moments was introduced in
Eqs.~(\ref{moment-transf1}), (\ref{moment-transf2}), and
(\ref{mass_transf}), and the ``horizon calibration'' of these 
moments was introduced in Eqs.~(\ref{Hcal1}) and (\ref{Hcal3}). For
our purposes in Sec.~\ref{sec:PN}, it is convenient to adopt an
alternative ``post-Newtonian calibration'' that ensures that all
radial functions begin an expansion in powers of $M/r$ with the
largest power possible. For example, with a generic value for
$\ce{q}{1}$, the function $\eedot{q}{tt}$ possesses an expansion that
begins at order $M/r$. Setting $\ce{q}{1} = -52/15$ eliminates this
leading term, as well as terms of order $(M/r)^2$ and $(M/r)^3$, and
leaves an expansion that begins at order $(M/r)^6$. This choice of
integration constant therefore defines the post-Newtonian calibration
for this radial function.   

Adapting the procedure to all other radial functions, we find that the
post-Newtonian calibration is achieved with  
\begin{equation} 
\ce{q}{1} = -\frac{52}{15}, \qquad 
\ce{o}{1} = -\frac{30}{7}, \qquad 
\cb{q}{1} = -\frac{29}{10}, \qquad 
\cb{o}{1} = -\frac{293}{70}, \qquad   
\cg{o} = 0.  
\label{PNcal1} 
\end{equation} 
The values of $\cp{h}$, $\cq{h}$, and $\ch{h}$ continue to be fixed by 
Eq.~(\ref{Hcal2}), 
\begin{equation} 
\cp{h} = \frac{355}{3}, \qquad   
\cq{h} = -\frac{5}{3}, \qquad  
\ch{h} = -10. 
\label{PNcal2} 
\end{equation} 
And because the remaining constants do not affect the leading terms in
expansions of the radial functions in powers of $M/r$, we set them to
zero for simplicity: 
\begin{equation} 
\ce{q}{2} = \cb{q}{2} = \cp{m} = \cp{q} = \cq{m} = \cq{q} = \ch{q} 
= 0. 
\label{PNcal3} 
\end{equation} 

\section{Relation between tidal moments} 
\label{sec:calibrations} 

Tidal moments corresponding to a calibration $\bm{\mu}$, where
$\bm{\mu}$ collectively denotes a specific set of integration constants 
$\{\ce{q}{1}, \ce{q}{2}, \cdots, \ch{q}, \ch{h}\}$, are obtained
from bare tidal moments through the redefinitions of
Eqs.~(\ref{moment-transf1}). We have that  
\begin{subequations} 
\begin{align}  
\E_{ab}(\bm{\mu}) &= \E_{ab}(\bm{0}) 
+ \ce{q}{1}\, M \dot{\E}_{ab}(\bm{0}) 
+ \ce{q}{2}\, M^2 \ddot{\E}_{ab}(\bm{0}) 
+ \cp{q}\, M^2 \E_{p\langle a} \E^p_{\ b\rangle}(\bm{0}) 
+ \cq{p}\, M^2 \B_{p\langle a} \B^p_{\ b\rangle}(\bm{0}), \\  
\B_{ab}(\bm{\mu}) &= \B_{ab}(\bm{0}) + \cb{q}{1}\, M \dot{\B}_{ab}(\bm{0})  
+ \cb{q}{2}\, M^2 \ddot{\B}_{ab}(\bm{0}) 
+ \ch{q}\, M^2 \E_{p\langle a} \B^p_{\ b\rangle}(\bm{0}),\\ 
\E_{abc}(\bm{\mu}) &= \E_{abc}(\bm{0}) + \ce{o}{1}\, M \dot{\E}_{abc}(\bm{0}) 
+ \cg{o}\, M \epsilon_{pq\langle a} \E^p_{\ b} \B^q_{\ c\rangle}(\bm{0}),\\ 
\B_{abc}(\bm{\mu}) &= \B_{abc}(\bm{0}) + \cb{o}{1}\, M \dot{\B}_{abc}(\bm{0}),  
\end{align} 
\end{subequations} 
where $\E_{ab}(\bm{\mu})$ and so on stand for the calibrated moments,
while $\E_{ab}(\bm{0})$ and so on stand for the bare moments. It follows
from these relations that tidal moments corresponding to different
calibrations can be related to each other, independently of the bare
moments. We have that 
\begin{subequations} 
\begin{align} 
\E_{ab}(\bm{\mu}_2) &= \E_{ab}(\bm{\mu}_1) 
+ [ \ce{q}{1}(2)-\ce{q}{1}(1) ] M \dot{\E}_{ab}(\bm{\mu}_1) 
+ \bigl\{ \ce{q}{2}(2)-\ce{q}{2}(1) 
- \ce{q}{1}(1) [ \ce{q}{1}(2)-\ce{q}{1}(1) ] \bigr\} 
  M^2 \ddot{\E}_{ab}(\bm{\mu}_1)
\nonumber \\ & \quad \mbox{} 
+ [ \cp{q}(2)-\cp{q}(1) ] M^2 
   \E_{p\langle a} \E^p_{\ b\rangle}(\bm{\mu}_1)   
+ [ \cq{q}(2)-\cp{q}(1) ] M^2 
   \B_{p\langle a} \B^p_{\ b\rangle}(\bm{\mu}_1), \\ 
\B_{ab}(\bm{\mu}_2) &= \B_{ab}(\bm{\mu}_1) 
+ [ \cb{q}{1}(2)-\cb{q}{1}(1) ] M \dot{\B}_{ab}(\bm{\mu}_1) 
+ \bigl\{ \cb{q}{2}(2)-\cb{q}{2}(1) 
- \cb{q}{1}(1) [ \cb{q}{1}(2)-\cb{q}{1}(1) ] \bigr\} 
  M^2 \ddot{\B}_{ab}(\bm{\mu}_1)
\nonumber \\ & \quad \mbox{} 
+ [ \ch{q}(2)-\ch{q}(1) ] M^2 
   \E_{p\langle a} \B^p_{\ b\rangle}(\bm{\mu}_1), \\ 
\E_{abc}(\bm{\mu}_2) &= \E_{abc}(\bm{\mu}_1) 
+ [ \ce{o}{1}(2)-\ce{o}{1}(1) ] M \dot{\E}_{abc}(\bm{\mu}_1) 
+ [ \cg{o}(2)-\cg{o}(1) ] M 
  \epsilon_{pq\langle a} \E^p_{\ b} \B^q_{\ c\rangle}(\bm{\mu}_1), \\ 
\B_{abc}(\bm{\mu}_2) &= \B_{abc}(\bm{\mu}_1) 
+ [ \cb{o}{1}(2)-\cb{o}{1}(1) ] M \dot{\B}_{abc}(\bm{\mu}_1), 
\end{align}
\end{subequations} 
where constants such as $\ce{q}{1}(1)$ are elements of $\bm{\mu}_1$,
while constants such as $\ce{q}{1}(2)$ are elements of
$\bm{\mu}_2$. With these rules it follows from Eqs.~(\ref{Hcal1}),
(\ref{Hcal2}), (\ref{Hcal3}), (\ref{PNcal1}), (\ref{PNcal2}), and
(\ref{PNcal3}) that the tidal moments in the horizon and
post-Newtonian calibrations are related by  
\begin{subequations} 
\label{H-PNcalibration} 
\begin{align} 
\E_{ab}(\mbox{H}) &= \E_{ab}(\mbox{PN})  
- \frac{8}{3} M \dot{\E}_{ab}(\mbox{PN}) 
+ \frac{1163}{75} M^2 \ddot{\E}_{ab}(\mbox{PN})
- \frac{2}{7} M^2 \E_{p\langle a} \E^p_{\ b\rangle}(\mbox{PN})   
+ \frac{18}{7} M^2 \B_{p\langle a} \B^p_{\ b\rangle}(\mbox{PN}), \\ 
\B_{ab}(\mbox{H}) &= \B_{ab}(\mbox{PN}) 
- \frac{13}{6} M \dot{\B}_{ab}(\mbox{PN}) 
+ \frac{23911}{2100} M^2 \ddot{\B}_{ab}(\mbox{PN})
- \frac{44}{7} M^2 \E_{p\langle a} \B^p_{\ b\rangle}(\mbox{PN}), \\ 
\E_{abc}(\mbox{H}) &= \E_{abc}(\mbox{PN}) 
- \frac{14}{3} M \dot{\E}_{abc}(\mbox{PN})
- 10 M \epsilon_{pq\langle a} \E^p_{\ b} \B^q_{\ c\rangle}(\mbox{PN}), \\ 
\B_{abc}(\mbox{H}) &= \B_{abc}(\mbox{PN}) 
- \frac{137}{30} M \dot{\B}_{abc}(\mbox{PN}). 
\end{align}
\end{subequations} 
We recall that $\E_{abcd}$ and $\B_{abcd}$ have a fixed calibration
determined by the definitions of Eqs.~(\ref{tidalmoment_4}).  

\section{Post-Newtonian potentials of a tidally deformed black hole}  
\label{sec:PN} 

In this section we obtain the post-Newtonian limit of the metric of
a tidally deformed black hole. We shall show that with a shift of the
radial coordinate and a transformation to Cartesian coordinates, it
can be expressed in the standard form (see, for example, Sec.~8.2 of
Ref.~\cite{poisson-will:14})
\begin{subequations} 
\label{metric_PN_cartesian1} 
\begin{align} 
g_{tt} &= -1 + 2U + 2(\Psi-U^2) + 2\pn, \\ 
g_{ta} &= -4U_a + 2\pn, \\ 
g_{ab} &= (1 + 2U) \delta_{ab} + 2\pn, 
\end{align} 
\end{subequations} 
in terms of a Newtonian potential $U$, a vector potential $U_a$, and a
post-Newtonian potential $\Psi$. This exercise prepares the way for
the matching calculation of Sec.~\ref{sec:matching}. 

The integration constants of Eqs.~(\ref{PNcal1}), (\ref{PNcal2}), and
(\ref{PNcal3}) --- which define the ``post-Newtonian calibration'' of
the tidal moments --- are inserted in the metric of
Eqs.~(\ref{blackhole_metric_tr}), which is then expressed as a
post-Newtonian expansion truncated after the first post-Newtonian
($1\pn$) order. The gravitoelectric tidal moments are themselves given
post-Newtonian expansions of the form 
\begin{subequations} 
\begin{align} 
\E_{ab} &= \E_{ab}(0\pn) + \E_{ab}(1\pn) + 2\pn, \\ 
\E_{abc} &= \E_{abc}(0\pn) + \E_{abc}(1\pn) + 2\pn,  \\ 
\E_{abcd} &= \E_{abcd}(0\pn) + \E_{abcd}(1\pn) + 2\pn, 
\end{align} 
\end{subequations} 
while the gravitomagnetic moments are truncated at their leading,
$1\pn$ order. The metric is expressed in terms of the harmonic radial
coordinate 
\begin{equation} 
\bar{r} = r - M, 
\end{equation} 
where $r$ is the usual areal coordinate employed previously; this
shift by $-M$ is a post-Newtonian correction. 

With all this the metric becomes 
\begin{subequations}
\label{metric_PN_angular} 
\begin{align} 
g_{tt} &= -1 + \biggl[ \frac{2M}{\bar{r}} - \bar{r}^2\, \EE{q}(0\pn) 
- \frac{1}{3} \bar{r}^3\, \EE{o}(0\pn) 
- \frac{1}{12} \bar{r}^4\, \EE{h}(0\pn) \biggr] 
\nonumber \\ & \quad \mbox{} 
+ \biggl[ -\frac{2M^2}{\bar{r}^2} + 2M\bar{r}\, \EE{q}(0\pn) 
+ \frac{2}{3} M \bar{r}^2\, \EE{o}(0\pn) 
+ \frac{1}{6} M \bar{r}^3\, \EE{h}(0\pn) 
- \frac{11}{42} \bar{r}^4\, \EEdd{q}(0\pn) 
\nonumber \\ & \quad \mbox{} 
- \bar{r}^2\, \EE{q}(1\pn) 
- \frac{1}{3} \bar{r}^3\, \EE{o}(1\pn) 
- \frac{1}{12} \bar{r}^4\, \EE{h}(1\pn)  
- \frac{1}{15} \bar{r}^4\, \PP{m} - \frac{2}{7} \bar{r}^4\, \PP{q} 
+ \frac{1}{3} \bar{r}^4\, \PP{h} \biggr] + O(\bar{r}^5, 2\pn), \\ 
g_{tr} &= -\frac{2}{3} \bar{r}^3\, \EEd{q}(0\pn) 
- \frac{1}{6} \bar{r}^4\, \EEd{o}(0\pn) + O(\bar{r}^5, 2\pn), \\ 
g_{rr} &= 1 + \biggl[ \frac{2M}{\bar{r}} - \bar{r}^2\, \EE{q}(0\pn)  
- \frac{1}{3} \bar{r}^3\, \EE{o}(0\pn) 
- \frac{1}{12} \bar{r}^4\, \EE{h}(0\pn) \biggr] 
+ O(\bar{r}^5, 2\pn), \\ 
g_{tA} &= \frac{2}{3} \bar{r}^3\, \BB{q}_A 
+ \frac{1}{4} \bar{r}^4\, \BB{o}_A 
+ \frac{1}{15} \bar{r}^5\, \BB{h}_A + O(\bar{r}^6, 2\pn), \\ 
g_{rA} &= 2\pn, \\ 
g_{AB} &= r^2 \Omega_{AB} + r^2 \Omega_{AB} 
\biggl[ \frac{2M}{\bar{r}} - \bar{r}^2\, \EE{q}(0\pn)  
- \frac{1}{3} \bar{r}^3\, \EE{o}(0\pn) 
- \frac{1}{12} \bar{r}^4\, \EE{h}(0\pn) \biggr] 
+ O(\bar{r}^7, 2\pn). 
\end{align} 
\end{subequations} 
In $g_{tt}$, the first set of terms within square brackets are of
Newtonian ($0\pn$) order, while the second set of terms are $1\pn$ 
corrections. It is understood that in the post-Newtonian terms, the
tidal potentials $\PP{m}$, $\PP{q}$, and $\PP{h}$ are constructed from
$\E_{ab}(0\pn)$. 

Conversion to Cartesian coordinates $x^a = \bar{r} \Omega^a(\theta^A)$
gives rise to the standard post-Newtonian form of the metric, as given
by Eqs.~(\ref{metric_PN_cartesian1}), with 
\begin{subequations} 
\label{metric_PN_cartesian2} 
\begin{align} 
U &= \frac{M}{\bar{r}} - \frac{1}{2} \bar{r}^2\, \EE{q}(0\pn)  
- \frac{1}{6} \bar{r}^3\, \EE{o}(0\pn) 
- \frac{1}{24} \bar{r}^4\, \EE{h}(0\pn) + O(\bar{r}^5) , \\ 
\Psi-U^2 &= -\frac{M^2}{\bar{r}^2} + M\bar{r}\, \EE{q}(0\pn) 
+ \frac{1}{3} M \bar{r}^2\, \EE{o}(0\pn) 
+ \frac{1}{12} M \bar{r}^3\, \EE{h}(0\pn) 
- \frac{11}{84} \bar{r}^4\, \EEdd{q}(0\pn) 
\nonumber \\ & \quad \mbox{} 
- \frac{1}{2} \bar{r}^2\, \EE{q}(1\pn) 
- \frac{1}{6} \bar{r}^3\, \EE{o}(1\pn) 
- \frac{1}{24} \bar{r}^4\, \EE{h}(1\pn)  
- \frac{1}{30} \bar{r}^4\, \PP{m} - \frac{1}{7} \bar{r}^4\, \PP{q} 
+ \frac{1}{6} \bar{r}^4\, \PP{h} + O(\bar{r}^5), \\ 
U_a &= -\frac{1}{6} \bar{r}^2\, \BB{q}_a 
- \frac{1}{16} \bar{r}^3\, \BB{o}_a 
- \frac{1}{60} \bar{r}^4\, \BB{h}_a
+ \frac{1}{6} \bar{r}^3\Omega_a\, \EEd{q}(0\pn) 
+ \frac{1}{24} \bar{r}^4\Omega_a\, \EEd{o}(0\pn)
+ O(\bar{r}^5). 
\end{align} 
\end{subequations} 
That the standard post-Newtonian form of the metric is recovered in
our calculation is the reason for employing $\bar{r}$ as a radial
coordinate; this form would {\it not} be achieved in the original
radial coordinate $r$. It should be noted that in spite of our use of
$\bar{r}$, the coordinates $x^a$ are {\it not} harmonic: the
potentials do not satisfy the harmonic condition $\partial_t U
+ \partial_a U^a = 0$. The reason for this, of course, is that the
metric perturbation is presented in Regge-Wheeler gauge instead of a
harmonic gauge. 

The square of the Newtonian potential $U$ can be calculated with the
help of the identity $\EE{q} \EE{q} = \frac{2}{15} \PP{m} 
+ \frac{4}{7} \PP{q} + \PP{h}$, and this reveals that the
post-Newtonian potential is given by 
\begin{equation} 
\Psi = -\frac{1}{2} \bar{r}^2\, \EE{q}(1\pn) 
- \frac{1}{6} \bar{r}^3\, \EE{o}(1\pn) 
- \frac{1}{24} \bar{r}^4\, \EE{h}(1\pn)  
-\frac{11}{84} \bar{r}^4\, \EEdd{q}(0\pn)
+ \frac{5}{12} \bar{r}^4\, \PP{h} + O(\bar{r}^5). 
\end{equation} 
Notice that $\Psi$ does not involve $\PP{m}$ and $\PP{q}$, which
have cancelled out in these manipulations. 

To obtain the final form of the potentials we insert the definitions
of the tidal potentials in terms of the tidal moments. This yields 
\begin{subequations} 
\label{PNpotentials} 
\begin{align} 
\bar{U} &= \frac{M}{\bar{r}} 
- \frac{1}{2} \bar{\E}_{ab}(0\pn)\, \bar{x}^a \bar{x}^b 
- \frac{1}{6} \bar{\E}_{abc}(0\pn)\, \bar{x}^a \bar{x}^b \bar{x}^c 
- \frac{1}{12} \bar{\E}_{abcd}(0\pn)\, 
  \bar{x}^a \bar{x}^b \bar{x}^c \bar{x}^d + O(\bar{r}^5), \\ 
\bar{U}_a &= -\frac{1}{6} \epsilon_{apq} \bar{x}^p \bar{\B}^q_{\ b} \bar{x}^b 
- \frac{1}{12} \epsilon_{apq} \bar{x}^p \bar{\B}^q_{\ bc} \bar{x}^b \bar{x}^c 
- \frac{1}{18} \epsilon_{apq} \bar{x}^p \bar{\B}^q_{\ bcd} 
    \bar{x}^b \bar{x}^c \bar{x}^d 
\nonumber \\ & \quad \mbox{} 
+ \frac{1}{6} \bar{x}_a \dot{\bar{\E}}_{bc}(0\pn) \bar{x}^b \bar{x}^c 
+ \frac{1}{24} \bar{x}_a \dot{\bar{\E}}_{bcd}(0\pn) \bar{x}^b \bar{x}^c \bar{x}^d 
+ O(\bar{r}^5), \\ 
\bar{\Psi} &= - \frac{1}{2} \bar{\E}_{ab}(1\pn)\, \bar{x}^a \bar{x}^b 
- \frac{1}{6} \bar{\E}_{abc}(1\pn)\, \bar{x}^a \bar{x}^b \bar{x}^c 
- \frac{1}{12} \bar{\E}_{abcd}(1\pn)\, 
    \bar{x}^a \bar{x}^b \bar{x}^c \bar{x}^d 
\nonumber \\ & \quad \mbox{} 
-\frac{11}{84} \bar{r}^2 \ddot{\bar{\E}}_{ab}(0\pn) \bar{x}^a \bar{x}^b 
+ \frac{5}{12} \bar{\E}_{\langle ab}\bar{\E}_{cd \rangle}(0\pn) 
    \bar{x}^a \bar{x}^b \bar{x}^c \bar{x}^d + O(\bar{r}^5). 
\end{align} 
\end{subequations} 
We have placed overbars on the potentials, tidal moments, and
coordinates in anticipation of the developments of
Sec.~\ref{sec:matching}; in this notation the tidal moments are
functions of time $\bar{t}$. The overbars indicate that the potentials,
tidal moments, and coordinates refer to the black hole's reference
frame, which is to be distinguished from the post-Newtonian
barycentric frame to be introduced below.  

It is straightforward to show that the potentials satisfy the
post-Newtonian field equations  
\begin{subequations} 
\begin{align} 
0 &= \bar{\nabla}^2 \bar{U}, \\ 
0 &= \bar{\nabla}^2 \bar{\Psi} 
+ 3 \partial_{\bar{t}\bar{t}} \bar{U} 
+ 4 \partial_{\bar{t}\bar{a}} \bar{U}^a, \\ 
0 &= \bar{\nabla}^2 \bar{U}_a 
- \partial_{\bar{t}\bar{a}} \bar{U} 
- \partial_{\bar{a}\bar{b}} \bar{U}^b. 
\end{align} 
\end{subequations} 
These reduce to the standard post-Newtonian equations (Sec.~8.2 of 
Ref.~\cite{poisson-will:14}) when the potentials satisfy the harmonic
condition $\partial_{\bar t} \bar{U} + \partial_{\bar{a}} \bar{U}^a = 0$.   

\section{Matching to a post-Newtonian metric} 
\label{sec:matching} 

The metric of a tidally deformed black hole was constructed in
Secs.~\ref{sec:metric1} and \ref{sec:metric2}, and it was expressed
in terms of tidal moments $\bar{\E}_{ab}$, $\bar{\E}_{abc}$,
$\bar{\E}_{abcd}$, $\bar{\B}_{ab}$, $\bar{\B}_{abc}$,
$\bar{\B}_{abcd}$. (See the remark regarding the overbar at the end of
Sec.~\ref{sec:PN}.) The tidal moments occur in the metric as freely specifiable
functions of time, and these cannot be determined by solving the Einstein 
field equations in a domain limited to the black hole's immediate
neighborhood. Their determination must instead rely on matching the
black-hole metric to another metric that incorporates all relevant
information regarding the black hole's remote environment. We achieve
this in this section, taking the black hole to be a member of a
post-Newtonian system containing any number of external bodies. In
this treatment, the black hole's internal gravity is allowed to be
strong, but the mutual gravity between black hole and external bodies
is assumed to be sufficiently weak to be adequately described by a
post-Newtonian expansion of the metric. The matching between the
black-hole and post-Newtonian metrics will determine the tidal
moments, which are calculated accurately through the first
post-Newtonian ($1\pn)$ approximation.  

\subsection{Barycentric potentials} 
\label{subsec:barycentric} 

The black hole is taken to be a member of a post-Newtonian system of
gravitating bodies. The metric is given by
Eq.~(\ref{metric_PN_cartesian1}), in terms of harmonic coordinates
$(t,x^a)$ attached to the system's barycenter, and in terms of
barycentric potentials $U$, $U_j$, and $\Psi$. The post-Newtonian
metric is valid in a domain that contains all bodies but leaves out a 
sphere of radius $\bar{r} \gg M$ surrounding the black hole; this
region is excluded because the black hole's internal gravity is
too strong to be adequately captured by a post-Newtonian
expansion. As illustrated in Fig.~\ref{fig:domains}, there exists an
overlap region in which the black-hole metric of
Sec.~\ref{sec:metric2} and the post-Newtonian metric are 
both valid; the matching of the metrics takes place in this region.  

It is helpful to define new post-Newtonian potentials $\psi$ and $X$
by the relation $\Psi = \psi + \frac{1}{2} \partial_{tt} X$. In the 
vacuum region between bodies, the potentials satisfy the
post-Newtonian field equations 
\begin{equation} 
\nabla^2 U = 0, \qquad \nabla^2 U_j = 0, \qquad 
\nabla^2 \psi = 0, \qquad \nabla^2 X = 2 U, 
\label{PNeqns} 
\end{equation} 
as well as the harmonic condition  
\begin{equation} 
\partial_t U + \partial_j U^j = 0. 
\label{harmonic} 
\end{equation} 
Each equation is linear, and a solution describing a black hole and a
collection of external bodies can be obtained by linear
superposition. We model the black hole as a post-Newtonian monopole of
mass $M$ at position $\bm{x} = \bm{r}(t)$ in the barycentric frame. 
We let $\bm{v} := d\bm{r}/dt$ be the black hole's velocity vector, and
$\bm{a} := d\bm{v}/dt$ be its acceleration vector. The potentials are
written as   
\begin{subequations} 
\label{PNpotentials1} 
\begin{align} 
U(t,\bm{x}) &= \frac{M}{s} + U_{\rm ext}(t,\bm{x}), \\ 
U^j(t,\bm{x}) &= \frac{M v^j}{s} + U^j_{\rm ext}(t,\bm{x}), \\ 
\psi(t,\bm{x}) &= \frac{M \mu}{s} + \psi_{\rm ext}(t,\bm{x}), \\ 
X(t,\bm{x}) &= M s + X_{\rm ext}(t,\bm{x}), 
\end{align}
\end{subequations} 
where $s := |\bm{x}-\bm{r}|$ is the Euclidean distance
between the black hole and a field point at $\bm{x}$, while 
$U_{\rm ext}$, $U^j_{\rm ext}$, $\psi_{\rm ext}$, and 
$X_{\rm ext}$ are the potentials created by the external
bodies; these separately satisfy the field equations of
Eq.~(\ref{PNeqns}) and the harmonic condition of
Eq.~(\ref{harmonic}). The arbitrary function $\mu(t)$ represents a
post-Newtonian correction to the mass parameter. It cannot be
determined by integrating the post-Newtonian field equations in a
domain that excludes the black hole, and must instead be obtained by 
matching the post-Newtonian metric with the black-hole metric of
Sec.~\ref{sec:metric2}.     

Inserting $\psi$ and $X$ in the expression for $\Psi$ returns 
\begin{equation} 
\Psi(t,\bm{x}) = -\frac{M}{2s^3} (\bm{v} \cdot \bm{s})^2 
+ \frac{M}{s} \biggl( \mu + \frac{1}{2} v^2 \biggr) 
- \frac{M}{2s} \bm{a} \cdot \bm{s} 
+ \Psi_{\rm ext}(t,\bm{x}), 
\label{PNpotentials2} 
\end{equation} 
where $\bm{s} := \bm{x}-\bm{r}$, $v^2 := \bm{v} \cdot \bm{v}$, and
$\Psi_{\rm ext} := \psi_{\rm ext} 
+ \frac{1}{2} \partial_{tt} X_{\rm ext}$. 

\subsection{Transformation to the black-hole frame} 

The post-Newtonian metric of the preceding subsection was presented in
coordinates $(t, x^a)$ attached to the barycentric frame of the entire
post-Newtonian system. On the other hand, the black-hole metric of 
Sec.~\ref{sec:metric2} was presented in
coordinates $(\bar{t},\bar{x}^a)$ that are attached to the black
hole's own reference frame, which is moving relative to the
barycentric frame. To match the metrics we must
transform the post-Newtonian potentials to the coordinates
$(\bar{t},\bar{x}^a)$, and compare the expressions with
Eqs.~(\ref{PNpotentials}).  

The systems $(t,x^a)$ and $(\bar{t},\bar{x}^a)$ are both
compatible with the standard form of the post-Newtonian metric, and
the coordinate transformation relating two such systems is presented
in Sec.~8.3 of Ref.~\cite{poisson-will:14}  (a summary of work
previously carried out in Ref.~\cite{racine-flanagan:05}). It should be
noted that while $(t, x^a)$ is a system of harmonic coordinates, the
$(\bar{t},\bar{x}^a)$ coordinates are not harmonic. The transformation
described in Ref.~\cite{poisson-will:14} applies to such situations, but
the reader should be warned that the summary of Box 8.2 refers
strictly to two systems of harmonic coordinates.   

The coordinate transformation is characterized by arbitrary functions 
$A(\bar{t})$, $H^a(\bar{t})$, $R^a(\bar{t})$, and
$\beta(\bar{t},\bar{x}^a)$ in addition to the black-hole position
vector $r^a(\bar{t})$, here expressed as a function of the barred time 
coordinate. The transformation is given by 
\begin{equation} 
t = \bar{t} + \bigl( A + v_a \bar{x}^a \bigr) + \beta + 3\pn, \qquad 
x^a = \bar{x}^a + r^a + \Bigl( H^a + H^a_{\ b}\, \bar{x}^b  
+ \frac{1}{2} H^a_{\ bc}\, \bar{x}^b \bar{x}^c \Bigr) + 2\pn, 
\end{equation} 
where $H_{ab} := \epsilon_{abc} R^c + \frac{1}{2} v_a v_b - (\dot{A} -
\frac{1}{2} v^2) \delta_{ab}$ and $H_{abc} := -\delta_{ab} a_c 
- \delta_{ac} a_b + \delta_{bc} a_a$, with $v^a := dr^a/d\bar{t}$ and 
$a^a := dv^a/d\bar{t}$; an overdot indicates differentiation
with respect to $\bar{t}$. The bracketed terms in the equation for $t$
represent a $1\pn$ adjustment to the time coordinate that impacts
the Newtonian potential $U$, while $\beta$ represents a $2\pn$
adjustment that impacts the vector and post-Newtonian potentials. The 
bracketed terms in the equation for $x^a$ are of $1\pn$ order. In the
following the acceleration vector is decomposed into Newtonian
and post-Newtonian terms, 
\begin{equation}     
a^a = a^a(0\pn) + a^a(1\pn) + 2\pn; 
\end{equation} 
the acceleration that appears in $H_{abc}$ can be truncated at
Newtonian order. 

The transformed potentials $\bar{U}$, $\bar{U}^j$, and $\bar{\Psi}$,
those of the black-hole frame, are expressed in terms of ``hatted
potentials'' $\hat{U}$, $\hat{U}^j$, and $\hat{\Psi}$, related to the
original potentials $U$, $U^j$, and $\Psi$ by equations of the form  
\begin{equation} 
\hat{U}(\bar{t},\bm{\bar{x}}) := 
U \bigl(t = \bar{t}, \bm{x}=\bm{\bar{x}} + \bm{r}(\bar{t}) \bigr); 
\end{equation} 
the hatted potentials are therefore equal to the original potentials
evaluated at the time $\bar{t}$ and position $\bm{\bar{x}} +
\bm{r}$. Because of the time dependence contained in
$\bm{r}(\bar{t})$, the time derivative of a hatted potential is
related by 
\begin{equation} 
\partial_{\bar{t}} \hat{U} = \partial_t U + v^a \partial_a U 
\label{diff_rule} 
\end{equation} 
to derivatives of the original potential; it is understood that the
right-hand side of this equation is evaluated at $t=\bar{t}$ and
$x^a = \bar{x}^a + r^a(\bar{t})$. Spatial derivatives are related by
$\partial_{\bar{a}} \hat{U} = \partial_a U$, where again the
right-hand side is evaluated at the new time and position. 

By virtue of these differentiation rules, the harmonic condition
satisfied by the hatted potentials takes the form  
\begin{equation} 
\partial_{\bar{t}} \hat{U} - v^a \partial_{\bar{a}} \hat{U} 
+ \partial_{\bar{a}} \hat{U}^a = 0. 
\label{harmonic_hatted} 
\end{equation} 
The field equations satisfied by the Newtonian and vector potentials
become 
\begin{equation} 
\bar{\nabla}^2 \hat{U} = 0, \qquad 
\bar{\nabla}^2 \hat{U}^j = 0,
\label{PNeqn_hatted1} 
\end{equation} 
while the field equation satisfied by $\Psi$, $\nabla^2 \Psi
- \partial_{tt} U = 0$, takes the new form   
\begin{equation} 
\bar{\nabla}^2 \hat{\Psi} - \partial_{\bar{t}\bar{t}} \hat{U} 
+ 2 v^a \partial_{\bar{t}\bar{a}} \hat{U}  
+ a^a \partial_{\bar{a}} \hat{U} 
- v^a v^b \partial_{\bar{a}\bar{b}} \hat{U} = 0. 
\label{PNeqn_hatted2} 
\end{equation} 
We recall that the harmonic condition and the field equations are
satisfied separately by the external potentials $\hat{U}_{\rm ext}$,
$\hat{U}^j_{\rm ext}$, and $\hat{\Psi}_{\rm ext}$.  

We take advantage of the large separation between the black hole and
all external bodies to express each external potential as a Taylor
expansion about $\bar{x}^a = 0$. We write, for example, 
\begin{align} 
\hat{U}_{\rm ext}(\bar{t},\bm{\bar{x}}) &= 
\hat{U}_{\rm ext}(\bar{t},\bm{0}) 
+ \bar{x}^a\, \partial_{\bar{a}} \hat{U}_{\rm ext}(\bar{t},\bm{0}) 
+ \frac{1}{2} \bar{x}^a \bar{x}^b\, 
   \partial_{\bar{a}\bar{b}} \hat{U}_{\rm ext}(\bar{t},\bm{0})  
+ \frac{1}{6} \bar{x}^a \bar{x}^b \bar{x}^c\, 
   \partial_{\bar{a}\bar{b}\bar{c}} \hat{U}_{\rm ext}(\bar{t},\bm{0})  
\nonumber \\ & \quad \mbox{} 
+ \frac{1}{24} \bar{x}^a \bar{x}^b \bar{x}^c \bar{x}^d\,   
   \partial_{\bar{a}\bar{b}\bar{c}\bar{d}} \hat{U}_{\rm ext}(\bar{t},\bm{0})   
+ O(\bar{r}^5),  
\end{align} 
where $\bar{r}:= |\bm{\bar{x}}|$. In all equations that appear
below, the external potentials and their derivatives shall be
evaluated at $\bar{x}^a = 0$. In a similar way we express the function
$\beta$ that appears in the coordinate transformation as a Taylor
expansion of the form  
\begin{equation} 
\beta(\bar{t},\bm{\bar{x}}) = 
\mbox{}_0 \beta(\bar{t}) 
+ \mbox{}_1 \beta_a(\bar{t})\, \bar{x}^a 
+ \frac{1}{2} \mbox{}_2 \beta_{ab}(\bar{t})\, \bar{x}^a \bar{x}^b 
+ \frac{1}{6} \mbox{}_3 \beta_{abc}(\bar{t})\, 
    \bar{x}^a \bar{x}^b \bar{x}^c 
+ \frac{1}{24} \mbox{}_4 \beta_{abcd}(\bar{t})\, 
    \bar{x}^a \bar{x}^b \bar{x}^c \bar{x}^d 
+ O(\bar{r}^5), 
\end{equation} 
in which the expansion coefficients are fully symmetric tensors; the
number that appears before each tensor symbol indicates the associated
power of $\bar{r}$ in the expansion. 

With all these ingredients in place, a long but straightforward
computation reveals that the equations listed in Secs.~8.3.2 and 8.3.3
of Ref.~\cite{poisson-will:14} become  
\begin{subequations} 
\label{PNpotentials_transf} 
\begin{align} 
\bar{U} &= \frac{M}{\bar{r}} + \mbox{}_0 \bar{U} 
+ \mbox{}_1 \bar{U}_a\, \bar{x}^a
+ \frac{1}{2} \mbox{}_2 \bar{U}_{ab}\, \bar{x}^a \bar{x}^b 
+ \frac{1}{6} \mbox{}_3 \bar{U}_{abc}\, 
    \bar{x}^a \bar{x}^b \bar{x}^c 
+ \frac{1}{24} \mbox{}_4 \bar{U}_{abcd}\, 
    \bar{x}^a \bar{x}^b \bar{x}^c \bar{x}^d 
+ O(\bar{r}^5), \\ 
\bar{U}_j &= \mbox{}_0 \bar{U}_j  
+ \mbox{}_1 \bar{U}_{ja}\, \bar{x}^a
+ \frac{1}{2} \mbox{}_2 \bar{U}_{jab}\, \bar{x}^a \bar{x}^b 
+ \frac{1}{6} \mbox{}_3 \bar{U}_{jabc}\, 
    \bar{x}^a \bar{x}^b \bar{x}^c 
+ \frac{1}{24} \mbox{}_4 \bar{U}_{jabcd}\, 
    \bar{x}^a \bar{x}^b \bar{x}^c \bar{x}^d 
+ O(\bar{r}^5), \\
\bar{\Psi} &= -\frac{M}{\bar{r}^3} F_a \bar{x}^a 
+ \frac{M}{\bar{r}} ( \dot{A} - 2v^2 + \mu ) 
+ \mbox{}_0 \bar{\Psi} 
+ \mbox{}_1 \bar{\Psi}_a\, \bar{x}^a
+ \frac{1}{2} \mbox{}_2 \bar{\Psi}_{ab}\, \bar{x}^a \bar{x}^b 
+ \frac{1}{6} \mbox{}_3 \bar{\Psi}_{abc}\, 
    \bar{x}^a \bar{x}^b \bar{x}^c 
\nonumber \\ & \quad \mbox{} 
+ \frac{1}{24} \mbox{}_4 \bar{\Psi}_{abcd}\, 
    \bar{x}^a \bar{x}^b \bar{x}^c \bar{x}^d 
+ O(\bar{r}^5),
\end{align}
\end{subequations} 
where $F_a := H_a - A v_a$ and the remaining expansion coefficients
are given by 
\begin{subequations} 
\begin{align} 
\mbox{}_0 \bar{U} &= \hat{U}_{\rm ext} - \dot{A} + \frac{1}{2} v^2, \\ 
\mbox{}_1 \bar{U}_a &= \partial_{\bar{a}} \hat{U}_{\rm ext} 
- a_a(0\pn), \\ 
\mbox{}_2 \bar{U}_{ab} &= \partial_{\bar{a}\bar{b}} 
   \hat{U}_{\rm ext}, \\  
\mbox{}_3 \bar{U}_{abc} &= \partial_{\bar{a}\bar{b}\bar{c}} 
   \hat{U}_{\rm ext}, \\  
\mbox{}_4 \bar{U}_{abcd} &= \partial_{\bar{a}\bar{b}\bar{c}\bar{d}} 
   \hat{U}_{\rm ext},  
\end{align}  
\end{subequations} 
\begin{subequations} 
\begin{align} 
4\, \mbox{}_0 \bar{U}_j &= 4 \hat{P}_j + (2\dot{A} - v^2) v_j 
- \dot{H}_j + \epsilon_{jpq} v^p R^q + \mbox{}_1 \beta_j, \\ 
4\, \mbox{}_1 \bar{U}_{ja} &= 4 \partial_{\bar{a}} \hat{P}_j 
+ \frac{3}{2} v_j a_a + \frac{1}{2} a_j v_a 
+ (\ddot{A} - 2 v_p a^p) \delta_{ja} 
- \epsilon_{jap} \dot{R}^p 
+ \mbox{}_2 \beta_{ja}, \\ 
4\, \mbox{}_2 \bar{U}_{jab} &= 4 \partial_{\bar{a}\bar{b}} \hat{P}_j  
+ 2\delta_{j(a} \dot{a}_{b)} - \delta_{ab} \dot{a}_j 
+ \mbox{}_3 \beta_{jab}, \\ 
4\, \mbox{}_3 \bar{U}_{jabc} &= 
4 \partial_{\bar{a}\bar{b}\bar{c}} \hat{P}_j  
+ \mbox{}_4 \beta_{jabc}, \\ 
4\, \mbox{}_4 \bar{U}_{jabcd} &= 
4 \partial_{\bar{a}\bar{b}\bar{c}\bar{d}} \hat{P}_j  
+ \mbox{}_5 \beta_{jabcd},
\end{align} 
\end{subequations} 
with $\hat{P}_j := \hat{U}^{\rm ext}_j - v_j \hat{U}^{\rm ext}$ and
$a_a \equiv a_a(0\pn)$, and  
\begin{subequations} 
\begin{align} 
\mbox{}_0 \bar{\Psi} &= \hat{P} + A \partial_{\bar{t}} \hat{U}_{\rm ext} 
+ F^p \partial_{\bar{p}} \hat{U}_{\rm ext} 
+ \frac{1}{2} \dot{A}^2 - \dot{A} v^2 + \frac{1}{4} v^4 
+ \dot{H}_p v^p - \mbox{}_0 \dot{\beta}, \\ 
\mbox{}_1 \bar{\Psi}_a &= 
\partial_{\bar{a}} \hat{P} 
+ A \partial_{\bar{t}\bar{a}} \hat{U}_{\rm ext} 
+ v_a \partial_{\bar{t}} \hat{U}_{\rm ext} 
+ F^p \partial_{\bar{p}\bar{a}} \hat{U}_{\rm ext} 
+ \Bigl( \dot{A} - \frac{1}{2} v^2 \Bigr) 
    \bigl[ a_a(0\pn) - \partial_{\bar{a}} \hat{U}_{\rm ext} \bigr] 
- \frac{1}{2} v_a v^p \partial_{\bar{p}} \hat{U}_{\rm ext} 
\nonumber \\ & \quad \mbox{}
+ \epsilon_{pqa} R^p \partial^{\bar{q}} \hat{U}_{\rm ext} 
- a_a(1\pn) - \Bigl[ \ddot{A} - \frac{3}{2} v_p a^p(0\pn) \Bigr] v_a 
- \epsilon_{apq} v^p \dot{R}^q - \mbox{}_1 \dot{\beta}_a, \\ 
\mbox{}_2 \bar{\Psi}_{ab} &= 
\partial_{\bar{a}\bar{b}} \hat{P} 
+ A \partial_{\bar{t}\bar{a}\bar{b}} \hat{U}_{\rm ext} 
+ 2 v_{(a} \partial_{\bar{t}\bar{b})} \hat{U}_{\rm ext} 
+ F^p \partial_{\bar{p}\bar{a}\bar{b}} \hat{U}_{\rm ext} 
- 2 \Bigl( \dot{A} - \frac{1}{2} v^2 \Bigr) 
   \partial_{\bar{a}\bar{b}} \hat{U}_{\rm ext} 
- v^p v_{(a} \partial_{\bar{b})\bar{p}} \hat{U}_{\rm ext} 
\nonumber \\ & \quad \mbox{}
+ 2\epsilon_{pq(a} R^p \partial^{\bar{q}}_{\ \bar{b})} \hat{U}_{\rm ext} 
- 2 a_{(a}(0\pn) \partial_{\bar{b})} \hat{U}_{\rm ext} 
+ \delta_{ab}\,  a^{p}(0\pn) \partial_{\bar{p}} \hat{U}_{\rm ext} 
+ a_a(0\pn) a_b(0\pn) 
\nonumber \\ & \quad \mbox{}
- 2 v_{(a} \dot{a}_{b)}(0\pn) 
+ \delta_{ab}\, v_p \dot{a}^p(0\pn) 
- \mbox{}_2 \dot{\beta}_{ab}, \\ 
\mbox{}_3 \bar{\Psi}_{abc} &= 
\partial_{\bar{a}\bar{b}\bar{c}} \hat{P} 
+ A \partial_{\bar{t}\bar{a}\bar{b}\bar{c}} \hat{U}_{\rm ext} 
+ 3 v_{(a} \partial_{\bar{t}\bar{b}\bar{c})} \hat{U}_{\rm ext} 
+ F^p \partial_{\bar{p}\bar{a}\bar{b}\bar{c}} \hat{U}_{\rm ext} 
- 3 \Bigl( \dot{A} - \frac{1}{2} v^2 \Bigr) 
   \partial_{\bar{a}\bar{b}\bar{c}} \hat{U}_{\rm ext} 
- \frac{3}{2} v^p v_{(a} \partial_{\bar{b}\bar{c})\bar{p}} \hat{U}_{\rm ext} 
\nonumber \\ & \quad \mbox{}
+ 3\epsilon_{pq(a} R^p \partial^{\bar{q}}_{\ \bar{b}\bar{c})} \hat{U}_{\rm ext} 
- 6 a_{(a}(0\pn) \partial_{\bar{b}\bar{c})} \hat{U}_{\rm ext} 
+ 3 \delta_{(ab}\,  a^{p}(0\pn) \partial_{\bar{c})\bar{p}} \hat{U}_{\rm ext} 
- \mbox{}_3 \dot{\beta}_{abc}, \\ 
\mbox{}_4 \bar{\Psi}_{abcd} &= 
\partial_{\bar{a}\bar{b}\bar{c}\bar{d}} \hat{P} 
+ A \partial_{\bar{t}\bar{a}\bar{b}\bar{c}\bar{d}} \hat{U}_{\rm ext} 
+ 4 v_{(a} \partial_{\bar{t}\bar{b}\bar{c}\bar{d})} \hat{U}_{\rm ext} 
+ F^p \partial_{\bar{p}\bar{a}\bar{b}\bar{c}\bar{d}} \hat{U}_{\rm ext} 
- 4 \Bigl( \dot{A} - \frac{1}{2} v^2 \Bigr) 
   \partial_{\bar{a}\bar{b}\bar{c}\bar{d}} \hat{U}_{\rm ext} 
- 2 v^p v_{(a} \partial_{\bar{b}\bar{c}\bar{d})\bar{p}} \hat{U}_{\rm ext} 
\nonumber \\ & \quad \mbox{}
+ 4\epsilon_{pq(a} R^p \partial^{\bar{q}}_{\ \bar{b}\bar{c}\bar{d})} \hat{U}_{\rm ext} 
- 12 a_{(a}(0\pn) \partial_{\bar{b}\bar{c}\bar{d})} \hat{U}_{\rm ext} 
+ 6 \delta_{(ab}\,  a^{p}(0\pn) \partial_{\bar{c}\bar{d})\bar{p}} \hat{U}_{\rm ext} 
- \mbox{}_4 \dot{\beta}_{abcd},
\end{align} 
\end{subequations} 
with $\hat{P} := \hat{\Psi}_{\rm ext} - 4 v_p \hat{U}^p_{\rm ext} 
+ 2 v^2 \hat{U}_{\rm ext}$. As stated previously, it is understood
that all external potentials and their derivatives are evaluated at
$\bar{x}^a = 0$.

\subsection{Matching} 
\label{subsec:matching} 

The potentials of Eq.~(\ref{PNpotentials_transf}) are the barycentric
potentials transformed to the black-hole frame, and they must agree
with the potentials of Eq.~(\ref{PNpotentials}), obtained in the
post-Newtonian expansion of the black-hole metric. A precise match
between the expressions shall reveal the details of the coordinate
transformation, the identity of the metric function $\mu(\bar{t})$,
and the tidal moments.  

The Newtonian potentials match at order $\bar{r}^{-1}$, and a match at
order $\bar{r}^0$ implies 
\begin{equation} 
\dot{A} = \hat{U}_{\rm ext} + \frac{1}{2} v^2. 
\label{Adot} 
\end{equation} 
A match at order $\bar{r}^1$ further reveals that $a_a(0\pn)
= \partial_{\bar{a}} \hat{U}^{\rm ext}$. Examining now the
post-Newtonian potentials $\bar{\Psi}$, a match at order
$\bar{r}^{-3}$ implies that $F_a = 0$, so that $H_a = A v_a$. A match
at order $\bar{r}^{-1}$ further produces $\mu = \frac{3}{2} v^2 
- \hat{U}_{\rm ext}$. A match at order $\bar{r}^0$ then yields 
\begin{equation} 
\mbox{}_0 \dot{\beta} = \hat{\Psi}_{\rm ext} 
- 4 v_p \hat{U}^p_{\rm ext} + \frac{1}{2} \hat{U}^2_{\rm ext} 
+ \frac{5}{2} v^2 \hat{U}_{\rm ext} + \frac{3}{8} v^4 
+ A \bigl[ \partial_{\bar{t}} \hat{U}_{\rm ext} + v_p a^p(0\pn)
\bigr]. 
\label{beta_0} 
\end{equation} 
Turning next to the vector potentials $\bar{U}_j$, a match at order
$\bar{r}^0$ reveals that 
\begin{equation} 
\mbox{}_1 \beta_a = -4 \hat{U}_a^{\rm ext} 
+ \Bigl( 3 \hat{U}_{\rm ext} + \frac{1}{2} v^2 \Bigr) v_a 
+ A a_a(0\pn) - \epsilon_{apq} v^q R^q. 
\label{beta_1} 
\end{equation} 
A match at order $\bar{r}^1$ produces 
\begin{equation} 
4 \partial_{\bar{b}} \hat{P}_a + \frac{3}{2} v_a a_b(0\pn) 
+ \frac{1}{2} a_a(0\pn) v_b 
- \delta_{ab}\, \partial_{\bar{p}} \hat{U}^p_{\rm ext} 
- \epsilon_{abp} \dot{R}^p 
+ \mbox{}_2 \beta_{ab} = 0
\end{equation} 
after making use of the harmonic condition of
Eq.~(\ref{harmonic_hatted}). Taking the symmetric part of this
equation yields 
\begin{equation} 
\mbox{}_2 \beta_{ab} = -4 \partial_{(\bar{a}} \hat{U}^{\rm ext}_{b)} 
+ 2 v_{(a} a_{b)}(0\pn) 
+ \delta_{ab}\, \partial_{\bar{p}} \hat{U}^p_{\rm ext}, 
\label{beta_2} 
\end{equation} 
while taking the antisymmetric part implies 
\begin{equation} 
\epsilon_{abp} \dot{R}^p 
= -4 \partial_{[\bar{a}} \hat{U}^{\rm ext}_{b]}   
- 3 v_{[a} a_{b]}(0\pn). 
\label{Rdot} 
\end{equation} 
Returning to $\bar{\Psi}$, we find that a match at order $\bar{r}^1$
requires  
\begin{equation} 
a_a(1\pn) = \partial_{\bar{a}} \hat{\Psi}_{\rm ext} 
- 4 v_p \partial_{\bar{a}} \hat{U}^p_{\rm ext} 
+ 4 \partial_{\bar{t}} \hat{U}^{\rm ext}_a 
+ (v^2 - 4 \hat{U}_{\rm ext}) a_a(0\pn) 
- \bigl[ 3\partial_{\bar{t}} \hat{U}_{\rm ext} 
+ v_p a^p(0\pn) \bigr] v_a. 
\end{equation} 
At this stage the acceleration of the black hole in the barycentric
frame, $a^a = a^a(0\pn) + a^a(1\pn)$, is determined, and some  
important details of the coordinate transformation are provided 
by Eqs.~(\ref{Adot}) and (\ref{Rdot}). 

We next match the terms in the potentials that occur at order 
$\bar{r}^2$ and beyond. The Newtonian potential returns 
\begin{equation} 
\bar{\E}_{ab}(0\pn) 
= -\partial_{\bar{a}\bar{b}} \hat{U}_{\rm ext}, 
\qquad 
\bar{\E}_{abc}(0\pn) 
= -\partial_{\bar{a}\bar{b}\bar{c}} \hat{U}_{\rm ext}, 
\qquad 
2\bar{\E}_{abcd}(0\pn) 
= -\partial_{\bar{a}\bar{b}\bar{c}\bar{d}} \hat{U}_{\rm ext},  
\label{E_0pn} 
\end{equation} 
where (as stated previously) the hatted potentials are evaluated at
$\bar{x}^a = 0$ after differentiation.  

From the vector potential at order $\bar{r}^2$ we get the matching
condition   
\begin{equation} 
-\frac{2}{3} \bigl( \epsilon_{ja}^{\ \ \, p} \bar{\B}_{pb} 
+ \epsilon_{jb}^{\ \ \, p} \bar{\B}_{pa} \bigr)  
= 4 \partial_{\bar{a}\bar{b}} \hat{P}_j 
+ \delta_{ja} \dot{a}_b + \delta_{jb} \dot{a}_a 
- \delta_{ab} \dot{a}_j + \mbox{}_3 \beta_{jab}, 
\end{equation} 
in which $a_a \equiv a_a(0\pn)$ --- the same shorthand will be employed
in all equations below. To extract the consequences of this
equation we decompose $\mbox{}_3 \beta_{jab}$ into irreducible pieces
according to Eq.~(\ref{Sdecomp3}), and the remaining terms of the
right-hand side are decomposed according to
Eq.~(\ref{Adecomp3}). Equating all this with the left-hand side
reveals that  
\begin{equation} 
\mbox{}_3 \beta_\stf{abc} = -4 \partial_{\langle \bar{a}\bar{b}} 
\bigl( \hat{U}^{\rm ext}_{c\rangle} - v_{c\rangle} \hat{U}^{\rm ext}
\bigr),\qquad 
\mbox{}_3 \beta_a = \dot{a}_a, 
\label{beta_3} 
\end{equation} 
and
\begin{equation} 
\bar{\B}_{ab} = 2 \epsilon^{pq}_{\ \ (a} \partial_{\bar{b})\bar{p}} 
\bigl( \hat{U}^{\rm ext}_{q} - v_{q} \hat{U}^{\rm ext} \bigr). 
\label{Bab} 
\end{equation} 
To arrive at these results we made use of the harmonic condition of
Eq.~(\ref{harmonic_hatted}) and the field equations
(\ref{PNeqn_hatted1}). At the next order we get 
\begin{equation} 
-\frac{2}{3} \bigl( \epsilon_{ja}^{\ \ \, p} \bar{\B}_{pbc} 
+ \epsilon_{jc}^{\ \ \, p} \bar{\B}_{pab} 
+ \epsilon_{jb}^{\ \ \, p} \bar{\B}_{pca} \bigr)
+ \frac{4}{3} \bigl( \delta_{ja}\, \dot{\bar{\E}}_{bc} 
+ \delta_{jc}\, \dot{\bar{\E}}_{ab} 
+ \delta_{jb}\, \dot{\bar{\E}}_{ca} \bigr)  
= 4\partial_{\bar{a}\bar{b}\bar{c}} \hat{P}_j
+ \mbox{}_4 \beta_{jabc},  
\end{equation} 
and we decompose $\mbox{}_4 \beta_{jabc}$ according to
Eq.~(\ref{Sdecomp4}), and $4\partial_{\bar{a}\bar{b}\bar{c}} \hat{P}_j$
according to Eq.~(\ref{Adecomp4}). After simplifying the results with
the harmonic condition, the field equations, and Eq.~(\ref{E_0pn}), we
arrive at 
\begin{equation} 
\mbox{}_4 \beta_\stf{abcd} = -4 \partial_{\langle \bar{a}\bar{b}\bar{c}} 
\bigl( \hat{U}^{\rm ext}_{d\rangle} - v_{d\rangle} \hat{U}^{\rm ext}
\bigr),
\qquad 
\mbox{}_4 \beta_\stf{ab} = \frac{8}{3} \dot{\bar{\E}}_{ab}, 
\qquad  
\mbox{}_4 \beta = 0, 
\label{beta_4} 
\end{equation} 
and
\begin{equation} 
\bar{\B}_{abc} = \frac{3}{2} \epsilon^{pq}_{\ \ (a} 
  \partial_{\bar{b}\bar{c})\bar{p}} 
  \bigl( \hat{U}^{\rm ext}_{q} - v_{q} \hat{U}^{\rm ext} \bigr). 
\label{Babc} 
\end{equation} 
Matching the vector potentials at order $\bar{r}^4$ produces 
\begin{equation} 
-\frac{4}{3} \bigl( \epsilon_{ja}^{\ \ \, p} \bar{\B}_{pbcd} 
+ \epsilon_{jd}^{\ \ \, p} \bar{\B}_{pabc} 
+ \epsilon_{jc}^{\ \ \, p} \bar{\B}_{pdab} 
+ \epsilon_{jb}^{\ \ \, p} \bar{\B}_{pcda} \bigr)
+ \bigl( \delta_{ja}\, \dot{\bar{\E}}_{bcd} 
+ \delta_{jd}\, \dot{\bar{\E}}_{abc} 
+ \delta_{jc}\, \dot{\bar{\E}}_{dab} 
+ \delta_{jb}\, \dot{\bar{\E}}_{cda} \bigr)  
= 4\partial_{\bar{a}\bar{b}\bar{c}\bar{d}} \hat{P}_j
+ \mbox{}_5 \beta_{jabcd}.   
\end{equation} 
After decomposing $\mbox{}_5 \beta_{jabcd}$ according to
Eq.~(\ref{Sdecomp5}), $4\partial_{\bar{a}\bar{b}\bar{c}\bar{d}} \hat{P}_j$
according to Eq.~(\ref{Adecomp4}), and simplifying, we arrive at 
\begin{equation} 
\mbox{}_5 \beta_\stf{abcde} 
= -4 \partial_{\langle \bar{a}\bar{b}\bar{c}\bar{d}} 
\bigl( \hat{U}^{\rm ext}_{e\rangle} - v_{e\rangle} \hat{U}^{\rm ext} \bigr),
\qquad 
\mbox{}_5 \beta_\stf{abc} = 2 \dot{\bar{\E}}_{abc}, 
\qquad  
\mbox{}_5 \beta_a = 0 
\label{beta_5} 
\end{equation} 
and 
\begin{equation} 
\bar{\B}_{abcd} = \frac{3}{5} \epsilon^{pq}_{\ \ (a} 
  \partial_{\bar{b}\bar{c}\bar{d})\bar{p}} 
  \bigl( \hat{U}^{\rm ext}_{q} - v_{q} \hat{U}^{\rm ext} \bigr). 
\label{Babcd} 
\end{equation} 
At this stage the gravitomagnetic tidal moments are all determined, as 
well as the details of the coordinate transformation contained in the
function $\beta(\bar{t},\bm{\bar{x}})$. 

We finally return to the post-Newtonian potential. Matching at order
$\bar{r}^2$ implies that 
\begin{equation} 
\bar{\E}_{ab}(1\pn) = -\partial_{\bar{a}\bar{b}} \hat{P} 
- 2 \hat{U}_{\rm ext}\, \bar{\E}_{ab} 
- v_{(a} \bar{\E}_{b)p} v^p 
+ a_a a_b 
- \delta_{ab} \bigl( a^2 + v_p \dot{a}^p \bigr) 
+ \mbox{}_2 \dot{\beta}_{ab}
+ A \dot{\bar{\E}}_{ab} 
+ 2 \epsilon_{pq(a} R^p \bar{\E}^q_{\ b)},  
\end{equation} 
where $\hat{P} := \hat{\Psi}_{\rm ext} - 4 v_p \hat{U}^p_{\rm ext} 
+ 2 v^2 \hat{U}_{\rm ext}$, $a_a \equiv a_a(0\pn)$, and 
$\bar{\E}_{ab} \equiv \bar{\E}_{ab}(0\pn)$ on the right-hand side of
the equation --- a similar shorthand will be employed in all equations
below. Making use of Eq.~(\ref{beta_2}) and decomposing all tensors
into irreducible pieces, we arrive at 
\begin{equation} 
\bar{\E}_{ab}(1\pn) = -\partial_\stf{\bar{a}\bar{b}} \hat{P} 
- 4 \partial_{\bar{t} \langle \bar{a}} \hat{U}^{\rm ext}_{b\rangle} 
- 2 \hat{U}_{\rm ext}\, \bar{\E}_{ab} 
- v_{\langle a} \bar{\E}_{b\rangle p} v^p 
+ 3 a_{\langle a} a_{b \rangle}  
+ 2 v_{\langle a} \dot{a}_{b\rangle} 
+ A \dot{\bar{\E}}_{ab} 
+ 2 \epsilon_{pq(a} R^p \bar{\E}^q_{\ b)}. 
\label{Eab} 
\end{equation} 
At order $\bar{r}^3$ we get  
\begin{equation} 
\bar{\E}_{abc}(1\pn) = -\partial_{\bar{a}\bar{b}\bar{c}} \hat{P} 
+ 3 v_{(a} \dot{\bar{\E}}_{bc)} 
- 3 \hat{U}_{\rm ext}\, \bar{\E}_{abc} 
- \frac{3}{2} v_{(a} \bar{\E}_{bc)p} v^p 
- 6 a_{(a} \bar{\E}_{bc)} 
+ 3 \delta_{(ab} \bar{\E}_{c)p} a^p  
+ \mbox{}_3 \dot{\beta}_{abc}
+ A \dot{\bar{\E}}_{abc} 
+ 3 \epsilon_{pq(a} R^p \bar{\E}^q_{\ bc)}, 
\end{equation} 
and substitution of Eq.~(\ref{beta_3}) and decomposition into
irreducible pieces yields 
\begin{equation} 
\bar{\E}_{abc}(1\pn) = -\partial_\stf{\bar{a}\bar{b}\bar{c}} \hat{P} 
- 4 \partial_{\bar{t} \langle \bar{a}\bar{b}} \hat{U}^{\rm ext}_{c\rangle} 
- v_{\langle a} \dot{\bar{\E}}_{bc\rangle} 
- 10 a_{\langle a} \bar{\E}_{bc\rangle} 
- 3 \hat{U}_{\rm ext}\, \bar{\E}_{abc} 
- \frac{3}{2} v_{\langle a} \bar{\E}_{bc\rangle p} v^p 
+ A \dot{\bar{\E}}_{abc} 
+ 3 \epsilon_{pq(a} R^p \bar{\E}^q_{\ bc)}. 
\label{Eabc} 
\end{equation} 
At order $\bar{r}^4$ the matching condition is 
\begin{align} 
& 2 \bar{\E}_{abcd}(1\pn) 
+ \frac{11}{21} \bigl( \delta_{ab} \ddot{\bar{\E}}_{cd} 
+ \mbox{all symmetric permutations} \bigr) 
- 10 \bar{\E}_{\langle ab} \bar{\E}_{cd \rangle}
= -\partial_{\bar{a}\bar{b}\bar{c}\bar{d}} \hat{P} 
+ 4 v_{(a} \dot{\bar{\E}}_{bcd)} 
- 8 \hat{U}_{\rm ext}\, \bar{\E}_{abcd} 
\nonumber \\ & \quad \mbox{} 
- 4 v_{(a} \bar{\E}_{bcd)p} v^p 
- 12 a_{(a} \bar{\E}_{bcd)} 
+ 6 \delta_{(ab} \bar{\E}_{cd)p} a^p  
+ \mbox{}_4 \dot{\beta}_{abcd}
+ 2A \dot{\bar{\E}}_{abcd} 
+ 8 \epsilon_{pq(a} R^p \bar{\E}^q_{\ bcd)}, 
\end{align} 
and this yields 
\begin{align} 
2 \bar{\E}_{abcd}(1\pn) &= 
-\partial_\stf{\bar{a}\bar{b}\bar{c}\bar{d}} \hat{P} 
- 4 \partial_{\bar{t} \langle \bar{a}\bar{b}\bar{c}} 
      \hat{U}^{\rm ext}_{d\rangle} 
+ 10 \bar{\E}_{\langle ab} \bar{\E}_{cd \rangle}
- 16 a_{\langle a} \bar{\E}_{bcd\rangle} 
\nonumber \\ & \quad \mbox{} 
- 8 \hat{U}_{\rm ext}\, \bar{\E}_{abcd} 
- 4 v_{\langle a} \bar{\E}_{bcd\rangle p} v^p 
+ 2 A \dot{\bar{\E}}_{abcd} 
+ 8 \epsilon_{pq(a} R^p \bar{\E}^q_{\ bcd)} 
\label{Eabcd} 
\end{align} 
after substituting Eq.~(\ref{beta_4}) and decomposing all terms into
irreducible pieces. At this stage the gravitoelectric tidal moments
are all determined, and the matching procedure has come to a close.

\subsection{Barycentric tidal moments} 

The tidal moments obtained in the preceding subsection are defined in
the black-hole frame $(\bar{t},\bm{x}^a)$. For our purposes in
Sec.~\ref{sec:2body}, it is convenient to follow Racine and Flanagan
\cite{racine-flanagan:05} and introduce barycentric versions of these
moments. We do so with the transformations    
\begin{subequations}
\label{moment_transf1} 
\begin{align} 
\E_{ab}(t) &:= {\cal M}_a^{\ j}(\bar{t}) {\cal M}_b^{\ k}(\bar{t})\,
  \bar{\E}_{jk}(\bar{t}), \\
\E_{abc}(t) &:= {\cal M}_a^{\ j}(\bar{t}) {\cal M}_b^{\ k}(\bar{t}) 
   {\cal M}_c^{\ m}(\bar{t})\, \bar{\E}_{jkm}(\bar{t}), \\ 
\E_{abcd}(t) &:= {\cal M}_a^{\ j}(\bar{t}) {\cal M}_b^{\ k}(\bar{t}) 
   {\cal M}_c^{\ m}(\bar{t}) {\cal M}_d^{\ n}(\bar{t})\,
   \bar{\E}_{jkmn}(\bar{t}),
\end{align}
\end{subequations} 
as well as similar ones relating $\B_{ab\cdots}(t)$ to
$\bar{\B}_{ab\cdots}(\bar{t})$. The transformation matrix is defined
by 
\begin{equation} 
{\cal M}_{aj}(\bar{t}) := \delta_{aj} + \epsilon_{ajp} R^p(\bar{t})
+ 2\pn, 
\label{moment_transf2} 
\end{equation} 
with $R^p$ determined by Eq.~(\ref{Rdot}). Because the tidal moments
are tensors defined at $\bar{x}^a=0$, the relation between the time
coordinates is given by 
\begin{equation} 
t = \bar{t} + A(\bar{t}) + 2\pn, 
\label{moment_transf3} 
\end{equation}
with $A$ determined by Eq.~(\ref{Adot}). 

We apply Eq.~(\ref{moment_transf1}) to the tidal moments obtained  
previously, and express the results in terms of the original
potentials $U$, $U_j$, and $\Psi$ (instead of the hatted
ones). Recalling the differentiation rule of Eq.~(\ref{diff_rule}), we
find that the barycentric version of the gravitoelectric tidal moments
are given by  
\begin{subequations} 
\label{Eab_bary} 
\begin{align} 
\E_{ab} &= \E_{ab}(0\pn) + \E_{ab}(1\pn) + 2\pn, \\ 
\E_{ab}(0\pn) &= -\partial_{ab} U_{\rm ext}, \\ 
\E_{ab}(1\pn) &= -\partial_\stf{ab} \Psi_{\rm ext} 
- 4 \partial_{t \langle a} U^{\rm ext}_{b\rangle} 
+ 4 \bigl( \partial_\stf{ab} U^{\rm ext}_p 
   - \partial_{p \langle a} U^{\rm ext}_{b\rangle} \bigr) v^p 
+ 2 (v^2 - U_{\rm ext}) \E_{ab} 
\nonumber \\ & \quad \mbox{} 
- v_{\langle a} \E_{b\rangle p} v^p 
+ 3 a_{\langle a} a_{b \rangle}  
+ 2 v_{\langle a} \dot{a}_{b\rangle}, 
\end{align} 
\end{subequations} 
\begin{subequations} 
\label{Eabc_bary} 
\begin{align} 
\E_{abc} &= \E_{abc}(0\pn) + \E_{abc}(1\pn) + 2\pn, \\ 
\E_{abc}(0\pn) &= -\partial_{abc} U_{\rm ext}, \\ 
\E_{abc}(1\pn) &= -\partial_\stf{abc} \Psi_{\rm ext} 
- 4 \partial_{t \langle ab} U^{\rm ext}_{c\rangle} 
+ 4 \bigl( \partial_\stf{abc} U^{\rm ext}_p 
   - \partial_{p \langle ab} U^{\rm ext}_{c\rangle} \bigr) v^p 
+ (2v^2 - 3U_{\rm ext}) \E_{abc} 
\nonumber \\ & \quad \mbox{} 
- \frac{3}{2} v_{\langle a} \E_{bc\rangle p} v^p 
- v_{\langle a} \dot{\E}_{bc\rangle} 
- 10 a_{\langle a} \E_{bc\rangle},  
\end{align} 
\end{subequations} 
and 
\begin{subequations} 
\label{Eabcd_bary} 
\begin{align} 
\E_{abcd} &= \E_{abcd}(0\pn) + \E_{abcd}(1\pn) + 2\pn, \\ 
2\E_{abcd}(0\pn) &= -\partial_{abcd} U_{\rm ext}, \\ 
2\E_{abcd}(1\pn) &= -\partial_\stf{abcd} \Psi_{\rm ext} 
- 4 \partial_{t \langle abc} U^{\rm ext}_{d\rangle} 
+ 4 \bigl( \partial_\stf{abcd} U^{\rm ext}_p 
   - \partial_{p \langle abc} U^{\rm ext}_{d\rangle} \bigr) v^p 
+ 4(v^2 - 2U_{\rm ext}) \E_{abcd} 
\nonumber \\ & \quad \mbox{} 
- 4 v_{\langle a} \E_{bcd\rangle p} v^p 
- 16 a_{\langle a} \E_{bcd \rangle}
+ 10 \E_{\langle ab} \E_{cd\rangle}.    
\end{align} 
\end{subequations} 
The external potentials are evaluated at $x^a = r^a(t)$ after
differentiation, and it is understood that in the expressions for the
post-Newtonian terms, $a_a \equiv a_a(0\pn) = \partial_a U_{\rm ext}$, 
$\E_{ab} \equiv \E_{ab}(0\pn)$, and so on.  

Because the gravitomagnetic moments are quantities of $1\pn$ order,
the transformations analogous to those of Eq.~(\ref{moment_transf1})
have a trivial effect on them. The barycentric version of these moments
are therefore given by 
\begin{subequations} 
\label{Bmoments_bary}  
\begin{align} 
\B_{ab} &= 2 \epsilon^{pq}_{\ \ (a} \partial_{b) p} 
\bigl( U^{\rm ext}_{q} - v_{q} U^{\rm ext} \bigr),  
\label{Bab_bary} \\ 
\B_{abc} &= \frac{3}{2} \epsilon^{pq}_{\ \ (a} \partial_{bc) p} 
\bigl( U^{\rm ext}_{q} - v_{q} U^{\rm ext} \bigr),  
\label{Babc_bary} \\ 
\B_{abcd} &= \frac{3}{5} \epsilon^{pq}_{\ \ (a} \partial_{bcd) p} 
\bigl( U^{\rm ext}_{q} - v_{q} U^{\rm ext} \bigr).   
\label{Babcd_bary} 
\end{align} 
\end{subequations} 
Here also the external potentials are evaluated at $x^a = r^a(t)$
after differentiation.    

\section{Tidal moments for a two-body system} 
\label{sec:2body} 

The tidal moments obtained in Sec.~\ref{sec:matching} are expressed in
terms of generic external potentials that could describe any
post-Newtonian spacetime. In this section we specialize them to the
specific case in which the black hole is a member of a two-body
system.   

\subsection{External potentials} 

We take the spacetime to contain a body of mass $M_2$
at position $\bm{r}_2(t)$ in addition to the black hole. We let 
$\bm{v}_2(t) := d\bm{r}_2/dt$ and $\bm{a}_2(t) := d\bm{v}_2/dt$.
We refine the notation employed in Sec.~\ref{sec:matching}: the
black hole's mass shall again be denoted $M_1$ instead of $M$, and its
position in the barycentric frame shall be $\bm{r}_1(t)$ instead of
$\bm{r}$; we also set $\bm{v}_1(t) := d\bm{r}_1/dt$ (previously
$\bm{v}$) and $\bm{a}_1(t) := d\bm{v}_1/dt$ (previously $\bm{a}$).  

We take the external body to be another post-Newtonian monopole, and
write the external potentials as 
\begin{equation} 
U_{\rm ext} = \frac{M_2}{s}, \qquad 
U^j_{\rm ext} = \frac{M_2 v_2^j}{s}, \qquad 
\psi_{\rm ext} = \frac{M_2 \mu_2}{s}, \qquad 
X_{\rm ext} = M_2 s, 
\end{equation} 
where $s$ now stands for $|\bm{x}-\bm{r}_2|$, the Euclidean distance
between the field point $\bm{x}$ and the external
body. We recall that $\Psi_{\rm ext} = \psi_{\rm ext} +
\frac{1}{2} \partial_{tt} X_{\rm ext}$. According to our findings in
Sec.~\ref{subsec:matching}, the post-Newtonian correction to the mass
parameter is given by $\mu_2 = \frac{3}{2} v_2^2 - M_1/b$, where $b$ is
the inter-body distance.  

We let $\bm{b} := \bm{r}_1 - \bm{r}_2$ be the separation between 
bodies, $b := |\bm{b}|$, and $\bm{n} := \bm{b}/b$ is a unit vector
directed from the external body to the black hole. The relative
velocity is $\bm{v} := \bm{v}_1 - \bm{v}_2$, and $\dot{b} = \bm{v}
\cdot \bm{n}$ is its radial component. The post-Newtonian
equations of motion imply that $\bm{a}_1 = -(M_2/b^2) \bm{n} + 1\pn$
and $\bm{a}_2 = (M_1/b^2) \bm{n} +1\pn$. They also imply that 
$\bm{v}_1 = q_2 \bm{v} + 1\pn$ and $\bm{v}_2 = -q_1 \bm{v} + 1\pn$,
where 
\begin{equation} 
q_1 := \frac{M_1}{M_1 + M_2}, \qquad 
q_2 := \frac{M_2}{M_1 + M_2}
\end{equation} 
are the mass ratios, constrained by the identity $q_1 + q_2 = 1$. 

The various derivatives of the external potentials are evaluated with
the help of results collected in Appendix~\ref{sec:distance}. The position
$\bm{x}$ is set equal to $\bm{r}_1$ after differentiation, and in this
limit the vector $\bm{s} := \bm{x} - \bm{r}_2$ becomes $\bm{b}$; 
similarly, $\bm{\hat{s}} \to \bm{n}$ and $s \to b$. The
equations of motion are used to eliminate the
accelerations $\bm{a}_1$ and $\bm{a}_2$ from all expressions, and to
express the velocities $\bm{v}_1$ and $\bm{v}_2$ in terms of the
relative velocity $\bm{v}$.  

\subsection{Barycentric tidal moments} 

A long but straightforward calculation returns 
\begin{subequations} 
\label{Eab_2body} 
\begin{align} 
\E_{ab} &= \E_{ab}(0\pn) + \E_{ab}(1\pn) + 2\pn, \\ 
\E_{ab}(0\pn) &= -3 \frac{M_2}{b^3} n_\stf{ab}, \\  
\E_{ab}(1\pn) &= -3 \frac{M_2}{b^3} \Biggl\{ 
\biggl[ 2 v^2 - \frac{5}{2} q_1^2\, \dot{b}^2 
- \frac{1}{2} (6-q_1) \frac{M}{b} \biggr] n_\stf{ab} 
- (3-q_1^2)\, \dot{b}\, v_{\langle a} n_{b \rangle} 
+ v_{\langle a} v_{b\rangle} \Biggr\}, 
\end{align} 
\end{subequations} 
\begin{subequations} 
\label{Eabc_2body} 
\begin{align} 
\E_{abc} &= \E_{abc}(0\pn) + \E_{abc}(1\pn) + 2\pn, \\ 
\E_{abc}(0\pn) &= 15\frac{M_2}{b^4} n_\stf{abc}, \\ 
\E_{abc}(1\pn) &= 15 \frac{M_2}{b^4} \Biggl\{ 
\biggl[ 2 v^2 - \frac{7}{2} q_1^2\, \dot{b}^2 
- (5-3q_1) \frac{M}{b} \biggr] n_\stf{abc} 
- \frac{1}{2} (5-3q_1^2)\, \dot{b}\, v_{\langle a} n_b n_{c \rangle} 
+ v_{\langle a} v_b n_{c\rangle} \Biggr\}, 
\end{align} 
\end{subequations} 
\begin{subequations} 
\label{Eabcd_2body} 
\begin{align} 
\E_{abcd} &= \E_{abcd}(0\pn) + \E_{abcd}(1\pn) + 2\pn, \\ 
\E_{abcd}(0\pn) &= -\frac{105}{2} \frac{M_2}{b^5} n_\stf{abcd}, \\ 
\E_{abcd}(1\pn) &= -\frac{15}{2} \frac{M_2}{b^5} \Biggl\{ 
\biggl[ 14 v^2 - \frac{63}{2} q_1^2\, \dot{b}^2 
- \frac{25}{2} (4-3q_1) \frac{M}{b} \biggr] n_\stf{abcd} 
- 14 (1-q_1^2)\, \dot{b}\, v_{\langle a} n_b n_c n_{d \rangle} 
+ 6 v_{\langle a} v_b n_c n_{d\rangle} \Biggr\}
\end{align} 
\end{subequations} 
for the barycentric version of the gravitoelectric tidal moments, 
where $n_\stf{ab\cdots c}$ is shorthand for $n_{\langle a} n_b \cdots
n_{c\rangle}$. The gravitomagnetic moments are given by 
\begin{subequations} 
\label{Bmoments_2body} 
\begin{align} 
\B_{ab} &= -6 \frac{M_2}{b^3} \epsilon_{pq(a} n_{b)} n^p v^q, \\ 
\B_{abc} &= \frac{45}{2} \frac{M_2}{b^4} \epsilon_{pq(a} 
\Bigl( n_b n_{c)} - \frac{1}{5} \delta_{bc)} \Bigr) n^p v^q, \\ 
\B_{abcd} &= -63 \frac{M_2}{b^5} \epsilon_{pq(a} 
\Bigl( n_b n_c n_{d)} - \frac{3}{7} \delta_{bc} n_{d)} \Bigr) n^p v^q.  
\end{align} 
\end{subequations} 

Alternative expressions are obtained if the relative velocity vector
is decomposed according to  
\begin{equation} 
\bm{v} = \dot{b}\, \bm{n} + v_\perp\, \bm{\lambda}, 
\end{equation} 
in terms of radial and perpendicular components;
$\bm{\lambda}$ is a unit vector orthogonal to $\bm{n}$. We recall that
the post-Newtonian motion takes place in a fixed orbital plane, with a
vanishing normal component for the velocity. The tidal moments become  
\begin{subequations} 
\label{Emoments_decomp} 
\begin{align} 
\E_{ab}(1\pn) &= 3 \frac{M_2}{b^3} \Biggl\{ 
\biggl[ \frac{3}{2} q_1^2\, \dot{b}^2 - 2 v_\perp^2  
+ \frac{1}{2} (6-q_1) \frac{M}{b} \biggr] n_\stf{ab} 
+ (1-q_1^2)\, \dot{b} v_\perp \, \lambda_{\langle a} n_{b \rangle} 
- v_\perp^2 \lambda_{\langle a} \lambda_{b\rangle} \Biggr\}, \\
\E_{abc}(1\pn) &= 15 \frac{M_2}{b^4} \Biggl\{ 
\biggl[ \frac{1}{2}(1- 4q_1^2)\, \dot{b}^2 + 2 v_\perp^2  
- (5-3q_1) \frac{M}{b} \biggr] n_\stf{abc} 
- \frac{1}{2} (1-3q_1^2)\, \dot{b} v_\perp\, \lambda_{\langle a} n_b n_{c \rangle} 
+ v_\perp^2 \lambda_{\langle a} \lambda_b n_{c\rangle} \Biggr\}, \\
\E_{abcd}(1\pn) &= -\frac{15}{2} \frac{M_2}{b^5} \Biggl\{ 
\biggl[ \frac{1}{2} (12 - 35 q_1^2)\, \dot{b}^2 + 14 v_\perp^2   
- \frac{25}{2} (4-3q_1) \frac{M}{b} \biggr] n_\stf{abcd} 
\nonumber \\ & \quad \mbox{} 
- 2 (1-7q_1^2)\, \dot{b} v_\perp\, \lambda_{\langle a} n_b n_c n_{d \rangle} 
+ 6 v_\perp^2 \lambda_{\langle a} \lambda_b n_c n_{d\rangle} \Biggr\}
\end{align} 
\end{subequations} 
and
\begin{subequations} 
\label{Bmoments_decomp} 
\begin{align} 
\B_{ab} &= -6 \frac{M_2 v_\perp}{b^3} \epsilon_{pq(a} n_{b)} n^p \lambda^q, \\ 
\B_{abc} &= \frac{45}{2} \frac{M_2 v_\perp}{b^4} \epsilon_{pq(a} 
\Bigl( n_b n_{c)} - \frac{1}{5} \delta_{bc)} \Bigr) n^p \lambda^q, \\ 
\B_{abcd} &= -63 \frac{M_2 v_\perp}{b^5} \epsilon_{pq(a} 
\Bigl( n_b n_c n_{d)} - \frac{3}{7} \delta_{bc} n_{d)} \Bigr) n^p \lambda^q.  
\end{align} 
\end{subequations} 

The tidal moments in the black-hole frame are obtained by inverting
the transformation of Eqs.~(\ref{moment_transf1}),
(\ref{moment_transf2}), and (\ref{moment_transf3}). Making the
relevant substitutions in Eq.~(\ref{Adot}), we find that $A$ is 
determined by 
\begin{equation} 
\frac{dA}{dt} = \frac{1}{2} (1-q_1)^2 
\bigl( \dot{b}^2 + v_\perp^2 \bigr) 
+ (1-q_1) \frac{M}{b},  
\label{Adot_decomp} 
\end{equation}  
and 
\begin{equation} 
\frac{dR_a}{dt} = -\frac{1}{2} (1-q_1)(3+q_1) \frac{M v_\perp}{b^2} 
\epsilon_{abc} n^b \lambda^c 
\label{Rdot_decomp} 
\end{equation}   
follows from Eq.~(\ref{Rdot}).  

\subsection{Circular motion} 

Setting $\dot{b} = 0$ specializes the motion to a circular orbit with
$b = \mbox{constant}$. The equations of motion imply 
(see, for example, Sec.~10.1.2 of Ref.~\cite{poisson-will:14}) 
\begin{equation} 
V \equiv v_\perp = \sqrt{\frac{M}{b}} \biggl[ 1 
- \frac{1}{2}(3-q_1 q_2) \frac{M}{b} + 2\pn \biggr], 
\end{equation} 
and this equation reveals that $M/b = V^2 + 1\pn$. Making the
substitutions within Eqs.~(\ref{Emoments_decomp}) produces  
\begin{subequations} 
\label{Emoments_circ} 
\begin{align} 
\E_{ab}(1\pn) &= 3 \frac{M_2 V^2}{b^3} \biggl[ 
\frac{1}{2} (2-q_1) n_\stf{ab} 
- \lambda_{\langle a} \lambda_{b\rangle} \Bigr], \\ 
\E_{abc}(1\pn) &= -15 \frac{M_2 V^2}{b^4} \Bigl[
3(1-q_1) n_\stf{abc} 
- \lambda_{\langle a} \lambda_b n_{c\rangle} \Bigr], \\
\E_{abcd}(1\pn) &= \frac{15}{2} \frac{M_2 V^2}{b^5} \biggl[ 
\frac{1}{2} (72 - 75q_1) n_\stf{abcd} 
- 6 \lambda_{\langle a} \lambda_b n_c n_{d\rangle} \biggr]. 
\end{align} 
\end{subequations} 
The equations (\ref{Bmoments_decomp}) stay unchanged. 

In the case of a circular orbit the basis vectors can be given the
explicit representation 
\begin{equation} 
\bm{n} = [ \cos(\omega t), \sin(\omega t), 0 ], \qquad 
\bm{\lambda} = [-\sin(\omega t), \cos(\omega t), 0], 
\label{vectorbasis1} 
\end{equation} 
where 
\begin{equation} 
\omega := \frac{V}{b} = \sqrt{\frac{M}{b^3}} \biggl[ 1 
- \frac{1}{2}(3-q_1 q_2) \frac{M}{b} + 2\pn \biggr] 
\end{equation}
is the orbital angular velocity in the barycentric frame. It is useful
to complete the vector basis with 
\begin{equation} 
\bm{\ell} := [0, 0, 1], 
\label{vectorbasis2} 
\end{equation} 
a unit vector normal to the orbital plane. 

Equations (\ref{Adot_decomp}) and (\ref{Rdot_decomp}) become 
\begin{equation} 
\frac{dA}{dt} =\frac{1}{2} (1-q_1)(3-q_1) V^2, \qquad 
\frac{dR^a}{dt} = -\frac{1}{2}(1-q_1)(3+q_1) \sqrt{\frac{M}{b^3}}
V^2\,\ell^a 
\end{equation} 
in the case of circular motion. The equations integrate to 
\begin{equation} 
A = k t, \qquad k = \frac{1}{2} (1-q_1)(3-q_1) V^2 
\label{A_circular} 
\end{equation} 
and 
\begin{equation} 
R^a = -\Omega t\, \ell^a, \qquad 
\Omega := \frac{1}{2}(1-q_1)(3+q_1) \sqrt{\frac{M}{b^3}} V^2, 
\label{R_circular} 
\end{equation} 
with $\Omega$ denoting the precessional angular velocity of the
black-hole frame relative to the barycentric frame. 

\subsection{Transformation to the black-hole frame} 

The transformation from the barycentric frame of the post-Newtonian
spacetime to the black hole's moving frame is effected by inverting
Eqs.~(\ref{moment_transf1}), (\ref{moment_transf2}), and 
(\ref{moment_transf3}). Under the inverse transformation, a vector
$p_a(t)$ defined in the barycentric frame becomes 
\begin{equation} 
\bar{p}_a(\bar{t}) = \bigl[ {\cal M}^{-1}(t) \bigr]_a^{\ j}\, p_j(t) 
\end{equation} 
in the black-hole frame, with 
\begin{equation} 
\bigl[ {\cal M}^{-1}(t) \bigr]_{aj} = \delta_{aj} 
- \epsilon_{ajp} R^p(t) + 2\pn  
\end{equation} 
and $\bar{t} = t - A(t) + 2\pn$. In the case of circular motion,
Eqs.~(\ref{A_circular}) and (\ref{R_circular}) imply that the
transformation takes the form of
\begin{equation} 
\bm{\bar{p}} = \bm{p} - (\Omega t)\, \bm{\ell} \times \bm{p},  
\end{equation} 
with $t = (1+k) \bar{t}$ inserted onto the right-hand
side. Applying this rule to our vectorial basis yields 
\begin{equation} 
\bm{\bar{n}} = \bm{n} - (\Omega t)\, \bm{\lambda}, \qquad 
\bm{\bar{\lambda}} = \bm{\lambda} 
+ (\Omega t)\, \bm{n}, \qquad
\bm{\bar{\ell}} = \bm{\ell}. 
\end{equation} 
With the representation of Eqs.~(\ref{vectorbasis1}) and
(\ref{vectorbasis2}), and with $\Omega$ recognized as a
post-Newtonian correction to the angular velocity $\omega$, we have
that 
\begin{equation} 
\bm{\bar{n}} 
= [\cos(\bar{\omega} \bar{t}), \sin(\bar{\omega} \bar{t}), 0], \qquad 
\bm{\bar{\lambda}} 
= [-\sin(\bar{\omega} \bar{t}), \cos(\bar{\omega} \bar{t}), 0], \qquad 
\bm{\bar{\ell}} = [0,0,1],  
\end{equation}   
where $\bar{\omega} = (1+k)\omega - \Omega$, or 
\begin{equation}  
\bar{\omega} = \sqrt{\frac{M}{b^3}} \biggl[ 1 
- \frac{1}{2}(3+q_1 q_2) V^2 + 2\pn \biggr]. 
\end{equation} 
This is the angular frequency of the tidal field as measured in the
black-hole frame. It differs from $\omega$, the orbital angular
velocity in the barycentric frame, because of the mismatch in the time
coordinates (measured by $k$), and also because of the relative
precession of the two frames (measured by $\Omega$). 

The tidal moments in the black-hole frame are obtained directly from
the barycentric moments by replacing $\bm{n}$ with $\bm{\bar{n}}$, and 
$\bm{\lambda}$ with $\bm{\bar{\lambda}}$. In this transcription,
however, we shall also take the opportunity to modify our convention
for the basis vectors. We recall that $\bm{n}$ is proportional to
$\bm{r}_1 - \bm{r}_2$, and is therefore directed from the external
body to the black hole. In the black-hole frame it is convenient to
reverse this direction, and we therefore let 
$\bm{\bar{n}} \to -\bm{\bar{n}}$ in our expressions for
the tidal moments. Similarly, we recall $\bm{\lambda}$ is proportional
to $\bm{v}_1 - \bm{v}_2$, and choose to reverse this direction as well
by letting $\bm{\bar{\lambda}} \to -\bm{\bar{\lambda}}$. 

With these changes accounted for, we find that the tidal moments in
the black-hole frame are given by 
\begin{subequations} 
\label{Eab_BH} 
\begin{align} 
\bar{\E}_{ab} &= \bar{\E}_{ab}(0\pn) + \bar{\E}_{ab}(1\pn) + 2\pn, \\ 
\bar{\E}_{ab}(0\pn) &= -3 \frac{M_2}{b^3} \bar{n}_\stf{ab}, \\  
\bar{\E}_{ab}(1\pn) &= 3 \frac{M_2 V^2}{b^3} \biggl[ 
\frac{1}{2} (2-q_1) \bar{n}_\stf{ab} 
- \bar{\lambda}_{\langle a} \bar{\lambda}_{b\rangle} \Bigr], \\ 
\end{align} 
\end{subequations} 
\begin{subequations} 
\label{Eabc_BH} 
\begin{align} 
\bar{\E}_{abc} &= \bar{\E}_{abc}(0\pn) + \bar{\E}_{abc}(1\pn) + 2\pn, \\ 
\bar{\E}_{abc}(0\pn) &= -15\frac{M_2}{b^4} \bar{n}_\stf{abc}, \\ 
\bar{\E}_{abc}(1\pn) &= 15 \frac{M_2 V^2}{b^4} \Bigl[
3(1-q_1) \bar{n}_\stf{abc} 
- \bar{\lambda}_{\langle a} \bar{\lambda}_b \bar{n}_{c\rangle} \Bigr], \\
\end{align} 
\end{subequations} 
\begin{subequations} 
\label{Eabcd_BH} 
\begin{align} 
\bar{\E}_{abcd} &= \bar{\E}_{abcd}(0\pn) + \bar{\E}_{abcd}(1\pn) + 2\pn, \\ 
\bar{\E}_{abcd}(0\pn) &= -\frac{105}{2} \frac{M_2}{b^5} \bar{n}_\stf{abcd}, \\ 
\bar{\E}_{abcd}(1\pn) &= \frac{15}{2} \frac{M_2 V^2}{b^5} \biggl[ 
\frac{1}{2} (72 - 75q_1) \bar{n}_\stf{abcd} 
- 6 \bar{\lambda}_{\langle a} \bar{\lambda}_b \bar{n}_c \bar{n}_{d\rangle} \biggr]. 
\end{align} 
\end{subequations} 
and 
\begin{subequations} 
\label{Bmoments_BH} 
\begin{align} 
\bar{\B}_{ab} &= 6 \frac{M_2 V}{b^3} \epsilon_{pq(a} 
\bar{n}_{b)} \bar{n}^p \bar{\lambda}^q, \\ 
\bar{\B}_{abc} &= \frac{45}{2} \frac{M_2 V}{b^4} \epsilon_{pq(a} 
\Bigl( \bar{n}_b \bar{n}_{c)} - \frac{1}{5} \delta_{bc)} \Bigr) 
\bar{n}^p \bar{\lambda}^q, \\ 
\bar{\B}_{abcd} &= 63 \frac{M_2 V}{b^5} \epsilon_{pq(a} 
\Bigl( \bar{n}_b \bar{n}_c \bar{n}_{d)} 
- \frac{3}{7} \delta_{bc} \bar{n}_{d)} 
\Bigr) \bar{n}^p \bar{\lambda}^q.  
\end{align} 
\end{subequations} 

\subsection{Harmonic components of the tidal moments} 
\label{subsec:harmonic} 

The harmonic components of the tidal moments are defined in
Tables~\ref{tab:E_ang} and \ref{tab:B_ang}. Making use of
Eqs.~(\ref{Eab_BH}), (\ref{Eabc_BH}), (\ref{Eabcd_BH}), and
(\ref{Bmoments_BH}), we find that in the case of circular motion, the
nonvanishing components are 
\begin{subequations} 
\label{Eab_harmonic} 
\begin{align} 
\EEbar{q}_0 &= -\frac{1}{2} \frac{M_2}{b^3} 
\Biggl[ 1 + \frac{1}{2} q_1 V^2 + 2\pn \Biggr], \\ 
\EEbar{q}_{2c} &= -\frac{3}{2} \frac{M_2}{b^3} 
\Biggl[ 1 + \frac{1}{2} (q_1 - 4) V^2 + 2\pn \Biggr] 
\cos(2\bar{\omega} \bar{t}), \\ 
\EEbar{q}_{2s} &= -\frac{3}{2} \frac{M_2}{b^3} 
\Biggl[ 1 + \frac{1}{2} (q_1 - 4) V^2 + 2\pn \Biggr] 
\sin(2\bar{\omega} \bar{t}),
\end{align} 
\end{subequations} 
\begin{subequations} 
\label{Eabc_harmonic} 
\begin{align} 
\EEbar{o}_{1c} &= -\frac{3}{2} \frac{M_2}{b^4} 
\biggl[ 1 + \frac{1}{3} (9q_1 - 8) V^2 + 2\pn \biggr] 
\cos(\bar{\omega} \bar{t}), \\ 
\EEbar{o}_{1s} &= -\frac{3}{2} \frac{M_2}{b^4} 
\biggl[ 1 + \frac{1}{3} (9q_1 - 8) V^2 + 2\pn \biggr] 
\sin(\bar{\omega} \bar{t}), \\ 
\EEbar{o}_{3c} &= -\frac{15}{4} \frac{M_2}{b^4} 
\biggl[ 1 + ( 3q_1 - 4 ) V^2 + 2\pn \biggr] 
\cos(3\bar{\omega} \bar{t}), \\ 
\EEbar{o}_{3s} &= -\frac{15}{4} \frac{M_2}{b^4} 
\biggl[ 1 + ( 3q_1 - 4 ) V^2 + 2\pn \biggr] 
\sin(3\bar{\omega} \bar{t}), 
\end{align} 
\end{subequations} 
\begin{subequations} 
\label{Eabcd_harmonic} 
\begin{align} 
\EEbar{h}_{0} &= -\frac{9}{4} \frac{M_2}{b^5} 
\biggl[ 1 + \frac{1}{14}(75q_1 - 68) V^2 + 2\pn \biggr], \\ 
\EEbar{h}_{2c} &= -\frac{15}{2} \frac{M_2}{b^5} 
\biggl[ 1 + \frac{3}{14}(25q_1 - 24) V^2
+ 2\pn \biggr] \cos(2\bar{\omega} \bar{t}), \\ 
\EEbar{h}_{2s} &= -\frac{15}{2} \frac{M_2}{b^5} 
\biggl[ 1 + \frac{3}{14}(25q_1 - 24) V^2
+ 2\pn \biggr] \sin(2\bar{\omega} \bar{t}), \\ 
\EEbar{h}_{4c} &= -\frac{105}{8} \frac{M_2}{b^5} 
\biggl[ 1 + \frac{3}{14}(25q_1 - 28) V^2 
+ 2\pn \biggr] \cos(4\bar{\omega} \bar{t}), \\ 
\EEbar{h}_{4s} &= -\frac{105}{8} \frac{M_2}{b^5} 
\biggl[ 1 + \frac{3}{14}(25q_1 - 28) V^2 
+ 2\pn \biggr] \sin(4\bar{\omega} \bar{t}),
\end{align} 
\end{subequations} 
and 
\begin{subequations} 
\label{Bab_harmonic} 
\begin{align} 
\BBbar{q}_{1c} &= 3 \frac{M_2 V}{b^3} \cos(\bar{\omega} \bar{t}), \\ 
\BBbar{q}_{1s} &= 3 \frac{M_2 V}{b^3} \sin(\bar{\omega} \bar{t}), 
\end{align} 
\end{subequations} 
\begin{subequations} 
\label{Babc_harmonic} 
\begin{align} 
\BBbar{o}_{0} &= 3 \frac{M_2 V}{b^4}, \\ 
\BBbar{o}_{2c} &= 5 \frac{M_2 V}{b^4} \cos(2\bar{\omega} \bar{t}), \\ 
\BBbar{o}_{2s} &= 5 \frac{M_2 V}{b^4} \sin(2\bar{\omega} \bar{t}),
\end{align} 
\end{subequations} 
\begin{subequations} 
\label{Babcd_harmonic} 
\begin{align} 
\BBbar{h}_{1c} &= \frac{45}{4} \frac{M_2 V}{b^5} \cos(\bar{\omega} \bar{t}), \\ 
\BBbar{h}_{1s} &= \frac{45}{4} \frac{M_2 V}{b^5} \sin(\bar{\omega} \bar{t}), \\ 
\BBbar{h}_{3c} &= \frac{105}{8} \frac{M_2 V}{b^5} \cos(3\bar{\omega} \bar{t}), \\ 
\BBbar{h}_{3s} &= \frac{105}{8} \frac{M_2 V}{b^5} \sin(3\bar{\omega} \bar{t}). 
\end{align} 
\end{subequations} 

The harmonic components of the gravitoelectric moments can be
substituted within the tidal potentials introduced in
Sec.~\ref{sec:potentials}. With $\bar{\Omega}^a = [\sin\bar{\theta}
\cos\bar{\phi}, \sin\bar{\theta}\sin\bar{\phi},\cos\bar{\theta}]$, we
have that 
\begin{subequations} 
\label{Epotentials} 
\begin{align} 
\EEbar{q} &:= \bar{\E}_{ab} \bar{\Omega}^a \bar{\Omega}^b 
= \frac{1}{2} \frac{M_2}{b^3} 
  \biggl[ 1 + \frac{1}{2} q_1 V^2 + 2\pn \biggr] 
  (3\cos^2\bar{\theta} - 1) 
- \frac{3}{2} \frac{M_2}{b^3} 
  \biggl[ 1 + \frac{1}{2} (q_1 - 4) V^2 + 2\pn\biggr] 
  \sin^2\bar{\theta} \cos(2\bar{\psi}), \\ 
\EEbar{o} &:= 
\bar{\E}_{abc} \bar{\Omega}^a \bar{\Omega}^b \bar{\Omega}^c 
= \frac{9}{4} \frac{M_2}{b^4} 
  \biggl[ 1 + \frac{1}{3} (9q_1 - 8) V^2 + 2\pn \biggr] 
  \sin\bar{\theta} (5\cos^2\bar{\theta}-1) \cos(\bar{\psi}) 
\nonumber \\ & \quad \mbox{} 
- \frac{15}{4} \frac{M_2}{b^4} 
  \biggl[ 1 + (3q_1 - 4) V^2 + 2\pn \biggr] 
  \sin^3\bar{\theta} \cos(3\bar{\psi}), \\ 
\EEbar{h} &:= 
\bar{\E}_{abcd} \bar{\Omega}^a \bar{\Omega}^b \bar{\Omega}^c \bar{\Omega}^d 
= -\frac{9}{16} \frac{M_2}{b^5} 
  \biggl[ 1 + \frac{1}{14} (75q_1 - 68) V^2 + 2\pn \biggr] 
  (35\cos^4\bar{\theta} - 30\cos^2\bar{\theta} + 3)   
\nonumber \\ & \quad \mbox{} 
+ \frac{15}{4} \frac{M_2}{b^5} 
  \biggl[ 1 + \frac{3}{14} (25q_1 - 24) V^2 + 2\pn \biggr] 
  \sin^2\bar{\theta} (7\cos^2\bar{\theta}-1) \cos(2\bar{\psi}),
\nonumber \\ & \quad \mbox{} 
- \frac{105}{16} \frac{M_2}{b^5} 
  \biggl[ 1 + \frac{3}{14} (25q_1 - 28) V^2 + 2\pn \biggr] 
  \sin^4\bar{\theta} \cos(4\bar{\psi}),
\end{align} 
\end{subequations} 
where $\bar{\psi} := \bar{\phi} - \bar{\omega} \bar{t}$. 

\section{Geometry of a tidally deformed horizon} 
\label{sec:horizon} 

The induced metric on the event horizon of a tidally deformed black
hole was constructed in Sec.~IV B of Ref.~\cite{poisson-vlasov:10}  --- refer
to their Eq.~(4.3). This metric is expressed in terms of tidal moments
defined in a ``horizon calibration'' that differs from the
``post-Newtonian calibration'' adopted in this work. The
transformation between these calibrations was detailed in
Sec.~\ref{sec:calibrations}, and Eqs.~(\ref{H-PNcalibration}) reveal
that the tidal moments are identical up to corrections of order $v^3$ 
and beyond, 
\begin{equation} 
\E_{ab}(\mbox{H}) = \E_{ab}(\mbox{PN}) + 1.5\pn, \qquad 
\E_{abc}(\mbox{H}) = \E_{abc}(\mbox{PN}) + 1.5\pn, 
\end{equation} 
and 
\begin{equation} 
\B_{ab}(\mbox{H}) = \B_{ab}(\mbox{PN}) + 1.5\pn, \qquad 
\B_{abc}(\mbox{H}) = \B_{abc}(\mbox{PN}) + 1.5\pn. 
\end{equation} 
These corrections are of no concern in the determination of the
horizon's geometry through $1\pn$ order. We recall that 
$\E_{abcd}$ and $\B_{abcd}$ have a fixed calibration, and are
therefore the same in the horizon and post-Newtonian calibrations. 

We begin with an examination of the quadrupole, octupole, and
hexadecapole contributions to the induced metric, neglecting all
contributions that are bilinear in the tidal moments. We have  
\begin{equation} 
\gamma_{AB} = (2M_1)^2 \biggl[ \Omega_{AB} 
- \frac{2}{3} M_1^2 \bigl( \EE{q}_{AB} + \BB{q}_{AB} \bigr) 
- \frac{2}{15} M_1^3 \bigl( \EE{o}_{AB} + \BB{o}_{AB} \bigr) 
- \frac{2}{105} M_1^4 \bigl( \EE{h}_{AB} + \BB{h}_{AB} \bigr) 
\biggr], 
\end{equation} 
where the tidal potentials are given in the horizon calibration. This
metric reflects a choice of coordinates on the horizon: $\theta^A$ is
defined to be constant on the horizon's null generators, and the tidal
moments are given as functions of the advanced-time coordinate $v$. To
keep the notation uncluttered, we no longer make use of the overbar on
the coordinates and tidal moments; it is now understood that all
expressions refer to the black-hole frame. 

Continuing to neglect all terms that are bilinear in the tidal
moments, the Ricci scalar associated with the induced metric is given
by   
\begin{equation} 
{\cal R} = \frac{1}{2M_1^2} \biggl( 1 
- 4 M_1^2 \EE{q} 
- \frac{4}{3} M_1^3 \EE{o}  
- \frac{2}{7} M_1^4 \EE{h} \biggr), 
\end{equation} 
where $\EE{q}$, $\EE{o}$, and $\EE{h}$ are the tidal potentials
introduced in Sec.~\ref{sec:potentials} and evaluated for circular
motion in Sec.~\ref{subsec:harmonic}. There is no gravitomagnetic
scalar potential, and $\B_{ab}$, $\B_{abc}$ and $\B_{abcd}$ do not
appear in the Ricci scalar. 

It is convenient to relate the
horizon's intrinsic geometry to that of a fictitious two-dimensional
surface embedded in a three-dimensional flat space. The surface is
described by   
\begin{equation} 
r = 2M_1 \bigl[ 1 + \varepsilon(\theta^A) \bigr],  
\end{equation} 
where $\varepsilon$ is a displacement function that represents the
tidal deformation. With 
\begin{equation} 
\varepsilon = -M_1^2 \EE{q} - \frac{2}{15} M_1^3 \EE{o}
- \frac{1}{63} M_1^4 \EE{h},  
 \end{equation} 
the embedded surface possesses the same intrinsic Ricci scalar as the
horizon.   

Looking separately at each multipole component of the displacement
function, we substitute Eqs.~(\ref{Epotentials}) and obtain 
\begin{subequations} 
\begin{align} 
\varepsilon[\ell=2] &= -\frac{M_1^2 M_2}{2b^3} \biggl\{ 
\biggl[ 1 + \frac{1}{2} q_1 V^2 + 2\pn \biggr] 
  (3\cos^2\theta - 1) 
\nonumber \\ & \quad \mbox{} 
- 3 \biggl[ 1 + \frac{1}{2} (q_1 - 4) V^2 + 2\pn\biggr] 
  \sin^2\theta \cos(2\psi) + 1.5\pn \biggr\}, \\ 
\varepsilon[\ell=3] &= -\frac{M_1^3 M_2}{10 b^4} \biggl\{ 
3\biggl[ 1 + \frac{1}{3} (9q_1 - 8) V^2 + 2\pn \biggr] 
  \sin\theta (5\cos^2\theta-1) \cos(\psi) 
\nonumber \\ & \quad \mbox{} 
- 5\biggl[ 1 + (3q_1 - 4) V^2 + 2\pn \biggr] 
  \sin^3\theta \cos(3\psi) + 1.5\pn \biggr\}, \\ 
\varepsilon[\ell=4] &= -\frac{M_1^4 M_2}{336 b^5} \biggl\{ 
-3  \biggl[ 1 + \frac{1}{14} (75q_1 - 68) V^2 + 2\pn \biggr] 
  (35\cos^4\theta - 30\cos^2\theta + 3)   
\nonumber \\ & \quad \mbox{} 
 + 20\biggl[ 1 + \frac{3}{14} (25q_1 - 24) V^2 + 2\pn \biggr] 
  \sin^2\theta (7\cos^2\theta-1) \cos(2\psi),
\nonumber \\ & \quad \mbox{} 
-35 \biggl[ 1 + \frac{3}{14} (25q_1 - 28) V^2 + 2\pn \biggr] 
  \sin^4\theta \cos(4\psi) + 2\pn \biggr\},
\end{align} 
\end{subequations} 
where $\psi :=\phi - \bar{\omega} v$. 
The quadrupole displacement scales as $M_1^2 M_2/b^3$, which we take
to be of Newtonian order, and it features a $1\pn$ correction of fractional
order $V^2$. The octupole displacement comes with an additional factor
of order $M_1/b = q_1 V^2$ and therefore represents a $1\pn$
correction to the quadrupole deformation. The hexadecapole
displacement, with its additional factor of order $(M_1/b)^2$,
represents a $2\pn$ correction to the quadrupole displacement. The
neglected bilinear terms would scale as $M_1^4 M_2^2/b^6$ and
therefore represent a $3\pn$ correction to the leading term.  

If we truncate the displacement function to an overall accuracy of
$1\pn$ order, we obtain     
\begin{align} 
\varepsilon &= \frac{M_1^2 M_2}{2 b^3} \Biggl\{
-\biggl[ 1 + \frac{1}{2} q_1 V^2 \biggr] (3\cos^2\theta-1) 
+ 3 \biggl[ 1 + \frac{1}{2} (q_1 - 4)V^2 \biggr] \sin^2\theta \cos(2\psi) 
\nonumber \\ & \quad \mbox{} 
- \frac{3}{5} q_1 V^2 \sin\theta(5\cos^2\theta-1) \cos(\psi) 
+ q_1 V^2 \sin^3\theta\cos(3\psi) + 1.5\pn \Biggr\}. 
\end{align} 
As was discussed further in Sec.~\ref{sec:intro}, the displacement
function describes a tidal bulge aligned with 
$\phi = \bar{\omega} v$.     

A consequence of the black hole's tidal interaction is the fact that
its mass slowly increases. The equation that describes this tidal
heating is derived in Sec.~IV D of Ref.~\cite{poisson-vlasov:10}. 
According to their Eqs.~(4.14) and (4.16), the rate of change of the
mass is given by   
\begin{equation} 
\dot{M_1} = \dot{M_1}[\ell = 2] + \dot{M_1}[\ell = 3] 
+ \mbox{higher order}, 
\end{equation} 
where 
\begin{subequations} 
\begin{align} 
\dot{M_1}[\ell = 2] &= \frac{16}{45} M_1^6 
\bigl( \dot{\E}_{ab} \dot{\E}^{ab} 
+ \dot{\B}_{ab} \dot{\B}^{ab}\bigr), \\ 
\dot{M_1}[\ell = 3] &= \frac{16}{4725} M_1^8 
\biggr( \dot{\E}_{abc} \dot{\E}^{abc} 
+ \frac{16}{9} \dot{\B}_{abc} \dot{\B}^{abc} \biggr). 
\end{align} 
\end{subequations}  
The influence of the hexadecapole moments on the mass is not currently
known, and we therefore exclude it from the expression. We insert our
previous expressions for the quadrupole and octupole contributions,
and obtain 
\begin{subequations} 
\begin{align} 
\dot{M_1}[\ell = 2] &= \frac{32}{5} q_1^6 q_2^2 
\biggl( \frac{M}{b} \biggr)^9 
\Bigl[ 1 + (q_1^2 - 6) V^2 + 1.5\pn \Bigr], \\ 
\dot{M_1}[\ell = 3] &= \frac{64}{35} q_1^8 q_2^2 
\biggl( \frac{M}{b} \biggr)^{11}  
\biggl[ 1 +  \Bigl( q_1^2 + 5q_1 - \frac{175}{18} \Bigr) V^2 
+ 1.5\pn \biggr]. 
\end{align} 
\end{subequations} 
The leading, quadrupole term scales as $(M/b)^9 = V^{18}$, and it
represents a correction of $4\pn$ order to the energy lost to
gravitational waves. The octupole term scales as $(M/b)^{11} =
V^{22}$; it represents a $4\pn$ correction to the quadrupole
contribution, and a $8\pn$ correction to the radiated energy.   

\begin{acknowledgments} 
This work was supported by the Natural Sciences and Engineering
Research Council of Canada.     
\end{acknowledgments} 

\appendix 

\section{Decomposition of Cartesian tensors into irreducible pieces} 
\label{sec:irreducible} 

For our purposes in this Appendix, the irreducible pieces of a
Cartesian tensor consist of its traces, symmetric-tracefree (STF)
pieces, antisymmetric pieces, and all mixtures among these basic
building blocks. We need decompositions for fully symmetric tensors 
$S_{abc \cdots}$ and for tensors $A_{jabc \cdots}$ that are fully
symmetric with respect to the $abc \cdots$ indices. We use round
brackets $(a b c \cdots)$ to denote the complete symmetrization of
indices, square brackets $[a b c \cdots]$ to denote a complete
antisymmetrization, and angular brackets $\stf{a b c \cdots}$
to denote an STF combination of indices (complete symmetrization and
removal of all traces). 

The decomposition of a fully symmetric tensor into irreducible pieces
is accomplished with 
\begin{subequations} 
\label{Sdecomp} 
\begin{align} 
\mbox{}_2 S_{ab} &= \mbox{}_2 S_\stf{ab} 
+ \frac{1}{3} \delta_{ab}\, \mbox{}_2 S, 
\label{Sdecomp2} \\ 
\mbox{}_3 S_{abc} &= \mbox{}_3 S_\stf{abc} 
+ \frac{1}{5} \bigl( \delta_{ab}\, \mbox{}_3 S_c 
+ \delta_{ac}\, \mbox{}_3 S_b + \delta_{bc}\, \mbox{}_3 S_a \bigr), 
\label{Sdecomp3} \\ 
\mbox{}_4 S_{abcd} &= \mbox{}_4 S_\stf{abcd} + \frac{1}{7} \bigl( 
  \delta_{ab}\, \mbox{}_4 S_\stf{cd} + \delta_{ac}\, \mbox{}_4 S_\stf{bd} 
+ \delta_{ad}\, \mbox{}_4 S_\stf{bc} + \delta_{bc}\, \mbox{}_4 S_\stf{ad} 
+ \delta_{bd}\, \mbox{}_4 S_\stf{ac} + \delta_{cd}\, \mbox{}_4 S_\stf{ab} \bigr) 
\nonumber \\ & \quad \mbox{} 
+ \frac{1}{15} \bigl( \delta_{ab} \delta_{cd} + \delta_{ac}\delta_{bd}
+ \delta_{ad} \delta_{bc} \bigr)\, \mbox{}_4 S, 
\label{Sdecomp4} \\ 
\mbox{}_5 S_{abcde} &= \mbox{}_5 S_\stf{abcde} + \frac{1}{9} \bigl( 
   \delta_{ab}\, \mbox{}_5 S_\stf{cde} + \delta_{ac}\, \mbox{}_5 S_\stf{bde} 
+ \delta_{ad}\, \mbox{}_5 S_\stf{bce} + \delta_{ae}\, \mbox{}_5 S_\stf{bcd} 
\nonumber \\ & \quad \mbox{} 
+ \delta_{bc}\, \mbox{}_5 S_\stf{ade} + \delta_{bd}\, \mbox{}_5 S_\stf{ace} 
+ \delta_{be}\, \mbox{}_5 S_\stf{acd} + \delta_{cd}\, \mbox{}_5 S_\stf{abe} 
+ \delta_{ce}\, \mbox{}_5 S_\stf{abd} + \delta_{de}\, \mbox{}_5 S_\stf{abc} \bigr) 
\nonumber \\ & \quad \mbox{} 
+ \frac{1}{35} \Bigl[ \bigl( \delta_{ab} \delta_{cd} 
+ \delta_{ac} \delta_{bd} + \delta_{ad} \delta_{bc} \bigr)\, \mbox{}_5 S_e      
+ \bigl( \delta_{ea} \delta_{bc} 
+ \delta_{eb} \delta_{ac} + \delta_{ec} \delta_{ab} \bigr)\, \mbox{}_5 S_d
+ \bigl( \delta_{de} \delta_{ab} 
+ \delta_{da} \delta_{eb} + \delta_{db} \delta_{ea} \bigr)\, \mbox{}_5 S_c
\nonumber \\ & \quad \mbox{} 
+ \bigl( \delta_{cd} \delta_{ea} 
+ \delta_{ce} \delta_{da} + \delta_{ca} \delta_{de} \bigr)\, \mbox{}_5 S_b
+ \bigl( \delta_{bc} \delta_{de} 
+ \delta_{bd} \delta_{ce} + \delta_{be} \delta_{cd} \bigr)\, \mbox{}_5 S_a 
\Bigr], 
\label{Sdecomp5} 
\end{align}
\end{subequations} 
where $\mbox{}_2 S := \mbox{}_2 S_a^{\ a}$, 
$\mbox{}_3 S_a := \mbox{}_3 S_{ab}^{\ \ b}$, 
$\mbox{}_4 S_\stf{a b}$ is the STF piece of 
$\mbox{}_4 S_{abc}^{\ \ \ c}$, 
$\mbox{}_4 S := \mbox{}_4 S_{ab}^{\ \ ab}$, 
$\mbox{}_5 S_\stf{abc}$ is the STF piece of 
$\mbox{}_5 S_{abcd}^{\ \ \ \ d}$, and 
$\mbox{}_5 S_a := \mbox{}_5 S_{abc}^{\ \ \ bc}$. The number 
before each tensor symbol indicates the rank of the parent tensor. The
6 independent components of $\mbox{}_2 S_{ab}$ are packaged into the 5
independent components of $\mbox{}_2 S_\stf{ab}$ and the single
component of $\mbox{}_2 S$. The 10 independent components of
$\mbox{}_3 S_{abc}$ are packaged into the 7 independent components of
$\mbox{}_3 S_\stf{abc}$ and the 3 components of $\mbox{}_3 S_a$. The
15 independent components of $\mbox{}_4 S_{abcd}$ are packaged into
the 9 independent components of $\mbox{}_4 S_\stf{abcd}$, the 5
independent components of $\mbox{}_4 S_\stf{ab}$, and the single
component of $\mbox{}_4 S$. And finally, the 21 independent
components of $\mbox{}_5 S_{abcde}$ are packaged into the 11
independent components of $\mbox{}_5 S_\stf{abcde}$, the 7 
independent components of $\mbox{}_4 S_\stf{abc}$, and the 3 
components of $\mbox{}_5 S_a$.    

The decomposition of a tensor $A_{jab} = A_{j(ab)}$ is accomplished
with 
\begin{equation} 
\mbox{}_3 A_{jab} = \mbox{}_3 A_\stf{jab} 
+ \epsilon_{ja}^{\ \ \, p}\, \mbox{}_3 A_\stf{pb} 
+ \epsilon_{jb}^{\ \ \, p}\, \mbox{}_3 A_\stf{pa} 
+ \delta_{ja} \bigl( \mbox{}_3 \bar{A}_b - \mbox{}_3 A_b \bigr) 
+ \delta_{jb} \bigl( \mbox{}_3 \bar{A}_a - \mbox{}_3 A_a \bigr) 
+ \delta_{ab} \bigl( \mbox{}_3 \bar{A}_j + 2\, \mbox{}_3 A_j \bigr), 
\label{Adecomp3} 
\end{equation} 
where 
\begin{subequations} 
\label{Adecomp3_list} 
\begin{align} 
\mbox{}_3 \bar{A}_a &:= \frac{1}{15} \bigl( \mbox{}_3 A_{ap}^{\ \ \, p} 
+ 2\, \mbox{}_3 A^p_{\ pa} \bigr), \\ 
\mbox{}_3 A_a &:= \frac{1}{6} \bigl( \mbox{}_3 A_{ap}^{\ \ \, p} 
- \mbox{}_3 A^p_{\ pa} \bigr), \\
\mbox{}_3 A_\stf{ab} &:= 
\frac{1}{3} \epsilon_{pq(a}\, \mbox{}_3 A^{pq}_{\ \ b)}. 
\end{align} 
\end{subequations} 
The 18 independent components of $\mbox{}_3 A_{jab}$ are packaged into
the 7 independent components of $\mbox{}_3 A_\stf{jab}$, the 5
independent components of $\mbox{}_3 A_\stf{ab}$, the 3 components of
$\mbox{}_3 A_{a}$, and the 3 components of $\mbox{}_3 \bar{A}_a$.  

The decomposition of a tensor $A_{jabc} = A_{j(abc)}$ into irreducible
pieces is given by 
\begin{align} 
\mbox{}_4 A_{jabc} &= \mbox{}_4 A_\stf{jabc} 
+ \epsilon_{ja}^{\ \ \, p}\, \mbox{}_4 A_\stf{pbc} 
+ \epsilon_{jc}^{\ \ \, p}\, \mbox{}_4 A_\stf{pab} 
+ \epsilon_{jb}^{\ \ \, p}\, \mbox{}_4 A_\stf{pca} 
\nonumber \\ & \quad \mbox{} 
+ \delta_{ja} \bigl( \mbox{}_4 \bar{A}_\stf{bc} + \mbox{}_4 A_\stf{bc} \bigr) 
+ \delta_{jc} \bigl( \mbox{}_4 \bar{A}_\stf{ab} + \mbox{}_4 A_\stf{ab} \bigr) 
+ \delta_{jb} \bigl( \mbox{}_4 \bar{A}_\stf{ca} + \mbox{}_4 A_\stf{ca} \bigr) 
\nonumber \\ & \quad \mbox{} 
+ \delta_{ab} \bigl( \mbox{}_4 \bar{A}_\stf{jc} - \mbox{}_4 A_\stf{jc} \bigr) 
+ \delta_{ac} \bigl( \mbox{}_4 \bar{A}_\stf{jb} - \mbox{}_4 A_\stf{jb} \bigr) 
+ \delta_{bc} \bigl( \mbox{}_4 \bar{A}_\stf{ja} - \mbox{}_4 A_\stf{ja}\bigr) 
\nonumber \\ & \quad \mbox{} 
+ \bigl( \delta_{ab}\, \epsilon_{jc}^{\ \ \, p} 
+ \delta_{ac}\, \epsilon_{jb}^{\ \ \, p} 
+ \delta_{bc}\, \epsilon_{ja}^{\ \ \, p} \bigl)\, \mbox{}_4 A_p 
+ \bigl( \delta_{ja} \delta_{bc} + \delta_{jb} \delta_{ac} 
+ \delta_{jc} \delta_{ab} \bigr)\, \mbox{}_4 \bar{A}, 
\label{Adecomp4} 
\end{align} 
where 
\begin{subequations} 
\label{Adecomp4_list} 
\begin{align} 
\mbox{}_4 \bar{A} &:= \frac{1}{15}\, \mbox{}_4 A^{p\ \ q}_{\ pq}, \\ 
\mbox{}_4 \bar{A}_\stf{ab} &:= \frac{1}{28} \bigl( 
\mbox{}_4 A_{abp}^{\ \ \ \, p} + A_{bap}^{\ \ \ \, p} 
+ 2\, \mbox{}_4 A^{p}_{\ pab} \bigr) 
- \frac{1}{21} \delta_{ab}\, \mbox{}_4 A^{p\ \ q}_{\ pq}, \\ 
\mbox{}_4 A_a &:= \frac{1}{10} \epsilon_{apq}\, 
       \mbox{}_4 A^{pqr}_{\ \ \ r}, \\ 
\mbox{}_4 A_\stf{ab} &:= \frac{1}{12} \bigl( 
2\, \mbox{}_4 A^p_{\ pab} - \mbox{}_4 A_{a b p}^{\ \ \ \, p} 
- \mbox{}_4 A_{b a p}^{\ \ \ \, p} \bigr), \\
\mbox{}_4 A_\stf{abc} &:= 
\frac{1}{4} \epsilon_{pq(a}\, \mbox{}_4 A^{pq}_{\ \ bc)} 
- \frac{1}{6} \bigl( \delta_{ab}\, \mbox{}_4 A_c 
+ \delta_{ac}\, \mbox{}_4 A_b 
+ \delta_{bc}\, \mbox{}_4 A_a \bigr). 
\end{align} 
\end{subequations} 
The 30 independent components of $\mbox{}_4 A_{jabc}$ are packaged
into the 9 independent components of $\mbox{}_4 A_\stf{jabc}$, the 7
independent components of $\mbox{}_4 A_\stf{abc}$, the 5 independent
components of $\mbox{}_4 A_\stf{ab}$, the 5 independent
components of $\mbox{}_4 \bar{A}_\stf{ab}$, the 3 components of
$\mbox{}_4 A_a$, and the single component of 
$\mbox{}_4 \bar{A}$.   

The decomposition of a tensor $A_{jabcd} = A_{j(abcd)}$ is accomplished
with 
\begin{align} 
\mbox{}_5 A_{jabcd} &:= \mbox{}_5 A_\stf{jabcd} 
+ \epsilon_{ja}^{\ \ \, p}\, \mbox{}_5 A_\stf{pbcd} 
+ \epsilon_{jd}^{\ \ \, p}\, \mbox{}_5 A_\stf{pabc} 
+ \epsilon_{jc}^{\ \ \, p}\, \mbox{}_5 A_\stf{pdab} 
+ \epsilon_{jb}^{\ \ \, p}\, \mbox{}_5 A_\stf{pcda} 
\nonumber \\ & \quad \mbox{} 
+ \delta_{ja} \bigl( \mbox{}_5 \bar{A}_\stf{bcd} 
    - 3\, \mbox{}_5 A_\stf{bcd} \bigr) 
+ \delta_{jb} \bigl( \mbox{}_5 \bar{A}_\stf{cda} 
    - 3\, \mbox{}_5 A_\stf{cda} \bigr) 
+ \delta_{jc} \bigl( \mbox{}_5 \bar{A}_\stf{dab} 
    - 3\, \mbox{}_5 A_\stf{dab} \bigr) 
+ \delta_{jd} \bigl( \mbox{}_5 \bar{A}_\stf{abc} 
    - 3\, \mbox{}_5 A_\stf{abc} \bigr) 
\nonumber \\ & \quad \mbox{} 
+ \delta_{ab} \bigl( \mbox{}_5 \bar{A}_\stf{jcd} 
    + 2\, \mbox{}_5 A_\stf{jcd} \bigr) 
+ \delta_{ac} \bigl( \mbox{}_5 \bar{A}_\stf{jbd} 
    + 2\, \mbox{}_5 A_\stf{jbd} \bigr) 
+ \delta_{ad} \bigl( \mbox{}_5 \bar{A}_\stf{jbc} 
    + 2\, \mbox{}_5 A_\stf{jbc} \bigr) 
+ \delta_{bc} \bigl( \mbox{}_5 \bar{A}_\stf{jad} 
    + 2\, \mbox{}_5 A_\stf{jad} \bigr) 
\nonumber \\ & \quad \mbox{} 
+ \delta_{bd} \bigl( \mbox{}_5 \bar{A}_\stf{jac} 
    + 2\, \mbox{}_5 A_\stf{jac} \bigr) 
+ \delta_{cd} \bigl( \mbox{}_5 \bar{A}_\stf{jab} 
    + 2\, \mbox{}_5 A_\stf{jab} \bigr) 
\nonumber \\ & \quad \mbox{} 
+ \epsilon_{ja}^{\ \ \, p} \bigl( \delta_{bc}\, \mbox{}_5 A_\stf{pd}
    + \delta_{db}\, \mbox{}_5 A_\stf{pc} 
    + \delta_{cd}\, \mbox{}_5 A_\stf{pb} \bigr) 
+ \epsilon_{jd}^{\ \ \, p} \bigl( \delta_{ab}\, \mbox{}_5 A_\stf{pc}
    + \delta_{ca}\, \mbox{}_5 A_\stf{pb} 
    + \delta_{bc}\, \mbox{}_5 A_\stf{pa} \bigr) 
\nonumber \\ & \quad \mbox{} 
+ \epsilon_{jc}^{\ \ \, p} \bigl( \delta_{da}\, \mbox{}_5 A_\stf{pb}
    + \delta_{bd}\, \mbox{}_5 A_\stf{pa} 
    + \delta_{ab}\, \mbox{}_5 A_\stf{pd} \bigr) 
+ \epsilon_{jb}^{\ \ \, p} \bigl( \delta_{cd}\, \mbox{}_5 A_\stf{pa}
    + \delta_{ac}\, \mbox{}_5 A_\stf{pd} 
    + \delta_{da}\, \mbox{}_5 A_\stf{pc} \bigr)
\nonumber \\ & \quad \mbox{} 
+ \bigl( \delta_{ja} \delta_{bc} + \delta_{jc} \delta_{ab} +
    \delta_{jb} \delta_{ca} \bigr) 
    \bigl( \mbox{}_5 \bar{A}_d + \mbox{}_5 A_d \bigr) 
+ \bigl( \delta_{jd} \delta_{ab} + \delta_{jb} \delta_{da} +
    \delta_{ja} \delta_{bd} \bigr) 
    \bigl( \mbox{}_5 \bar{A}_c + \mbox{}_5 A_c \bigr) 
\nonumber \\ & \quad \mbox{} 
+ \bigl( \delta_{jc} \delta_{da} + \delta_{ja} \delta_{cd} +
    \delta_{jd} \delta_{ac} \bigr) 
    \bigl( \mbox{}_5 \bar{A}_b + \mbox{}_5 A_b \bigr) 
+ \bigl( \delta_{jb} \delta_{cd} + \delta_{jd} \delta_{bc} +
    \delta_{jc} \delta_{db} \bigr) 
    \bigl( \mbox{}_5 \bar{A}_a + \mbox{}_5 A_a \bigr) 
\nonumber \\ & \quad \mbox{} 
+ \bigl( \delta_{ab} \delta_{cd} + \delta_{ac} \delta_{bd} +
    \delta_{ad} \delta_{bc} \bigr) 
    \bigl( \mbox{}_5 \bar{A}_j - 4\, \mbox{}_5 A_j \bigr), 
\label{Adecomp5} 
\end{align} 
where 
\begin{subequations} 
\label{Adecomp5_list} 
\begin{align} 
\mbox{}_5 \bar{A}_a &:= \frac{1}{175} \bigl( 
    \mbox{}_5 A_{ap\ q}^{\ \ \, p\ q} + 4\, \mbox{}_5 A_{\ paq}^{p\ \ \ \, q}
    \bigr), \\
\mbox{}_5 \bar{A}_\stf{abc} &:= \frac{1}{45} \bigl( 
     \mbox{}_5 A_{abcp}^{\ \ \ \ \, p}  + \mbox{}_5 A_{cabp}^{\ \ \ \ \, p}  
    + \mbox{}_5 A_{bcap}^{\ \ \ \ \, p} + 2\, \mbox{}_5 A^p_{\ pabc} 
   \bigr) - \frac{7}{9} \bigl( \delta_{ab}\, \mbox{}_5 \bar{A}_c
   + \delta_{ac}\, \mbox{}_5 \bar{A}_b 
   + \delta_{bc}\, \mbox{}_5 \bar{A}_a \bigr), \\ 
\mbox{}_5 A_a &:= \frac{1}{50} \bigl( 
    \mbox{}_5 A_{\ paq}^{p\ \ \ \, q} - \mbox{}_5 A_{ap\ q}^{\ \ \, p\ q}  
    \bigr), \\
\mbox{}_5 A_\stf{ab} &:= \frac{1}{21} \epsilon_{pq(a}\, 
   \mbox{}_5 A^{pq\ \ \ r}_{\ \ \, b)r}, \\ 
\mbox{}_5 A_\stf{abc} &:= \frac{1}{60} \bigl( 
     \mbox{}_5 A_{abcp}^{\ \ \ \ \, p}  + \mbox{}_5 A_{cabp}^{\ \ \ \ \, p}  
    + \mbox{}_5 A_{bcap}^{\ \ \ \ \, p} - 3\, \mbox{}_5 A^p_{\ pabc} 
   \bigr) + \frac{1}{6} \bigl( \delta_{ab}\, \mbox{}_5 A_c
   + \delta_{ac}\, \mbox{}_5 A_b 
   + \delta_{bc}\, \mbox{}_5 A_a \bigr), \\ 
\mbox{}_5 A_\stf{abcd} &:= 
   \frac{1}{5} \epsilon_{pq(a}\, \mbox{}_5 A^{pq}_{\ \ bcd)} 
- \frac{3}{10} \bigl( \delta_{ab}\, \mbox{}_5 A_\stf{cd}
+ \delta_{ac}\, \mbox{}_5 A_\stf{bd}
+ \delta_{ad}\, \mbox{}_5 A_\stf{bc}
+ \delta_{bc}\, \mbox{}_5 A_\stf{ad}
+ \delta_{bd}\, \mbox{}_5 A_\stf{ac}
+ \delta_{cd}\, \mbox{}_5 A_\stf{ab} \bigr). 
\end{align} 
\end{subequations} 
The 45 independent components of $\mbox{}_5 A_{jabcd}$ are packaged
into the 11 independent components of $\mbox{}_5 A_\stf{jabcd}$, the 9
independent components of $\mbox{}_5 A_\stf{abcd}$, the 7 independent
components of $\mbox{}_5 A_\stf{abc}$, the 7 independent
components of $\mbox{}_5 \bar{A}_\stf{abc}$, the 5 independent
components of $\mbox{}_5 A_\stf{ab}$, the 3 components of
$\mbox{}_5 A_a$, and the 3 components of $\mbox{}_5 \bar{A}_a$. 

The decompositions of Eqs.~(\ref{Sdecomp}) are familiar (see, for
example, Sec.~1.5.3 of Ref.~\cite{poisson-will:14}), and there is no need
to provide a derivation. The decompositions of Eqs.~(\ref{Adecomp3}),
(\ref{Adecomp4}), and (\ref{Adecomp5}) are perhaps less familiar, and
we snow sketch their derivation. 

Equation (\ref{Adecomp3}) can be obtained from the identity $A_{jab} = 
A_{(jab)} + \frac{2}{3} (A_{[ja]b} + A_{[jb]a})$. The first
term is decomposed into trace and STF pieces, and the second and third
terms are re-expressed in terms of a general tensor $X_{pa}$ according
to $A_{[ja]b} = \epsilon_{ja}^{\ \ \, p} X_{pb}$. In the next step,
$X_{ab}$ is decomposed into trace, STF, and antisymmetric pieces
according to $X_{ab} = \frac{1}{3} \delta_{ab} X + X_\stf{ab} +
\epsilon_{ab}^{\ \ \, p} Y_p$, where $X := X_p^{\ p}$ and $Y_p$ is a
vector. Making the substitutions and simplifying
returns Eq.~(\ref{Adecomp3}) after introducing the definitions of
Eqs.~(\ref{Adecomp3_list}), with $\mbox{}_3 \bar{A}_a$ representing
the trace part of $A_{(jab)}$, $\mbox{}_3 A_a$ substituting for $Y_a$,
and $\mbox{}_3 A_\stf{ab}$ substituting for $X_\stf{ab}$; $X$ does not
appear in the final expression.  

Equation (\ref{Adecomp4}) is obtained in a similar way. The starting
identity is $A_{jabc} = A_{(jabc)} + \frac{1}{2} (A_{[ja]bc} 
+ A_{[jc]ab} + A_{[jb]ca})$, and the first term
is decomposed into trace and STF pieces, while the remaining terms are
rewritten in terms of a tensor $X_{pab}$ according to 
$A_{[ja]bc} = \epsilon_{ja}^{\ \ \, p} X_{pbc}$. Because $X_{pab}$ is
symmetric in the last two indices, it can be decomposed according to 
Eq.~(\ref{Adecomp3}). Making the substitutions, simplifying, and
introducing the definitions of Eqs.~(\ref{Adecomp4_list}) returns
Eq.~(\ref{Adecomp4}). 

For Eq.~(\ref{Adecomp5}) we begin with $A_{jabcd} = A_{(jabcd)} 
+ \frac{2}{5} (A_{[ja]bcd} + A_{[jd]abc} + A_{[jc]dab}  
+ A_{[jb]cda})$ and write 
$A_{[ja]bcd} = \epsilon_{ja}^{\ \ \, p} X_{pbcd}$, with $X_{pbcd}$
symmetric in the last three indices. This tensor can be decomposed
according to Eq.~(\ref{Adecomp4}), and Eq.~(\ref{Adecomp5}) eventually
follows. 

\section{Derivatives of distance function} 
\label{sec:distance}

We introduce the Cartesian vector $\bm{s} := \bm{x} - \bm{r}(t)$, in
which $\bm{x}$ denotes the variable position of a point in
three-dimensional flat space, and $\bm{r}(t)$ is a known vectorial function of
time. We let $\bm{v} := d\bm{r}/dt$ and $\bm{a} := d\bm{v}/dt$. The
``distance function'' of the title is defined by $s := |\bm{s}|$, and
we also introduce the unit vector $\bm{\hat{s}} := \bm{s}/s$. We let
$v_s := \bm{v} \cdot \bm{\hat{s}}$ and $a_s := \bm{a} \cdot
\bm{\hat{s}}$. Below we record a number of derivatives of the distance
function, which are needed in the main text. These results are easily
derived from the elementary identities  
\begin{subequations} 
\begin{align} 
\partial_a s &= \hat{s}_a, \\ 
\partial_a \hat{s}_b &= 
\frac{1}{s} (\delta_{ab} - \hat{s}_a \hat{s}_b ), \\ 
\partial_t s &= -v_s, \\ 
\partial_t \hat{s}_a &= -\frac{1}{s} (v_a - v_s \hat{s}_a), \\ 
\partial_a v_s &= \frac{1}{s} (v_a - v_s \hat{s}_a), \\ 
\partial_a a_s &= \frac{1}{s} (a_a - a_s \hat{s}_a), \\ 
\partial_t v_s &= a_s - \frac{1}{s}( v^2 - v_s^2 ), 
\end{align} 
\end{subequations} 
where $v^2 := \bm{v} \cdot \bm{v}$. 

Derivatives of $s^{-1}$ are given by 
\begin{subequations} 
\begin{align} 
\partial_a s^{-1} &= -\frac{1}{s^2} \hat{s}_a, \\ 
\partial_{ab} s^{-1} &= \frac{3}{s^3} \hat{s}_\stf{ab} 
= \frac{3}{s^3} \biggl( \hat{s}_a \hat{s}_b - \frac{1}{3} \delta_{ab}
    \biggr), \\ 
\partial_{abc} s^{-1} &= -\frac{15}{s^4} \hat{s}_\stf{abc} 
= -\frac{15}{s^4} \biggl[ \hat{s}_a \hat{s}_b \hat{s}_c 
- \frac{1}{5} \bigl( \delta_{ab} \hat{s}_c + \delta_{ac} \hat{s}_b
+ \delta_{bc} \hat{s}_a \bigr) \biggr], \\ 
\partial_{abcd} s^{-1} &= \frac{105}{s^5} \hat{s}_\stf{abcd} 
= \frac{105}{s^5} \biggl[ \hat{s}_a \hat{s}_b \hat{s}_c \hat{s}_d  
- \frac{1}{7} \bigl( \delta_{ab} \hat{s}_c \hat{s}_d 
+ \delta_{ac} \hat{s}_b \hat{s}_d + \delta_{ad} \hat{s}_b \hat{s}_c 
+ \delta_{bc} \hat{s}_a \hat{s}_d + \delta_{bd} \hat{s}_a \hat{s}_c  
+ \delta_{cd} \hat{s}_a \hat{s}_b \bigr) 
\nonumber \\ & \quad \mbox{}
+ \frac{1}{35} \bigl( \delta_{ab} \delta_{cd} 
+ \delta_{ac} \delta_{bd} + \delta_{ad} \delta_{bc} \bigr) \biggr],  
\end{align} 
\end{subequations} 
and 
\begin{subequations} 
\begin{align} 
\partial_{ta} s^{-1} &= \frac{1}{s^3} \Bigl( -3v_s \hat{s}_a + v_a \Bigr), \\ 
\partial_{tab} s^{-1} &= \frac{1}{s^4} \Bigl[ 
  15v_s \hat{s}_a \hat{s}_b - 3 (v_a \hat{s}_b + \hat{s}_a v_b) 
- 3 v_s \delta_{ab} \Bigr], \\ 
\partial_{tabc} s^{-1} &= \frac{1}{s^5} \Bigl[ 
  -105v_s \hat{s}_a \hat{s}_b \hat{s}_c 
  + 15 (v_a \hat{s}_b \hat{s}_c + \hat{s}_a v_b \hat{s}_c 
  + \hat{s}_a \hat{s}_b v_c) 
  + 15  v_s (\delta_{ab} \hat{s}_c + \delta_{ac} \hat{s}_b 
  + \delta_{bc} \hat{s}_a) 
\nonumber \\ & \quad \mbox{}
  - 3 (\delta_{ab} v_c + \delta_{ac} v_b + \delta_{bc} v_a)  
\Bigr]. 
\end{align} 
\end{subequations} 

Derivatives of $s$ are given by 
\begin{subequations} 
\begin{align} 
\partial_{ab} s &= 
\frac{1}{s} \Bigl( -\hat{s}_a \hat{s}_b + \delta_{ab} \Bigr), \\  
\partial_{abc} s &= 
\frac{1}{s^2} \Bigl[ 3 \hat{s}_a \hat{s}_b \hat{s}_c 
- ( \delta_{ab} \hat{s}_c + \delta_{ac} \hat{s}_b 
+ \delta_{bc} \hat{s}_a ) \Bigr], \\ 
\partial_{abcd} s &= 
\frac{1}{s^3} \Bigl[ -15 \hat{s}_a \hat{s}_b \hat{s}_c \hat{s}_d 
+ 3 ( \delta_{ab} \hat{s}_c \hat{s}_d 
+ \delta_{ac} \hat{s}_b \hat{s}_d + \delta_{ad} \hat{s}_b \hat{s}_c 
+ \delta_{bc} \hat{s}_a \hat{s}_d + \delta_{bd} \hat{s}_a \hat{s}_c
+ \delta_{cd} \hat{s}_a \hat{s}_b ) 
\nonumber \\ & \quad \mbox{}
- ( \delta_{ab} \delta_{cd} + \delta_{ac} \delta_{bd} 
+ \delta_{ad} \delta_{bc} ) \Bigr]
\end{align} 
\end{subequations} 
and 
\begin{subequations} 
\begin{align} 
\partial_{tab} s &= 
\frac{1}{s^3} \Bigl[ -3 v_s \hat{s}_a \hat{s}_b 
+ ( v_a \hat{s}_b + \hat{s}_a v_b ) 
+ v_s \delta_{ab} \Bigr], \\ 
\partial_{tabc} s &= 
\frac{1}{s^3} \Bigl[ 15 v_s \hat{s}_a \hat{s}_b \hat{s}_c 
- 3( v_a \hat{s}_b \hat{s}_c + \hat{s}_a v_b \hat{s}_c 
+ \hat{s}_a \hat{s}_b v_c )  
-3 v_s ( \delta_{ab} \hat{s}_c + \delta_{ac} \hat{s}_b 
+ \delta_{bc} \hat{s}_a ) 
\nonumber \\ & \quad \mbox{}
+ ( \delta_{ab} v_c + \delta_{ac} v_b + \delta_{bc} v_a ) 
\Bigr], \\ 
\partial_{tabcd} s &= 
\frac{1}{s^4} \Bigl[ -105 v_s \hat{s}_a \hat{s}_b \hat{s}_c \hat{s}_d 
+ 15( v_a \hat{s}_b \hat{s}_c \hat{s}_d 
+ \hat{s}_a v_b \hat{s}_c \hat{s}_d  
+ \hat{s}_a \hat{s}_b v_c \hat{s}_d 
+ \hat{s}_a \hat{s}_b \hat{s}_c v_d )  
\nonumber \\ & \quad \mbox{}
+ 15 v_s ( \delta_{ab} \hat{s}_c \hat{s}_d
+ \delta_{ac} \hat{s}_b \hat{s}_d + \delta_{ad} \hat{s}_b \hat{s}_c
+ \delta_{bc} \hat{s}_a \hat{s}_d + \delta_{bd} \hat{s}_a \hat{s}_c
+ \delta_{cd} \hat{s}_a \hat{s}_b )
\nonumber \\ & \quad \mbox{}
- 6( \delta_{ab} v_{(c} \hat{s}_{d)}
+ \delta_{ac} v_{(b} \hat{s}_{d)} + \delta_{ad} v_{(b} \hat{s}_{c)}
+ \delta_{bc} v_{(a} \hat{s}_{d)} + \delta_{bd} v_{(a} \hat{s}_{c)}
+ \delta_{cd} v_{(a} \hat{s}_{b)} ) 
\nonumber \\ & \quad \mbox{}
- 3 v_s ( \delta_{ab} \delta_{cd} + \delta_{ac} \delta_{bd} 
+ \delta_{ad} \delta_{bc} ) \Bigr]
\end{align} 
\end{subequations} 
and 
\begin{subequations} 
\begin{align} 
\partial_{ttab} s &= \frac{1}{s^2} \Bigl[ 
-3 a_s \hat{s}_a \hat{s}_b 
+ ( a_a \hat{s}_b + \hat{s}_a a_b ) + a_s \delta_{ab} \Bigr] 
\nonumber \\ & \quad \mbox{}
+ \frac{1}{s^3} \Bigl[ 
3(v^2 - 5v_s^2) \hat{s}_a \hat{s}_b 
+ 6 v_s ( v_a \hat{s}_b + \hat{s}_a v_b ) 
- 2v_a v_b - (v^2 - 3v_s^2) \delta_{ab} \Bigr], \\ 
\partial_{ttabc} s &= \frac{1}{s^3} \Bigl[ 
15 a_s \hat{s}_a \hat{s}_b \hat{s}_c 
- 3 ( a_a \hat{s}_b \hat{s}_c + \hat{s}_a a_b \hat{s}_c 
+ \hat{s}_a \hat{s}_b a_c ) 
- 3 a_s (\delta_{ab} \hat{s}_c + \delta_{ac} \hat{s}_b
+ \delta_{bc} \hat{s}_a) 
+ ( \delta_{ab} a_c + \delta_{ac} a_b
+ \delta_{bc} a_a) \Bigr] 
\nonumber \\ & \quad \mbox{}
+ \frac{1}{s^4} \Bigl[ 
-15(v^2 - 7v_s^2) \hat{s}_a \hat{s}_b \hat{s}_c  
- 30 v_s ( v_a \hat{s}_b \hat{s}_c + \hat{s}_a v_b \hat{s}_c 
+ \hat{s}_a \hat{s}_b v_c ) 
+ 3(v^2 - 5v_s^2) (\delta_{ab} \hat{s}_c + \delta_{ac} \hat{s}_b 
+ \delta_{bc} \hat{s}_a ) 
\nonumber \\ & \quad \mbox{}
+ 6( v_a v_b \hat{s}_c + v_a \hat{s}_b v_c + \hat{s}_a v_b v_c) 
+ 6 v_s ( \delta_{ab} v_c + \delta_{ac} v_b + \delta_{bc} v_a ) 
\Bigr], \\
\partial_{ttabcd} s &= \frac{1}{s^4} \Bigl[ 
-105 a_s \hat{s}_a \hat{s}_b \hat{s}_c \hat{s}_d  
+ 15 ( a_a \hat{s}_b \hat{s}_c \hat{s}_d 
+ \hat{s}_a a_b \hat{s}_c \hat{s}_d 
+ \hat{s}_a \hat{s}_b a_c \hat{s}_d 
+ \hat{s}_a \hat{s}_b \hat{s}_c a_d ) 
\nonumber \\ & \quad \mbox{}
+ 15 a_s (\delta_{ab} \hat{s}_c \hat{s}_d
+ \delta_{ac} \hat{s}_b \hat{s}_d + \delta_{ad} \hat{s}_b \hat{s}_c
+ \delta_{bc} \hat{s}_a \hat{s}_d + \delta_{bd} \hat{s}_a \hat{s}_c
+ \delta_{cd} \hat{s}_a \hat{s}_b ) 
\nonumber \\ & \quad \mbox{}
- 12 ( \delta_{ab} a_{(c} \hat{s}_{d)} 
+ \delta_{ac} a_{(b} \hat{s}_{d)} + \delta_{ad} a_{(b} \hat{s}_{c)} 
+ \delta_{bc} a_{(a} \hat{s}_{d)} + \delta_{bd} a_{(a} \hat{s}_{c)} 
+ \delta_{cd} a_{(a} \hat{s}_{b)} )  
- 3 a_s (\delta_{ab} \delta_{cd} + \delta_{ac} \delta_{bd} 
+ \delta_{ad} \delta_{bc} )\Bigr] 
\nonumber \\ & \quad \mbox{}
+ \frac{1}{s^5} \Bigl[ 
105(v^2 - 9v_s^2) \hat{s}_a \hat{s}_b \hat{s}_c \hat{s}_d   
+ 210 v_s ( v_a \hat{s}_b \hat{s}_c \hat{s}_d 
+ \hat{s}_a v_b \hat{s}_c \hat{s}_d  
+ \hat{s}_a \hat{s}_b v_c \hat{s}_d 
+ \hat{s}_a \hat{s}_b \hat{s}_c v_d ) 
\nonumber \\ & \quad \mbox{}
-30 (v_a v_b \hat{s}_c \hat{s}_d + v_a  \hat{s}_b v_c \hat{s}_d 
+ v_a \hat{s}_b \hat{s}_c v_d + \hat{s}_a v_b v_c \hat{s}_d 
+ \hat{s}_a v_b \hat{s}_c v_d + \hat{s}_a \hat{s}_b v_c v_d)  
\nonumber \\ & \quad \mbox{}
- 15(v^2 - 7v_s^2) (\delta_{ab} \hat{s}_c \hat{s}_d 
+ \delta_{ac} \hat{s}_b \hat{s}_d + \delta_{ad} \hat{s}_b \hat{s}_c
+ \delta_{bc} \hat{s}_a \hat{s}_d + \delta_{bd} \hat{s}_a \hat{s}_c
+ \delta_{cd} \hat{s}_a \hat{s}_b ) 
\nonumber \\ & \quad \mbox{}
- 60 v_s ( \delta_{ab} v_{(c} \hat{s}_{d)} 
+ \delta_{ac} v_{(b} \hat{s}_{d)} + \delta_{ad} v_{(b} \hat{s}_{c)} 
+ \delta_{bc} v_{(a} \hat{s}_{d)} + \delta_{bd} v_{(a} \hat{s}_{c)} 
+ \delta_{cd} v_{(a} \hat{s}_{b)} )  
\nonumber \\ & \quad \mbox{}
+ 6( \delta_{ab} v_c v_d + \delta_{ac} v_b v_d + \delta_{ad} v_b v_c 
+ \delta_{bc} v_a v_d + \delta_{bd} v_a v_c + \delta_{cd} v_a v_b ) 
\nonumber \\ & \quad \mbox{}
+ 3(v^2-5v_s^2) (\delta_{ab} \delta_{cd} + \delta_{ac} \delta_{bd} 
+ \delta_{ad} \delta_{bc} ) \Bigr]. 
\end{align} 
\end{subequations}

\section{Tables} 
\label{sec:tables} 

\begin{table}[h]
\caption{Irreducible tidal potentials. The superscripts $\sf q$, $\sf
  o$, and $\sf h$ stand for ``quadrupole,'' ``octupole,'' and
  ``hexadecapole,'' respectively. The factor of 2 in $\EE{h}$ is
  inserted to respect Zhang's normalization convention for
  $\E_{abcd}$; see Eqs.~(\ref{tidalmoment_4}). The factors of
  $\frac{4}{3}$ and $\frac{10}{3}$ in $\BB{o}_a$ and $\BB{h}_a$ 
  are inserted to respect Zhang's normalization convention for
  $\B_{abc}$  and $\B_{abcd}$; see Eqs.~(\ref{tidalmoment_3}) and
  (\ref{tidalmoment_4}).}   
\begin{ruledtabular} 
\begin{tabular}{l} 
$ \EE{q} = \E_{cd} \Omega^c \Omega^d $ \\ 
$ \EE{o} = \E_{cde} \Omega^c \Omega^d \Omega^e$ \\ 
$ \EE{h} = 2\E_{cdef} \Omega^c \Omega^d \Omega^e \Omega^f $ \\ 
\\ 
$ \BB{q}_a = \epsilon_{apq} \Omega^p \B^q_{\ c} \Omega^c $ \\ 
$ \BB{o}_a = \frac{4}{3} \epsilon_{apq} \Omega^p \B^q_{\ cd} \Omega^c
  \Omega^d $ \\ 
$ \BB{h}_a = \frac{10}{3} \epsilon_{apq} \Omega^p \B^q_{\ cde}
  \Omega^c \Omega^d \Omega^e $ \\ 
\\
$ \PP{m} = \E_{pq} \E^{pq} $ \\ 
$ \PP{q} = \E_{p \langle c} \E^p_{\ d \rangle} \Omega^c \Omega^d $ \\ 
$ \PP{h} = \E_{\langle cd} \E_{ef \rangle} \Omega^c \Omega^d \Omega^e
  \Omega^f $ \\ 
\\
$ \QQ{m} = \B_{pq} \B^{pq} $ \\ 
$ \QQ{q} = \B_{p \langle c} \B^p_{\ d \rangle} \Omega^c \Omega^d $ \\ 
$ \QQ{h} = \B_{\langle cd} \B_{ef \rangle} \Omega^c \Omega^d \Omega^e
  \Omega^f $ \\ 
\\
$ \GG{d} = \epsilon_{cpq} \E^p_{\ r} \B^{rq} \Omega^c $ \\ 
$ \GG{o} = \epsilon_{pq\langle c} \E^p_{\ d} \B^q_{\ e \rangle} 
  \Omega^c \Omega^d \Omega^e $ \\ 
\\
$ \HH{q}_a = \epsilon_a^{\ pq} \Omega_p \E_{r \langle q} 
  \B^r_{\ c \rangle} \Omega^c $ \\
$ \HH{h}_a = \epsilon_a^{\ pq} \Omega_p \E_{\langle qc} 
  \B_{de \rangle} \Omega^c \Omega^d \Omega^e $ \\ 
\end{tabular}
\end{ruledtabular} 
\label{tab:potentials} 
\end{table} 

\begingroup
\squeezetable
\begin{table}[h]
\caption{Spherical-harmonic functions $Y^{\ell m}$ and harmonic
  components $\A^{(\ell)}_m$ involved in the decomposition of
  $\A^{(\ell)}  := \A_{\langle k_1 \cdots k_\ell \rangle} 
  \Omega^{k_1} \cdots \Omega^{k_\ell} = \sum_m
  \A^{(\ell)}_m Y^{\ell m}$. The functions are real, and they are listed for
  the relevant modes $\ell=1$ (dipole), $\ell=2$, (quadrupole), $\ell=3$
  (octupole), and $\ell=4$ (hexadecapole). The abstract index $m$
  describes the dependence of these functions on the angle $\phi$; for
  example $Y^{\ell,2s}$ is proportional to $\sin2\phi$. We write 
  $C := \cos\theta$ and $S := \sin\theta$. The harmonic components are
  expressed in terms of the independent components of the STF tensor 
  $\A_{\langle k_1\cdots k_\ell \rangle}$.} 
\begin{ruledtabular} 
\begin{tabular}{ll} 
$ Y^{1,0} = C $ & 
$ \Aa{d}_{0} = \A_3 $ \\ 
$ Y^{1,1c} = S \cos\phi $ &
$ \Aa{d}_{1c} = \A_1 $ \\ 
$ Y^{1,1s} = S \sin\phi $ &  
$ \Aa{d}_{1s} = \A_2 $ \\ 
\\
$ Y^{2,0} = 1-3C^2 $ &
$ \Aa{q}_{0} = \frac{1}{2} (\A_{11} + \A_{22}) $ \\ 
$ Y^{2,1c} = 2SC\cos\phi $ &
$ \Aa{q}_{1c} = \A_{13} $ \\ 
$ Y^{2,1s} = 2SC\sin\phi $ &
$ \Aa{q}_{1s} = \A_{23} $ \\ 
$ Y^{2,2c} = S^2\cos 2\phi $ &
$ \Aa{q}_{2c} = \frac{1}{2} (\A_{11} - \A_{22}) $ \\ 
$ Y^{2,2s} = S^2\sin 2\phi $ & 
$ \Aa{q}_{2s} = \A_{12} $ \\ 
\\ 
$ Y^{3,0} = C(3-5C^2) $ &  
$ \Aa{o}_{0} = \frac{1}{2} (\A_{113}+\A_{223}) $ \\  
$ Y^{3,1c} = \frac{3}{2} S(1-5C^2)\cos\phi $ & 
$ \Aa{o}_{1c} = \frac{1}{2} (\A_{111}+\A_{122}) $ \\  
$ Y^{3,1s} = \frac{3}{2} S(1-5C^2)\sin\phi $ &
$ \Aa{o}_{1s} = \frac{1}{2} (\A_{112}+\A_{222}) $ \\  
$ Y^{3,2c} = 3S^2 C \cos 2\phi $ &
$ \Aa{o}_{2c} = \frac{1}{2} (\A_{113}-\A_{223}) $ \\  
$ Y^{3,2s} = 3S^2 C \sin 2\phi $ & 
$ \Aa{o}_{2s} = \A_{123} $ \\  
$ Y^{3,3c} = S^3 \cos 3\phi $ &
$ \Aa{o}_{3c} = \frac{1}{4} (\A_{111}-3\A_{122}) $ \\  
$ Y^{3,3s} = S^3 \sin 3\phi $ & 
$ \Aa{o}_{3s} = \frac{1}{4} (3\A_{112}-\A_{222}) $ \\  
\\
$ Y^{4,0} = \frac{1}{2} (3-30C^2+35C^4) $ &  
$ \Aa{h}_{0} = \frac{1}{4} (\A_{1111}+2\A_{1122}+\A_{2222}) $ \\ 
$ Y^{4,1c} = 2SC(3-7C^2) \cos\phi $ &
$ \Aa{h}_{1c} = \frac{1}{2} (\A_{1113}+\A_{1223}) $ \\ 
$ Y^{4,1s} = 2SC(3-7C^2) \sin\phi $ &
$ \Aa{h}_{1s} = \frac{1}{2} (\A_{1123}+\A_{2223}) $ \\ 
$ Y^{4,2c} = S^2(1-7C^2) \cos 2\phi $ &
$ \Aa{h}_{2c} = \frac{1}{2} (\A_{1111}-\A_{2222}) $ \\ 
$ Y^{4,2s} = S^2(1-7C^2) \sin 2\phi $ &
$ \Aa{h}_{2s} = \A_{1112}+\A_{1222} $ \\ 
$ Y^{4,3c} = 4S^3 C \cos 3\phi $ &
$ \Aa{h}_{3c} = \frac{1}{4} (\A_{1113}-3\A_{1223}) $ \\ 
$ Y^{4,3s} = 4S^3 C \sin 3\phi $ &
$ \Aa{h}_{3s} = \frac{1}{4} (3\A_{1123}-\A_{2223}) $ \\ 
$ Y^{4,4c} = S^4 \cos 4\phi $ &
$ \Aa{h}_{4c} = \frac{1}{8} (\A_{1111}-6\A_{1122}+\A_{2222}) $ \\ 
$ Y^{4,4s} = S^4 \sin 4\phi $ & 
$ \Aa{h}_{4s} = \frac{1}{2} (\A_{1112}-\A_{1222}) $ 
\end{tabular}
\end{ruledtabular} 
\label{tab:Ylm} 
\end{table} 
\endgroup

\begin{table}[h]
\caption{The first column lists the harmonic components of type-$\E$,
  even-parity tidal potentials, as defined in
  Table~\ref{tab:potentials}. The second column lists their expansions in
  scalar, vector, and tensor harmonics. The components of $\EE{h}$
  come with an additional factor of 2 to accommodate Zhang's choice of
  normalization; see Table~\ref{tab:potentials}.}  
\begin{ruledtabular} 
\begin{tabular}{ll} 
$ \EE{q}_{0} = \frac{1}{2} (\E_{11} + \E_{22}) $ & \\
$ \EE{q}_{1c} = \E_{13} $ &
$ \EE{q} = \sum_m \EE{q}_m Y^{2,m} $ \\ 
$ \EE{q}_{1s} = \E_{23} $ &
$ \EE{q}_A = \frac{1}{2} \sum_m \EE{q}_m Y_A^{2,m} $ \\ 
$ \EE{q}_{2c} = \frac{1}{2} (\E_{11} - \E_{22}) $ & 
$ \EE{q}_{AB} = \sum_m \EE{q}_m Y_{AB}^{2,m} $ \\ 
$ \EE{q}_{2s} = \E_{12} $ & \\ 
\\ 
$ \EE{o}_{0} = \frac{1}{2} (\E_{113}+\E_{223}) $ & \\ 
$ \EE{o}_{1c} = \frac{1}{2} (\E_{111}+\E_{122}) $ & \\ 
$ \EE{o}_{1s} = \frac{1}{2} (\E_{112}+\E_{222}) $ & 
$ \EE{o} = \sum_m \EE{o}_m Y^{3,m} $ \\ 
$ \EE{o}_{2c} = \frac{1}{2} (\E_{113}-\E_{223}) $ & 
$ \EE{o}_A = \frac{1}{3} \sum_m \EE{o}_m Y_A^{3,m} $ \\ 
$ \EE{o}_{2s} = \E_{123} $ &
$ \EE{o}_{AB} = \frac{1}{3} \sum_m \EE{o}_m Y_{AB}^{3,m} $ \\ 
$ \EE{o}_{3c} = \frac{1}{4} (\E_{111}-3\E_{122}) $ & \\ 
$ \EE{o}_{3s} = \frac{1}{4} (3\E_{112}-\E_{222}) $ & \\ 
\\ 
$ \EE{h}_{0} = \frac{1}{2} (\E_{1111}+2\E_{1122}+\E_{2222}) $ & \\ 
$ \EE{h}_{1c} = \E_{1113}+\E_{1223} $ & \\ 
$ \EE{h}_{1s} = \E_{1123}+\E_{2223} $ & \\ 
$ \EE{h}_{2c} = \E_{1111}-\E_{2222} $ & 
$ \EE{h} = \sum_m \EE{h}_m Y^{4,m} $ \\ 
$ \EE{h}_{2s} = 2(\E_{1112}+\E_{1222}) $ & 
$ \EE{h}_A = \frac{1}{4} \sum_m \EE{h}_m Y_A^{4,m} $ \\ 
$ \EE{h}_{3c} = \frac{1}{2} (\E_{1113}-3\E_{1223}) $ & 
$ \EE{h}_{AB} = \frac{1}{6} \sum_m \EE{h}_m Y_{AB}^{4,m} $ \\ 
 $ \EE{h}_{3s} = \frac{1}{2} (3\E_{1123}-\E_{2223}) $ & \\ 
$ \EE{h}_{4c} = \frac{1}{4} (\E_{1111}-6\E_{1122}+\E_{2222}) $ & \\ 
$ \EE{h}_{4s} = \E_{1112}-\E_{1222} $ & \\ 
\end{tabular}
\end{ruledtabular} 
\label{tab:E_ang} 
\end{table} 

\begin{table}[h]
\caption{The first column lists the harmonic components of type-$\B$,
  odd-parity tidal potentials, as defined in
  Table~\ref{tab:potentials}. The second column lists their expansions in
  scalar, vector, and tensor harmonics. The components of $\BB{o}$
  come with an additional factor of $\frac{4}{3}$ to accommodate
  Zhang's choice of normalization, and those of $\BB{h}$ come with an
  additional factor $\frac{10}{3}$; see Table~\ref{tab:potentials}.}  
\begin{ruledtabular} 
\begin{tabular}{ll} 
$ \BB{q}_{0} = \frac{1}{2} (\B_{11} + \B_{22}) $ & \\
$ \BB{q}_{1c} = \B_{13} $ &
$ \BB{q} = \sum_m \BB{q}_m Y^{2,m} $ \\ 
$ \BB{q}_{1s} = \B_{23} $ & 
$ \BB{q}_A = \frac{1}{2} \sum_m \BB{q}_m X_A^{2,m} $ \\ 
$ \BB{q}_{2c} = \frac{1}{2} (\B_{11} - \B_{22}) $ & 
$ \BB{q}_{AB} = \sum_m \BB{q}_m X_{AB}^{2,m} $ \\ 
$ \BB{q}_{2s} = \B_{12} $ & \\ 
\\ 
$ \BB{o}_{0} = \frac{2}{3} (\B_{113}+\B_{223}) $ & \\ 
$ \BB{o}_{1c} = \frac{2}{3} (\B_{111}+\B_{122}) $ & \\ 
$ \BB{o}_{1s} = \frac{2}{3} (\B_{112}+\B_{222}) $ & 
$ \BB{o} = \sum_m \BB{o}_m Y^{3,m} $ \\ 
$ \BB{o}_{2c} = \frac{2}{3} (\B_{113}-\B_{223}) $ & 
$ \BB{o}_A = \frac{1}{3} \sum_m \BB{o}_m X_A^{3,m} $ \\ 
$ \BB{o}_{2s} = \frac{4}{3} \B_{123} $ & 
$ \BB{o}_{AB} = \frac{1}{3} \sum_m \BB{o}_m X_{AB}^{3,m} $ \\ 
$ \BB{o}_{3c} = \frac{1}{3} (\B_{111}-3\B_{122}) $ & \\ 
$ \BB{o}_{3s} = \frac{1}{3} (3\B_{112}-\B_{222}) $ & \\ 
\\ 
$ \BB{h}_{0} = \frac{5}{6} (\B_{1111}+2\B_{1122}+\B_{2222}) $ & \\ 
$ \BB{h}_{1c} = \frac{5}{3}(\B_{1113}+\B_{1223}) $ & \\ 
$ \BB{h}_{1s} = \frac{5}{3}(\B_{1123}+\B_{2223}) $ & \\ 
$ \BB{h}_{2c} = \frac{5}{3}(\B_{1111}-\B_{2222}) $ &
$ \BB{h} = \sum_m \BB{h}_m Y^{4,m} $ \\ 
$ \BB{h}_{2s} = \frac{10}{3}(\B_{1112}+\B_{1222}) $ & 
$ \BB{h}_A = \frac{1}{4} \sum_m \BB{h}_m X_A^{4,m} $ \\ 
$ \BB{h}_{3c} = \frac{5}{6} (\B_{1113}-3\B_{1223}) $ & 
$ \BB{h}_{AB} = \frac{1}{6} \sum_m \BB{h}_m X_{AB}^{4,m} $ \\ 
$ \BB{h}_{3s} = \frac{5}{6} (3\B_{1123}-\B_{2223}) $ & \\ 
$ \BB{h}_{4c} = \frac{5}{12} (\B_{1111}-6\B_{1122}+\B_{2222}) $ & \\ 
$ \BB{h}_{4s} = \frac{5}{3}(\B_{1112}-\B_{1222}) $ & 
\end{tabular}
\end{ruledtabular} 
\label{tab:B_ang} 
\end{table} 

\begin{table}[h]
\caption{Harmonic components of type-$\E\E$, even-parity tidal
  potentials, as defined in Table~\ref{tab:potentials}. The
  spherical-harmonic decompositions are 
  $\PP{q} = \sum_m \PP{q}_m Y^{2,m}$, $\PP{q}_A =
  \frac{1}{2} \sum_m \PP{q}_m Y_A^{2,m}$, $\PP{q}_{AB} = \sum_m
  \PP{q}_m Y_{AB}^{2,m}$, $\PP{h} = \sum_m \PP{h}_m Y^{4,m}$,
  $\PP{h}_A = \frac{1}{4} \sum_m \PP{h}_m Y_A^{4,m}$, and $\PP{h}_{AB}
  = \frac{1}{6} \sum_m \PP{h}_m Y_{AB}^{4,m}$.}
\begin{ruledtabular} 
\begin{tabular}{l} 
$ \PP{m} = 6(\EE{q}_0)^2 + 2(\EE{q}_{1c})^2 + 2(\EE{q}_{1s})^2 
+ 2(\EE{q}_{2c})^2 + 2(\EE{q}_{2s})^2 $ \\
\\
$ \PP{q}_0 = -(\EE{q}_0)^2 - \frac{1}{6} (\EE{q}_{1c})^2 
- \frac{1}{6} (\EE{q}_{1s})^2 + \frac{1}{3} (\EE{q}_{2c})^2 
+ \frac{1}{3} (\EE{q}_{2s})^2 $ \\ 
$ \PP{q}_{1c} = -\EE{q}_0 \EE{q}_{1c} + \EE{q}_{1c} \EE{q}_{2c} 
+ \EE{q}_{1s} \EE{q}_{2s} $ \\  
$ \PP{q}_{1s} = -\EE{q}_0 \EE{q}_{1s} + \EE{q}_{1c} \EE{q}_{2s} 
- \EE{q}_{1s} \EE{q}_{2c} $ \\  
$ \PP{q}_{2c} = 2 \EE{q}_0 \EE{q}_{2c} + \frac{1}{2} (\EE{q}_{1c})^2 
- \frac{1}{2} (\EE{q}_{1s})^2 $ \\ 
$ \PP{q}_{2s} = 2 \EE{q}_0 \EE{q}_{2s} + \EE{q}_{1c} \EE{q}_{1s} $ \\ 
\\
$ \PP{h}_0 = \frac{18}{35} (\EE{q}_0)^2 - \frac{4}{35} (\EE{q}_{1c})^2 
- \frac{4}{35} (\EE{q}_{1s})^2 + \frac{1}{35} (\EE{q}_{2c})^2
+ \frac{1}{35} (\EE{q}_{2s})^2 $ \\ 
$ \PP{h}_{1c} = \frac{6}{7} \EE{q}_0 \EE{q}_{1c} 
+ \frac{1}{7} \EE{q}_{1c} \EE{q}_{2c} 
+ \frac{1}{7} \EE{q}_{1s} \EE{q}_{2s} $ \\ 
$ \PP{h}_{1s} = \frac{6}{7} \EE{q}_0 \EE{q}_{1s} 
+ \frac{1}{7} \EE{q}_{1c} \EE{q}_{2s}  
- \frac{1}{7} \EE{q}_{1s} \EE{q}_{2c} $ \\ 
$ \PP{h}_{2c} = \frac{6}{7} \EE{q}_0 \EE{q}_{2c} 
- \frac{2}{7} (\EE{q}_{1c})^2 
+ \frac{2}{7} (\EE{q}_{1s})^2 $ \\ 
$ \PP{h}_{2s} = \frac{6}{7} \EE{q}_0 \EE{q}_{2s} 
- \frac{4}{7} \EE{q}_{1c} \EE{q}_{1s} $ \\ 
$ \PP{h}_{3c} = \frac{1}{2} \EE{q}_{1c} \EE{q}_{2c} 
- \frac{1}{2} \EE{q}_{1s} \EE{q}_{2s} $ \\ 
$ \PP{h}_{3s} = \frac{1}{2} \EE{q}_{1c} \EE{q}_{2s} 
+ \frac{1}{2} \EE{q}_{1s} \EE{q}_{2c} $ \\ 
$ \PP{h}_{4c} = \frac{1}{2} (\EE{q}_{2c})^2 
- \frac{1}{2} (\EE{q}_{2s})^2 $ \\ 
$ \PP{h}_{4s} = \EE{q}_{2c} \EE{q}_{2s} $
\end{tabular}
\end{ruledtabular} 
\label{tab:EE_ang} 
\end{table} 

\begin{table}[h]
\caption{Harmonic components of type-$\B\B$, even-parity tidal 
  potentials, as defined in Table~\ref{tab:potentials}. The
  spherical-harmonic decompositions are 
  $\QQ{q} = \sum_m \QQ{q}_m Y^{2,m}$, $\QQ{q}_A =
  \frac{1}{2} \sum_m \QQ{q}_m Y_A^{2,m}$, $\QQ{q}_{AB} = \sum_m
  \QQ{q}_m Y_{AB}^{2,m}$, $\QQ{h} = \sum_m \QQ{h}_m Y^{4,m}$,
  $\QQ{h}_A = \frac{1}{4} \sum_m \QQ{h}_m Y_A^{4,m}$, and $\QQ{h}_{AB}
  = \frac{1}{6} \sum_m \QQ{h}_m Y_{AB}^{4,m}$.}
\begin{ruledtabular} 
\begin{tabular}{l} 
$ \QQ{m} = 6(\BB{q}_0)^2 + 2(\BB{q}_{1c})^2 + 2(\BB{q}_{1s})^2 
+ 2(\BB{q}_{2c})^2 + 2(\BB{q}_{2s})^2 $ \\
\\
$ \QQ{q}_0 = -(\BB{q}_0)^2 - \frac{1}{6} (\BB{q}_{1c})^2 
- \frac{1}{6} (\BB{q}_{1s})^2 + \frac{1}{3} (\BB{q}_{2c})^2 
+ \frac{1}{3} (\BB{q}_{2s})^2 $ \\ 
$ \QQ{q}_{1c} = -\BB{q}_0 \BB{q}_{1c} + \BB{q}_{1c} \BB{q}_{2c} 
+ \BB{q}_{1s} \BB{q}_{2s} $ \\  
$ \QQ{q}_{1s} = -\BB{q}_0 \BB{q}_{1s} + \BB{q}_{1c} \BB{q}_{2s} 
- \BB{q}_{1s} \BB{q}_{2c} $ \\  
$ \QQ{q}_{2c} = 2 \BB{q}_0 \BB{q}_{2c} + \frac{1}{2} (\BB{q}_{1c})^2 
- \frac{1}{2} (\BB{q}_{1s})^2 $ \\ 
$ \QQ{q}_{2s} = 2 \BB{q}_0 \BB{q}_{2s} + \BB{q}_{1c} \BB{q}_{1s} $ \\ 
\\
$ \QQ{h}_0 = \frac{18}{35} (\BB{q}_0)^2 - \frac{4}{35} (\BB{q}_{1c})^2 
- \frac{4}{35} (\BB{q}_{1s})^2 + \frac{1}{35} (\BB{q}_{2c})^2
+ \frac{1}{35} (\BB{q}_{2s})^2 $ \\ 
$ \QQ{h}_{1c} = \frac{6}{7} \BB{q}_0 \BB{q}_{1c} 
+ \frac{1}{7} \BB{q}_{1c} \BB{q}_{2c} 
+ \frac{1}{7} \BB{q}_{1s} \BB{q}_{2s} $ \\ 
$ \QQ{h}_{1s} = \frac{6}{7} \BB{q}_0 \BB{q}_{1s} 
+ \frac{1}{7} \BB{q}_{1c} \BB{q}_{2s}  
- \frac{1}{7} \BB{q}_{1s} \BB{q}_{2c} $ \\ 
$ \QQ{h}_{2c} = \frac{6}{7} \BB{q}_0 \BB{q}_{2c} 
- \frac{2}{7} (\BB{q}_{1c})^2 
+ \frac{2}{7} (\BB{q}_{1s})^2 $ \\ 
$ \QQ{h}_{2s} = \frac{6}{7} \BB{q}_0 \BB{q}_{2s} 
- \frac{4}{7} \BB{q}_{1c} \BB{q}_{1s} $ \\ 
$ \QQ{h}_{3c} = \frac{1}{2} \BB{q}_{1c} \BB{q}_{2c} 
- \frac{1}{2} \BB{q}_{1s} \BB{q}_{2s} $ \\ 
$ \QQ{h}_{3s} = \frac{1}{2} \BB{q}_{1c} \BB{q}_{2s} 
+ \frac{1}{2} \BB{q}_{1s} \BB{q}_{2c} $ \\ 
$ \QQ{h}_{4c} = \frac{1}{2} (\BB{q}_{2c})^2 
- \frac{1}{2} (\BB{q}_{2s})^2 $ \\ 
$ \QQ{h}_{4s} = \BB{q}_{2c} \BB{q}_{2s} $
\end{tabular}
\end{ruledtabular} 
\label{tab:BB_ang} 
\end{table} 

\begin{table}[h]
\caption{Harmonic components of type-$\E\B$, even-parity tidal 
  potentials, as defined in Table~\ref{tab:potentials}. The
  spherical-harmonic decompositions are 
  $\GG{d} = \sum_m \GG{d}_m Y^{1,m}$, $\GG{d}_A =
  \sum_m \GG{d}_m Y_A^{1,m}$, $\GG{o} = \sum_m \GG{o}_m Y^{3,m}$,
  $\GG{o}_A = \frac{1}{3} \sum_m \GG{o}_m Y_A^{3,m}$, and $\GG{o}_{AB}
  = \frac{1}{3} \sum_m \GG{o}_m Y_{AB}^{3,m}$.}
\begin{ruledtabular} 
\begin{tabular}{l} 
$ \GG{d}_0 = \EE{q}_{1c} \BB{q}_{1s} - \EE{q}_{1s} \BB{q}_{1c}
+ 2\EE{q}_{2c} \BB{q}_{2s} - 2\EE{q}_{2s} \BB{q}_{2c} $ \\ 
$ \GG{d}_{1c} = (3\EE{q}_0 - \EE{q}_{2c}) \BB{q}_{1s} 
- \EE{q}_{1c} \BB{q}_{2s} - \EE{q}_{1s} (3\BB{q}_{0} 
- \BB{q}_{2c}) + \EE{q}_{2s} \BB{q}_{1c} $ \\ 
$ \GG{d}_{1s} = -(3\EE{q}_0 + \EE{q}_{2c}) \BB{q}_{1c} 
+ \EE{q}_{1c} (3\BB{q}_{0} + \BB{q}_{2c})
+ \EE{q}_{1s} \BB{q}_{2s} - \EE{q}_{2s} \BB{q}_{1s} $ \\ 
\\
$ \GG{o}_0 = -\frac{2}{5} \EE{q}_{1c} \BB{q}_{1s}
+ \frac{2}{5} \EE{q}_{1s} \BB{q}_{1c} 
+ \frac{1}{5} \EE{q}_{2c} \BB{q}_{2s}
- \frac{1}{5} \EE{q}_{2s} \BB{q}_{2c} $ \\ 
$ \GG{o}_{1c} = -\frac{2}{5} \EE{q}_0 \BB{q}_{1s} 
- \frac{1}{5} \EE{q}_{1c} \BB{q}_{2s} 
+ \frac{1}{5} \EE{q}_{1s} (2\BB{q}_{0} + \BB{q}_{2c})  
- \frac{1}{5} \EE{q}_{2c} \BB{q}_{1s} 
+ \frac{1}{5} \EE{q}_{2s} \BB{q}_{1c} $ \\ 
$ \GG{o}_{1s} = \frac{2}{5} \EE{q}_0 \BB{q}_{1c} 
- \frac{1}{5} \EE{q}_{1c} (2\BB{q}_{0} - \BB{q}_{2c}) 
+ \frac{1}{5} \EE{q}_{1s} \BB{q}_{2s} 
- \frac{1}{5} \EE{q}_{2c} \BB{q}_{1c} 
- \frac{1}{5} \EE{q}_{2s} \BB{q}_{1s} $ \\ 
$ \GG{o}_{2c} = \EE{q}_0 \BB{q}_{2s} - \EE{q}_{2s} \BB{q}_0 $ \\ 
$ \GG{o}_{2s} = -\EE{q}_0 \BB{q}_{2c} + \EE{q}_{2c} \BB{q}_0 $ \\
$ \GG{o}_{3c} = -\frac{1}{2} \EE{q}_{1c} \BB{q}_{2s}  
- \frac{1}{2} \EE{q}_{1s} \BB{q}_{2c}
+ \frac{1}{2} \EE{q}_{2c} \BB{q}_{1s}
+ \frac{1}{2} \EE{q}_{2s} \BB{q}_{1c} $ \\ 
$ \GG{o}_{3s} = \frac{1}{2} \EE{q}_{1c} \BB{q}_{2c}  
- \frac{1}{2} \EE{q}_{1s} \BB{q}_{2s}
- \frac{1}{2} \EE{q}_{2c} \BB{q}_{1c}
+ \frac{1}{2} \EE{q}_{2s} \BB{q}_{1s} $ 
\end{tabular}
\end{ruledtabular} 
\label{tab:EBeven_ang} 
\end{table} 

\begin{table}[h]
\caption{Harmonic components of type-$\E\B$, odd-parity tidal 
  potentials, as defined in Table~\ref{tab:potentials}. The
  spherical-harmonic decompositions are 
  $\HH{q}_A = \frac{1}{2} \sum_m \HH{q}_m
  X_A^{2,m}$, $\HH{q}_{AB} = \sum_m \HH{q}_m X_{AB}^{2,m}$, $\HH{h}_A
  = \frac{1}{4} \sum_m \HH{h}_m X_A^{4,m}$, and $\HH{h}_{AB} =
  \frac{1}{6} \sum_m \HH{h}_m X_{AB}^{4,m}$.}
\begin{ruledtabular} 
\begin{tabular}{l} 
$ \HH{q}_0 = -\EE{q}_{0} \BB{q}_{0} 
- \frac{1}{6} \EE{q}_{1c} \BB{q}_{1c} 
- \frac{1}{6} \EE{q}_{1s} \BB{q}_{1s} 
+ \frac{1}{3} \EE{q}_{2c} \BB{q}_{2c}
+ \frac{1}{3} \EE{q}_{2s} \BB{q}_{2s} $ \\
$ \HH{q}_{1c} = -\frac{1}{2} \EE{q}_{0} \BB{q}_{1c} 
- \frac{1}{2} \EE{q}_{1c} ( \BB{q}_{0} - \BB{q}_{2c} ) 
+ \frac{1}{2} \EE{q}_{1s} \BB{q}_{2s} 
+ \frac{1}{2} \EE{q}_{2c} \BB{q}_{1c} 
+ \frac{1}{2} \EE{q}_{2s} \BB{q}_{1s} $ \\ 
$ \HH{q}_{1s} = -\frac{1}{2} \EE{q}_{0} \BB{q}_{1s} 
+ \frac{1}{2} \EE{q}_{1c} \BB{q}_{2s} 
- \frac{1}{2} \EE{q}_{1s} ( \BB{q}_{0} + \BB{q}_{2c} ) 
- \frac{1}{2} \EE{q}_{2c} \BB{q}_{1s} 
+ \frac{1}{2} \EE{q}_{2s} \BB{q}_{1c} $ \\ 
$ \HH{q}_{2c} = \EE{q}_{0} \BB{q}_{2c} 
+ \frac{1}{2} \EE{q}_{1c} \BB{q}_{1c} 
- \frac{1}{2} \EE{q}_{1s} \BB{q}_{1s} 
+ \EE{q}_{2c} \BB{q}_{0} $ \\  
$ \HH{q}_{2s} = \EE{q}_{0} \BB{q}_{2s} 
+ \frac{1}{2} \EE{q}_{1c} \BB{q}_{1s} 
+ \frac{1}{2} \EE{q}_{1s} \BB{q}_{1c} 
+ \EE{q}_{2s} \BB{q}_{0} $ \\  
\\ 
$ \HH{h}_0 = \frac{18}{35} \EE{q}_0 \BB{q}_0 
- \frac{4}{35} \EE{q}_{1c} \BB{q}_{1c} 
- \frac{4}{35} \EE{q}_{1s} \BB{q}_{1s} 
+ \frac{1}{35} \EE{q}_{2c} \BB{q}_{2c} 
+ \frac{1}{35} \EE{q}_{2s} \BB{q}_{2s} $ \\ 
$ \HH{h}_{1c} = \frac{3}{7} \EE{q}_0 \BB{q}_{1c} 
+ \frac{1}{14} \EE{q}_{1c} ( 6\BB{q}_{0} + \BB{q}_{2c} ) 
+ \frac{1}{14} \EE{q}_{1s} \BB{q}_{2s} 
+ \frac{1}{14} \EE{q}_{2c} \BB{q}_{1c}
+ \frac{1}{14} \EE{q}_{2s} \BB{q}_{1s} $ \\ 
$ \HH{h}_{1s} = \frac{3}{7} \EE{q}_0 \BB{q}_{1s} 
+ \frac{1}{14} \EE{q}_{1c} \BB{q}_{2s} 
+ \frac{1}{14} \EE{q}_{1s} ( 6\BB{q}_{0} - \BB{q}_{2c} ) 
- \frac{1}{14} \EE{q}_{2c} \BB{q}_{1s}
+ \frac{1}{14} \EE{q}_{2s} \BB{q}_{1c} $ \\ 
$ \HH{h}_{2c} = \frac{3}{7} \EE{q}_0 \BB{q}_{2c} 
- \frac{2}{7} \EE{q}_{1c} \BB{q}_{1c} 
+ \frac{2}{7} \EE{q}_{1s} \BB{q}_{1s} 
+ \frac{3}{7} \EE{q}_{2c} \BB{q}_{0} $ \\ 
$ \HH{h}_{2s} = \frac{3}{7} \EE{q}_0 \BB{q}_{2s} 
- \frac{2}{7} \EE{q}_{1c} \BB{q}_{1s} 
- \frac{2}{7} \EE{q}_{1s} \BB{q}_{1c} 
+ \frac{3}{7} \EE{q}_{2s} \BB{q}_{0} $ \\ 
$ \HH{h}_{3c} = \frac{1}{4} \EE{q}_{1c} \BB{q}_{2c} 
- \frac{1}{4} \EE{q}_{1s} \BB{q}_{2s} 
+ \frac{1}{4} \EE{q}_{2c} \BB{q}_{1c} 
- \frac{1}{4} \EE{q}_{2s} \BB{q}_{1s} $ \\
$ \HH{h}_{3s} = \frac{1}{4} \EE{q}_{1c} \BB{q}_{2s} 
+ \frac{1}{4} \EE{q}_{1s} \BB{q}_{2c} 
+ \frac{1}{4} \EE{q}_{2c} \BB{q}_{1s} 
+ \frac{1}{4} \EE{q}_{2s} \BB{q}_{1c} $ \\
$ \HH{h}_{4c} = \frac{1}{2} \EE{q}_{2c} \BB{q}_{2c} 
- \frac{1}{2} \EE{q}_{2s} \BB{q}_{2s} $ \\ 
$ \HH{h}_{4s} = \frac{1}{2} \EE{q}_{2c} \BB{q}_{2s} 
+ \frac{1}{2} \EE{q}_{2s} \BB{q}_{2c} $   
\end{tabular}
\end{ruledtabular} 
\label{tab:EBodd_ang} 
\end{table} 

\begin{table}[h] 
\caption{Radial functions in $(v,r,\theta^A)$ coordinates: 
type-$\cal E$ potentials. The functions are expressed in terms of 
$x := r/M$ and  $f := 1-2/x$. The dilogarithm function is defined as 
$\dilog(z) =  \int_1^z dy\, \ln(y)/(1-y)$, with $\ln(z)$ denoting
the natural logarithm. The constants $\ce{q}{1}$, $\ce{q}{2}$, and
$\ce{o}{1}$ are arbitrary constants of integration, whose meaning is
explained in the text. In $\eeddot{q}{vv}$, the function
$\eedot{q}{vv}$ is evaluated with $\ce{q}{1}=0$; a similar statement
holds for $\eeddot{q}{vr}$, $\eeddot{q}{rr}$, and $\eeddot{q}{}$.}    
\begin{ruledtabular} 
\begin{tabular}{l} 
$ \ee{q}{vv} = -f^2 $ \\ 
$ \ee{q}{vr} = f $ \\ 
$ \ee{q}{rr} = -2 $ \\ 
$ \ee{q}{} = -1 + \frac{2}{x^2} $ \\ 
$ \eedot{q}{vv} = \frac{1}{x} [\ce{q}{1} - 2\ln(x/2)] \ee{q}{vv} + 1 
- \frac{112}{15x} + \frac{208}{15x^2} - \frac{28}{15x^3} 
- \frac{16}{3x^4} - \frac{8}{3x^5} $ \\ 
$ \eedot{q}{vr} = \frac{1}{x} [\ce{q}{1} - 2\ln(x/2)] \ee{q}{vr} 
- \frac{5}{3}  + \frac{77}{15x} - \frac{8}{5x^2} - \frac{10}{3x^3} 
- \frac{4}{3x^4} $ \\ 
$ \eedot{q}{rr} = \frac{1}{x} [\ce{q}{1} - 2\ln(x/2)] \ee{q}{rr} 
+ \frac{10}{3}  - \frac{18}{5x} - \frac{4}{x^2} - \frac{4}{3x^3} $ \\ 
$ \eedot{q}{} = \frac{1}{x} [\ce{q}{1} - 2\ln(x/2)] \ee{q}{} + 1 
- \frac{52}{15x} - \frac{6}{x^2} + \frac{44}{15x^3} 
+ \frac{8}{3x^4} $ \\ 
$ \eeddot{q}{vv} = \frac{1}{x^2} [ \ce{q}{2} - 4 \dilog(x/2) ]
  \ee{q}{vv} + \frac{1}{x} \ce{q}{1}\, \eedot{q}{vv}(\ce{q}{1}=0) 
- f^2 \bigl( \frac{2}{x} - \frac{514}{105x^2} \bigr) \ln(x/2) $ \\
$ \qquad \quad \mbox{} 
- \frac{16}{21} + \frac{1517}{315x} 
- \frac{4616}{315x^2} + \frac{302}{45x^3} + \frac{2986}{105x^4} 
- \frac{4112}{315x^5} - \frac{2056}{315x^6} $ \\ 
$ \eeddot{q}{vr} = \frac{1}{x^2} [ \ce{q}{2} - 4 \dilog(x/2) ]
  \ee{q}{vr}  + \frac{1}{x} \ce{q}{1}\, \eedot{q}{vr}(\ce{q}{1}=0) 
+ \frac{1}{f} \bigl( \frac{10}{3x}  - \frac{1564}{105x^2} 
+ \frac{2476}{105x^3} - \frac{1636}{105x^4} \bigr) \ln(x/2) $\\
$ \qquad \quad \mbox{} 
+ \frac{10}{7} - \frac{332}{63x} + \frac{102}{35x^2} 
+ \frac{2956}{315x^3} - \frac{514}{63x^4} - \frac{1028}{315x^5} $ \\  
$ \eeddot{q}{rr} = \frac{1}{x^2} [ \ce{q}{2} - 4 \dilog(x/2) ]
  \ee{q}{rr} + \frac{1}{x} \ce{q}{1}\, \eedot{q}{rr}(\ce{q}{1}=0)
- \frac{1}{f^2} \bigl( \frac{20}{3x}  - \frac{3128}{105x^2} 
+ \frac{4952}{105x^3} - \frac{3272}{105x^4} \bigr) \ln(x/2) $ \\
$ \qquad \quad \mbox{} 
- \frac{1}{f} \bigl( \frac{20}{7} - \frac{664}{63x} 
+ \frac{204}{35x^2}  + \frac{5912}{315x^3} - \frac{1028}{63x^4} 
- \frac{2056}{315x^5} \bigr) $ \\ 
$ \eeddot{q}{} = \frac{1}{x^2} [ \ce{q}{2} - 4 \dilog(x/2) ]
  \ee{q}{} + \frac{1}{x} \ce{q}{1}\, \eedot{q}{}(\ce{q}{1}=0)
- \bigl(1 - \frac{2}{x^2}\bigr) \bigl( \frac{2}{x}  - \frac{514}{105x^2}
  \bigr) \ln(x/2) $ \\ 
$ \qquad \quad \mbox{} 
- \frac{25}{42} + \frac{842}{315x} - \frac{263}{315x^2}  
- \frac{1892}{105x^3} - \frac{338}{35x^4} + \frac{2056}{315x^5} $ \\ 
\\
$ \ee{o}{vv} = -\frac{1}{3} f^2 (1 - \frac{1}{x}) $ \\ 
$ \ee{o}{vr} = \frac{1}{3} f (1 - \frac{1}{x}) $ \\ 
$ \ee{o}{rr} = -\frac{2}{3} (1 - \frac{1}{x}) $ \\ 
$ \ee{o}{} = -\frac{1}{3} (1 - \frac{2}{x} + \frac{4}{5x^3}) $ \\ 
$ \eedot{o}{vv} = \frac{1}{x} [\ce{o}{1} - 2\ln(x/2)] \ee{o}{vv} 
+ \frac{1}{3} - \frac{65}{21x} + \frac{178}{21x^2} - \frac{52}{7x^3} 
- \frac{4}{63x^4} + \frac{8}{9x^5} + \frac{8}{45x^6} $ \\ 
$ \eedot{o}{vr} = \frac{1}{x} [\ce{o}{1} - 2\ln(x/2)] \ee{o}{vr} 
- \frac{1}{2} + \frac{167}{63x} - \frac{221}{63x^2} + \frac{46}{315x^3} 
+ \frac{22}{45x^4} + \frac{4}{45x^5} $ \\ 
$ \eedot{o}{rr} = \frac{1}{x} [\ce{o}{1} - 2\ln(x/2)] \ee{o}{rr} 
+ 1 - \frac{208}{63x} + \frac{26}{63x^2} + \frac{8}{15x^3} 
+ \frac{4}{45x^4} $ \\ 
$ \eedot{o}{} = \frac{1}{x} [\ce{o}{1} - 2\ln(x/2)] \ee{o}{} 
+ \frac{1}{3} - \frac{44}{21x} + \frac{32}{21x^2} + \frac{8}{5x^3} 
- \frac{16}{63x^4} - \frac{8}{45x^5} $ \\ 
\\ 
$ \ee{h}{vv} = -\frac{1}{12} f^2 (1 - \frac{2}{x} + \frac{6}{7x^2}) $ \\   
$ \ee{h}{vr} = \frac{1}{12} f (1 - \frac{2}{x} + \frac{6}{7x^2}) $ \\   
$ \ee{h}{rr} = -\frac{1}{6} (1 - \frac{2}{x} + \frac{6}{7x^2}) $ \\   
$ \ee{h}{} = -\frac{1}{12}( 1 - \frac{10}{3x} + \frac{20}{7x^2} 
- \frac{8}{21x^4} ) $ 
\end{tabular}
\end{ruledtabular} 
\label{tab:e_functions_vr} 
\end{table} 

\begin{table}[h] 
\caption{Radial functions in $(v,r,\theta^A)$ coordinates: 
type-$\cal B$ potentials. The functions are expressed in terms of 
$x := r/M$ and  $f := 1-2/x$. The dilogarithm function is defined as 
$\dilog(z) =  \int_1^z dy\, \ln(y)/(1-y)$, with $\ln(z)$ denoting
the natural logarithm. The constants $\cb{q}{1}$, $\cb{q}{2}$, and
$\cb{o}{1}$ are arbitrary constants of integration, whose meaning is
explained in the text. In $\bbddot{q}{v}$, the function
$\bbdot{q}{v}$ is evaluated with $\cb{q}{1}=0$; a similar statement
holds for $\bbddot{q}{r}$.}    
\begin{ruledtabular} 
\begin{tabular}{l} 
$ \bb{q}{v} = \frac{2}{3} f $ \\ 
$ \bb{q}{r} = -\frac{2}{3}  $ \\ 
$ \bbdot{q}{v} = \frac{1}{x} [ \cb{q}{1} - 2\ln(x/2)] \bb{q}{v} 
- \frac{2}{3} + \frac{49}{15x} - \frac{6}{5x^2} - \frac{8}{3x^3} 
- \frac{16}{9x^4} - \frac{16}{9x^5} $ \\ 
$ \bbdot{q}{r} = \frac{1}{x} [ \cb{q}{1} - 2\ln(x/2)] \bb{q}{r} 
+ \frac{5}{6} - \frac{8}{5x} - \frac{2}{x^2} - \frac{4}{3x^3} 
- \frac{8}{9x^4} $ \\ 
$ \bbddot{q}{v} = \frac{1}{x^2} [ \cb{q}{2} - 4 \dilog(x/2) ]
  \bb{q}{v} + \frac{1}{x} \cb{q}{1}\, \bbdot{q}{v}(\cb{q}{1}=0) 
+ f \bigl( \frac{4}{3x} - \frac{158}{63x^2} \bigr) \ln(x/2) $ \\
$ \qquad \quad \mbox{} 
+ \frac{17}{42} - \frac{1403}{630x} + \frac{479}{126x^2} 
+ \frac{38}{7x^3} - \frac{344}{63x^4} - \frac{632}{189x^5} 
- \frac{632}{189x^6} $ \\ 
$ \bbddot{q}{r} = \frac{1}{x^2} [ \cb{q}{2} - 4 \dilog(x/2) ]
  \bb{q}{r} + \frac{1}{x} \cb{q}{1}\, \bbdot{q}{r}(\cb{q}{1}=0) 
- \frac{1}{f} \bigl( \frac{5}{3x} - \frac{326}{63x^2} 
+ \frac{316}{63x^3} \bigr) \ln(x/2) $ \\
$ \qquad \quad \mbox{} 
- \frac{4}{7} + \frac{395}{252x} - \frac{100}{21x^3} 
- \frac{200}{63x^4} - \frac{316}{189x^5} $ \\ 
\\ 
$ \bb{o}{v} = \frac{1}{4} f \bigl( 1 - \frac{4}{3x} \bigr) $ \\ 
$ \bb{o}{r} = -\frac{1}{4} \bigl( 1 - \frac{4}{3x} \bigr) $ \\ 
$ \bbdot{o}{v} = \frac{1}{x} [ \cb{o}{1} - 2\ln(x/2)] \bb{o}{v} 
- \frac{1}{4} + \frac{1579}{840x} - \frac{265}{84x^2} 
+ \frac{16}{35x^3} + \frac{2}{3x^4} + \frac{2}{9x^5} 
+ \frac{4}{45x^6}$ \\ 
$ \bbdot{o}{r} = \frac{1}{x} [ \cb{o}{1} - 2\ln(x/2)] \bb{o}{r} 
+ \frac{3}{10} - \frac{229}{168x} + \frac{3}{7x^2} 
+ \frac{2}{5x^3} + \frac{2}{15x^4} + \frac{2}{45x^5} $ \\
\\ 
$ \bb{h}{v} = \frac{1}{15} f \bigl(1 - \frac{5}{2x} + \frac{10}{7x^2}
  \bigr) $ \\ 
$ \bb{h}{r} = -\frac{1}{15} \bigl(1 - \frac{5}{2x} + \frac{10}{7x^2}
  \bigr) $ 
\end{tabular}
\end{ruledtabular} 
\label{tab:b_functions_vr} 
\end{table} 

\begin{table}[h] 
\caption{Radial functions in $(v,r,\theta^A)$ coordinates: 
type-${\cal P}$ and type-${\cal Q}$ potentials. The functions are
expressed in terms of $x := r/M$ and $f := 1-2/x$. The constants
$\cp{m}$, $\cp{q}$, $\cp{h}$, $\cq{m}$, $\cq{q}$, and $\cq{h}$ 
are arbitrary constants of integration, whose meaning is explained in
the text.}      
\begin{ruledtabular} 
\begin{tabular}{ll} 
$ \pp{m}{vv} = \frac{2}{x^5} \cp{m} - f^2 \bigl( \frac{1}{15} 
+ \frac{2}{75x} - \frac{4}{25x^2} + \frac{32}{225x^3} 
+ \frac{32}{225x^4} \bigr) $ &
$ \qq{m}{vv} = \frac{2}{x^5} \cq{m} + f^3 \bigl( \frac{1}{15} 
+ \frac{2}{25x} + \frac{2}{25x^2} + \frac{4}{75x^3} \bigr) $ \\
$ \pp{m}{vr} = -\frac{2}{x^5 f} \cp{m} + f \bigl( \frac{1}{15} 
+ \frac{2}{75x} - \frac{4}{25x^2} + \frac{32}{225x^3} 
+ \frac{32}{225x^4} \bigr) $ & 
$ \qq{m}{vr} = -\frac{2}{x^5 f} \cq{m} - f^2 \bigl( \frac{1}{15} 
+ \frac{2}{25x} + \frac{2}{25x^2} + \frac{4}{75x^3} \bigr) $ \\
$ \pp{m}{rr} = \frac{4}{x^5 f^2} \cp{m} - 2 \bigl( \frac{1}{15} 
+ \frac{2}{75x} - \frac{4}{25x^2} + \frac{32}{225x^3} 
+ \frac{32}{225x^4} \bigr) $ & 
$ \qq{m}{rr} = \frac{4}{x^5 f^2} \cq{m} + 2f \bigl( \frac{1}{15} 
+ \frac{2}{25x} + \frac{2}{25x^2} + \frac{4}{75x^3} \bigr) $ \\
$ \pp{m}{} = \frac{2}{75} - \frac{8}{45x^2} + \frac{128}{225x^5}  $ & 
$ \qq{m}{} = \frac{2}{225} - \frac{2}{45x} + \frac{32}{75x^5} $ \\ 
& \\
$ \pp{q}{vv} = \frac{1}{x^2} \cp{q}\, \ee{q}{vv} 
- f^2 \bigl( \frac{2}{7} + \frac{6}{7x} - \frac{8}{7x^2} \bigr) $ &
$ \qq{q}{vv} = \frac{1}{x^2} \cq{q}\, \ee{q}{vv} 
+ f \bigl( \frac{2}{7} - \frac{2}{3x} \bigr) $ \\ 
$ \pp{q}{vr} = \frac{1}{x^2} \cp{q}\, \ee{q}{vr} 
+ f \bigl( \frac{2}{7} + \frac{6}{7x} - \frac{8}{7x^2} \bigr) $ &
$ \qq{q}{vr} = \frac{1}{x^2} \cq{q}\, \ee{q}{vr} 
- \frac{2}{7} + \frac{2}{3x} $ \\ 
$ \pp{q}{rr} = \frac{1}{x^2} \cp{q}\, \ee{q}{rr} 
- \frac{6}{7} - \frac{8}{7x} + \frac{16}{7x^2} $ &
$ \qq{q}{rr} = \frac{1}{x^2} \cq{q}\, \ee{q}{rr} 
+ \frac{34}{21} $ \\ 
$ \pp{q}{} = \frac{1}{x^2} \cp{q}\, \ee{q}{} 
- \frac{1}{14} + \frac{12}{7x^4} $ & 
$ \qq{q}{} = \frac{1}{x^2} \cq{q}\, \ee{q}{} 
+ \frac{1}{2} - \frac{4}{7x} $ \\ 
& \\ 
$ \pp{h}{vv} = \cp{h}\, \ee{h}{vv} 
+ f^2 \bigl( \frac{367}{36} - \frac{164}{9x} + \frac{7}{x^2} \bigr) $&
$ \qq{h}{vv} = \cq{h}\, \ee{h}{vv} 
- f \bigl( \frac{17}{36} - \frac{49}{54x} \bigr) $ \\ 
$ \pp{h}{vr} = \cp{h}\, \ee{h}{vr} 
- f \bigl( \frac{367}{36} - \frac{164}{9x} + \frac{7}{x^2} \bigr) $ &
$ \qq{h}{vr} = \cq{h}\, \ee{h}{vr} 
+ \frac{17}{36} - \frac{49}{54x} $ \\ 
$ \pp{h}{rr} = \cp{h}\, \ee{h}{rr} 
+ \frac{379}{18} - \frac{340}{9x} + \frac{14}{x^2} $ &
$ \qq{h}{rr} = \cq{h}\, \ee{h}{rr} 
- \frac{29}{54} $ \\ 
$ \pp{h}{} = \cp{h}\, \ee{h}{} 
+ \frac{391}{36} - \frac{868}{27x} + \frac{206}{9x^2} $ &
$ \qq{h}{} = \cq{h}\, \ee{h}{} 
- \frac{7}{108} + \frac{1}{27x} $ 
\end{tabular}
\end{ruledtabular} 
\label{tab:pq_functions_vr} 
\end{table} 

\begin{table}[h] 
\caption{Radial functions in $(v,r,\theta^A)$ coordinates: 
type-${\cal G}$ and type-${\cal H}$ potentials. The functions are
expressed in terms of $x := r/M$ and $f := 1-2/x$. The constants
$\cg{o}$, $\ch{q}$, and $\ch{h}$ are arbitrary constants of
integration, whose meaning is explained in the text.}     
\begin{ruledtabular} 
\begin{tabular}{ll} 
$ \g{d}{vv} = 0 $ & \\ 
$ \g{d}{vr} = f \bigl( \frac{2}{15} - \frac{8}{15x} \bigr) $ & 
$ \hh{q}{v} = \frac{1}{x^2} \ch{q}\, \bb{q}{v} 
+ f \bigl( \frac{4}{21} + \frac{16}{21x} + \frac{32}{21x^2} \bigr) $ \\ 
$ \g{d}{rr} = -\frac{4}{15} + \frac{16}{15x} $ & 
$ \hh{q}{r} = \frac{1}{x^2} \ch{q}\, \bb{q}{r} 
- \bigl( \frac{4}{21} + \frac{16}{21x} + \frac{32}{21x^2} \bigr) $ \\ 
$ \g{d}{} = 0 $ & \\ 
& \\ 
$ \g{o}{vv} = \frac{1}{x} \cg{o}\, \ee{o}{vv} $ & \\
$ \g{o}{vr} = \frac{1}{x} \cg{o}\, \ee{o}{vr}  
+ f \bigl( \frac{1}{9} - \frac{4}{9x} \bigr) $ & 
$ \hh{h}{v} = \ch{h}\, \bb{h}{v} 
- f \bigl( \frac{8}{3x} - \frac{64}{21x^2} \bigr) $ \\ 
$ \g{o}{rr} = \frac{1}{x} \cg{o}\, \ee{o}{rr}  
- 2 \bigl( \frac{1}{9} - \frac{4}{9x} \bigr) $ & 
$ \hh{h}{r} = \ch{h}\, \bb{h}{r} 
+ \frac{8}{3x} - \frac{64}{21x^2} $ \\ 
$ \g{o}{} = \frac{1}{x} \cg{o}\, \ee{o}{} $ & 
\end{tabular}
\end{ruledtabular} 
\label{tab:gh_functions_vr} 
\end{table} 

\begin{table}[h] 
\caption{Radial functions in $(t,r,\theta^A)$ coordinates: 
type-$\cal E$ potentials. The functions are expressed in terms of 
$x := r/M$ and  $f := 1-2/x$.}    
\begin{ruledtabular} 
\begin{tabular}{l} 
$ \ee{q}{tt} = -f^2$ \\ 
$ \ee{q}{rr} = -1 $ \\ 
$ \ee{q}{} = -1 + \frac{2}{x^2} $ \\ 
$ \eedot{q}{tt} = \frac{1}{x} [\ce{q}{1} + 2\ln(f)] \ee{q}{tt} 
- \frac{52}{15x} + \frac{148}{15x^2} - \frac{28}{15x^3} 
- \frac{16}{3x^4} - \frac{8}{3x^5} $ \\ 
$ \eedot{q}{tr} = -\frac{1}{f} \bigl( \frac{2}{3}  - \frac{1}{x} 
- \frac{2}{x^2} + \frac{2}{x^3} \bigr) $ \\ 
$ \eedot{q}{rr} = \frac{1}{x} [\ce{q}{1} + 2\ln(f)] \ee{q}{rr} 
- \frac{1}{f^2} \bigl( \frac{52}{15x} - \frac{148}{15x^2} 
+ \frac{28}{15x^3} + \frac{16}{3x^4} + \frac{8}{3x^5} \bigr) $ \\ 
$ \eedot{q}{} = \frac{1}{x} [\ce{q}{1} + 2\ln(f)] \ee{q}{} 
- \frac{52}{15x} - \frac{4}{x^2} + \frac{44}{15x^3} + \frac{8}{3x^4} $ \\ 
$ \eeddot{q}{tt} = \frac{1}{x^2} [ \ce{q}{2} - 4 \dilog(x/2) 
- 2 \ln^2(x/2) + 2 \ln^2(f)] \ee{q}{tt} 
+ \frac{1}{x} \ce{q}{1}\, \eedot{q}{tt}(\ce{q}{1}=0) 
- \bigl( \frac{214}{105x^2} - \frac{16}{105x^3} - \frac{1664}{105x^4} 
+ \frac{32}{3x^5} + \frac{16}{3x^6} \bigr) \ln(x/2) $ \\ 
$ \qquad \quad \mbox{} 
- \bigl( \frac{104}{15x^2} - \frac{296}{15x^3} + \frac{56}{15x^4} 
+ \frac{32}{3x^5} + \frac{16}{3x^6} \bigr) \ln(f) 
- \frac{11}{42} - \frac{41}{63x} 
- \frac{878}{315x^2} + \frac{218}{45x^3} + \frac{2426}{105x^4} 
- \frac{4952}{315x^5} - \frac{2056}{315x^6} $ \\ 
$ \eeddot{q}{tr} = \frac{1}{x} [ \ce{q}{1} + 2 \ln(f) ] \eedot{q}{tr} 
- \frac{1}{f} \bigl( \frac{104}{45x}  - \frac{4}{5x^2} 
- \frac{124}{15x^3} - \frac{68}{45x^4} \bigr) $ \\
$ \eeddot{q}{rr} = \frac{1}{x^2} [ \ce{q}{2} - 4 \dilog(x/2) 
- 2 \ln^2(x/2) + 2 \ln^2(f)] \ee{q}{rr} 
+ \frac{1}{x} \ce{q}{1}\, \eedot{q}{rr}(\ce{q}{1}=0) 
- \frac{1}{f^2} \bigl( \frac{214}{105x^2} - \frac{16}{105x^3} - \frac{1664}{105x^4} 
+ \frac{32}{3x^5} + \frac{16}{3x^6} \bigr) \ln(x/2) $ \\ 
$ \qquad \quad \mbox{} 
- \frac{1}{f^2} \bigl( \frac{104}{15x^2} - \frac{296}{15x^3} + \frac{56}{15x^4} 
+ \frac{32}{3x^5} + \frac{16}{3x^6} \bigr) \ln(f) 
- \frac{1}{f^2} \Bigl( \frac{11}{42} + \frac{41}{63x} 
+ \frac{878}{315x^2} - \frac{218}{45x^3} - \frac{2426}{105x^4} 
+ \frac{4952}{315x^5} + \frac{2056}{315x^6} \Bigr) $ \\ 
$ \eeddot{q}{} = \frac{1}{x^2} [ \ce{q}{2} - 4 \dilog(x/2) 
- 2 \ln^2(x/2) + 2 \ln^2(f)] \ee{q}{} 
+ \frac{1}{x} \ce{q}{1}\, \eedot{q}{}(\ce{q}{1}=0) 
- \bigl( \frac{214}{105x^2} + \frac{8}{x^3} + \frac{412}{105x^4} 
- \frac{16}{3x^5} \bigr) \ln(x/2) $ \\ 
$ \qquad \quad \mbox{} 
- \bigl( \frac{104}{15x^2} + \frac{8}{x^3} - \frac{88}{15x^4} 
- \frac{16}{3x^5} \bigr) \ln(f) 
- \frac{2}{21} - \frac{50}{63x} - \frac{1838}{315x^2} 
- \frac{528}{35x^3} - \frac{734}{105x^4} + \frac{2056}{315x^5}  $ \\ 
\\
$ \ee{o}{tt} = -\frac{1}{3} f^2 (1 - \frac{1}{x}) $ \\ 
$ \ee{o}{rr} = -\frac{1}{3} (1 - \frac{1}{x}) $ \\ 
$ \ee{o}{} = -\frac{1}{3} (1 - \frac{2}{x} + \frac{4}{5x^3}) $ \\ 
$ \eedot{o}{tt} = \frac{1}{x} [\ce{o}{1} + 2\ln(f)] \ee{o}{tt} 
- \frac{10}{7x} + \frac{122}{21x^2} - \frac{128}{21x^3} 
- \frac{4}{63x^4} + \frac{8}{9x^5} + \frac{8}{45x^6} $ \\ 
$ \eedot{o}{tr} = - \frac{1}{f} \bigl( \frac{1}{6} - \frac{5}{9x} 
+ \frac{1}{3x^2} + \frac{4}{15x^3} - \frac{2}{15x^4} \bigr) $ \\ 
$ \eedot{o}{rr} = \frac{1}{x} [\ce{o}{1} + 2\ln(f)] \ee{o}{rr} 
- \frac{1}{f^2} \bigl( \frac{10}{7x} - \frac{122}{21x^2} 
+ \frac{128}{21x^3} + \frac{4}{63x^4} - \frac{8}{9x^5} 
- \frac{8}{45x^6} \bigr) $ \\ 
$ \eedot{o}{} = \frac{1}{x} [\ce{o}{1} + 2\ln(f)] \ee{o}{} 
- \frac{10}{7x} + \frac{32}{21x^2} + \frac{4}{3x^3} 
- \frac{16}{63x^4} - \frac{8}{45x^5} $ \\ 
\\ 
$ \ee{h}{tt} = -\frac{1}{12} f^2 (1 - \frac{2}{x} + \frac{6}{7x^2}) $ \\   
$ \ee{h}{rr} = -\frac{1}{12}\bigl( 1 - \frac{2}{x} + \frac{6}{7x^2} \bigr) $ \\   
$ \ee{h}{} = -\frac{1}{12} \bigl( 1 -  \frac{10}{3x} + \frac{20}{7x^2} 
- \frac{8}{21x^4} \bigr) $ 
\end{tabular}
\end{ruledtabular} 
\label{tab:e_functions_tr} 
\end{table} 

\begin{table}[h] 
\caption{Radial functions in $(t,r,\theta^A)$ coordinates: 
type-$\cal B$ potentials. The functions are expressed in terms of 
$x := r/M$ and  $f := 1-2/x$.}    
\begin{ruledtabular} 
\begin{tabular}{l} 
$ \bb{q}{t} = \frac{2}{3} f $ \\ 
$ \bbdot{q}{t} = \frac{1}{x} [ \cb{q}{1} + 2\ln(f)] \bb{q}{t} 
+ \frac{29}{15x} - \frac{6}{5x^2} - \frac{8}{3x^3} 
- \frac{16}{9x^4} - \frac{16}{9x^5} $ \\ 
$ \bbdot{q}{r} = \frac{1}{6f} $ \\ 
$ \bbddot{q}{t} = \frac{1}{x^2} [ \cb{q}{2} - 4 \dilog(x/2) 
- 2 \ln^2(x/2) + 2 \ln^2(f)] \bb{q}{t} 
+ \frac{1}{x} \cb{q}{1}\, \bbdot{q}{t}(\cb{q}{1}=0) 
+ \bigl( \frac{428}{315x^2} + \frac{824}{315x^3} - \frac{16}{3x^4} 
- \frac{32}{9x^5} - \frac{32}{9x^6} \bigr) \ln(x/2) $ \\ 
$ \qquad \quad \mbox{} 
+ \bigl( \frac{58}{15x^2} - \frac{12}{5x^3} - \frac{16}{3x^4} 
- \frac{32}{9x^5} - \frac{32}{9x^6} \bigr) \ln(f) 
+ \frac{1}{14} + \frac{47}{126x} 
+ \frac{1639}{630x^2} + \frac{58}{21x^3} - \frac{152}{21x^4} 
- \frac{968}{189x^5} - \frac{632}{189x^6} $ \\ 
$ \bbddot{q}{r} = \frac{1}{x} [ \cb{q}{1} + 2 \ln(f) ] \bbdot{q}{r} 
+ \frac{1}{f} \bigl( \frac{29}{60x}  + \frac{2}{3x^2} 
+ \frac{2}{3x^3} + \frac{8}{9x^4} + \frac{4}{3x^5} \bigr) $ \\
\\ 
$ \bb{o}{t} = \frac{1}{4} f \bigl( 1 - \frac{4}{3x} \bigr) $ \\ 
$ \bbdot{o}{t} = \frac{1}{x} [ \cb{o}{1} + 2\ln(f)] \bb{o}{t} 
+ \frac{293}{280x} - \frac{209}{84x^2} 
+ \frac{16}{35x^3} + \frac{2}{3x^4} + \frac{2}{9x^5} 
+ \frac{4}{45x^6}$ \\ 
$ \bbdot{o}{r} = \frac{1}{f} \bigl( \frac{1}{20} 
- \frac{1}{12x} \bigr) $ \\ 
\\ 
$ \bb{h}{t} = \frac{1}{15} f \bigl(1 - \frac{5}{2x} + \frac{10}{7x^2}
  \bigr) $ 
\end{tabular}
\end{ruledtabular} 
\label{tab:b_functions_tr} 
\end{table} 

\begin{table}[h] 
\caption{Radial functions in $(t,r,\theta^A)$ coordinates: 
type-${\cal P}$ and type-${\cal Q}$ potentials. The functions are
expressed in terms of $x := r/M$ and $f := 1-2/x$.}      
\begin{ruledtabular} 
\begin{tabular}{ll} 
$ \pp{m}{tt} = \frac{2}{x^5} \cp{m} - f^2 \bigl( \frac{1}{15} 
+ \frac{2}{75x} - \frac{4}{25x^2} + \frac{32}{225x^3} 
+ \frac{32}{225x^4} \bigr) $ &
$ \qq{m}{tt} = \frac{2}{x^5} \cq{m} + f^3 \bigl( \frac{1}{15} 
+ \frac{2}{25x} + \frac{2}{25x^2} + \frac{4}{75x^3} \bigr) $ \\
$ \pp{m}{rr} = \frac{2}{x^5 f^2} \cp{m} - \frac{1}{15} 
- \frac{2}{75x} + \frac{4}{25x^2} - \frac{32}{225x^3} 
- \frac{32}{225x^4} \bigr) $ & 
$ \qq{m}{rr} = \frac{2}{x^5 f^2} \cq{m} + \frac{1}{15} 
- \frac{4}{75x} - \frac{2}{25x^2} - \frac{8}{75x^3} 
- \frac{8}{75x^4}  $ \\
$ \pp{m}{} = \frac{2}{75} - \frac{8}{45x^2} + \frac{128}{225x^5}  $ & 
$ \qq{m}{} = \frac{2}{225} - \frac{2}{45x} + \frac{32}{75x^5} $ \\ 
& \\
$ \pp{q}{tt} = \frac{1}{x^2} \cp{q}\, \ee{q}{tt} 
- f^2 \bigl( \frac{2}{7} + \frac{6}{7x} - \frac{8}{7x^2} \bigr) $ &
$ \qq{q}{tt} = \frac{1}{x^2} \cq{q}\, \ee{q}{tt} 
+ f \bigl( \frac{2}{7} - \frac{2}{3x} \bigr) $ \\ 
$ \pp{q}{rr} = \frac{1}{x^2} \cp{q}\, \ee{q}{rr} 
- \frac{4}{7} - \frac{2}{7x} + \frac{8}{7x^2} $ &
$ \qq{q}{rr} = \frac{1}{x^2} \cq{q}\, \ee{q}{rr} 
+ \frac{1}{f} \bigl( \frac{4}{3} - \frac{18}{7x} \bigr) $ \\ 
$ \pp{q}{} = \frac{1}{x^2} \cp{q}\, \ee{q}{} 
- \frac{1}{14} + \frac{12}{7x^4} $ & 
$ \qq{q}{} = \frac{1}{x^2} \cq{q}\, \ee{q}{} 
+ \frac{1}{2} - \frac{4}{7x} $ \\ 
& \\ 
$ \pp{h}{tt} = \cp{h}\, \ee{h}{tt} 
+ f^2 \bigl( \frac{367}{36} - \frac{164}{9x} + \frac{7}{x^2} \bigr) $&
$ \qq{h}{tt} = \cq{h}\, \ee{h}{tt} 
- f \bigl( \frac{17}{36} - \frac{49}{54x} \bigr) $ \\ 
$ \pp{h}{rr} = \cp{h}\, \ee{h}{rr} 
+ \frac{391}{36} - \frac{176}{9x} + \frac{7}{x^2} $ &
$ \qq{h}{rr} = \cq{h}\, \ee{h}{rr} 
- \frac{1}{f} \bigl( \frac{7}{108} - \frac{1}{6x} \bigr) $ \\ 
$ \pp{h}{} = \cp{h}\, \ee{h}{} 
+ \frac{391}{36} - \frac{868}{27x} + \frac{206}{9x^2} $ &
$ \qq{h}{} = \cq{h}\, \ee{h}{} 
- \frac{7}{108} + \frac{1}{27x} $ 
\end{tabular}
\end{ruledtabular} 
\label{tab:pq_functions_tr} 
\end{table} 

\begin{table}[h] 
\caption{Radial functions in $(t,r,\theta^A)$ coordinates: 
type-${\cal G}$ and type-${\cal H}$ potentials. The functions are
expressed in terms of $x := r/M$ and $f := 1-2/x$.}     
\begin{ruledtabular} 
\begin{tabular}{ll} 
$ \g{d}{tt} = 0 $ & \\ 
$ \g{d}{tr} = f \bigl( \frac{2}{15} - \frac{8}{15x} \bigr) $ & 
$ \hh{q}{t} = \frac{1}{x^2} \ch{q}\, \bb{q}{t} 
+ f \bigl( \frac{4}{21} + \frac{16}{21x} + \frac{32}{21x^2} \bigr) $ \\ 
$ \g{d}{rr} = 0 $ & 
$ \hh{q}{r} = 0 $ \\ 
$ \g{d}{} = 0 $ & \\ 
& \\ 
$ \g{o}{tt} = \frac{1}{x} \cg{o}\, \ee{o}{tt} $ & \\
$ \g{o}{tr} = f \bigl( \frac{1}{9} - \frac{4}{9x} \bigr) $ & 
$ \hh{h}{t} = \ch{h}\, \bb{h}{t} 
- f \bigl( \frac{8}{3x} - \frac{64}{21x^2} \bigr) $ \\ 
$ \g{o}{rr} = \frac{1}{x} \cg{o}\, \ee{o}{rr} $ & 
$ \hh{h}{r} = 0 $ \\ 
$ \g{o}{} = \frac{1}{x} \cg{o}\, \ee{o}{} $ & 
\end{tabular}
\end{ruledtabular} 
\label{tab:gh_functions_tr} 
\end{table} 

\clearpage
\bibliography{../bib/master}

\end{document}